\documentclass[aps,11pt,eqsecnum, preprint,nofootinbib,superscriptaddress]{revtex4}
\usepackage{amssymb,amsmath,amsthm,graphicx,amscd}
\usepackage{enumerate,color,changepage,ulem,bm}
\usepackage[hidelinks]{hyperref}

\begin{document}

\title{Fluctuation-Dissipation Relation for a Quantum Brownian Oscillator \\ in a Parametrically Squeezed Thermal Field}

\author{Jen-Tsung Hsiang}
\email{cosmology@gmail.com}
\affiliation{Center for High Energy and High Field Physics, National Central University, Taoyuan 320317, Taiwan, ROC}
\author{Bei-Lok Hu}
\email{blhu@umd.edu}
\affiliation{Maryland Center for Fundamental Physics and Joint Quantum Institute,  University of Maryland, College Park, Maryland 20742, USA}

\begin{abstract}
In this paper we study the nonequilibrium evolution of a quantum Brownian oscillator,  modeling  the internal degree of freedom of  a harmonic atom or an Unruh-DeWitt detector, coupled to a nonequilibrium and nonstationary quantum field bath  and inquire whether a fluctuation-dissipation relation (FDR) can exist after/if  it approaches equilibration.  This is a nontrivial issue because  a squeezed field bath cannot reach equilibration and yet, as this work shows,  the system oscillator indeed can, which is a necessary condition for FDRs.  We discuss three different settings:  A) The bath field essentially remains in a squeezed thermal state throughout, whose squeeze parameter is a mode- and time-independent constant.  This situation  is often encountered in quantum optics and quantum thermodynamics.  B)  The bath field is initially in a thermal state, but is subjected to a parametric process  leading to  mode- and time-dependent squeezing.  This scenario is   encountered in cosmology and dynamical Casimir effects. The squeezing in the bath in both types of processes will affect  the oscillator's nonequilibrium evolution. We show that at late times it approaches equilibration and this stationarity condition warrants the existence of a FDR.  The trait of squeezing is marked by the oscillator's effective equilibrium temperature, and the proportionality factor   in  the FDR is only related to the stationary component of the noise kernel of the bath field.  Setting C)  is more subtle:  A finite system-bath coupling strength can set the  oscillator in a squeezed state  even though the bath field is stationary and does not engage in any parametric process. The squeezing of the system in this case is in general time-dependent but becomes constant when the internal dynamics is fully relaxed.  We begin with comments on the broad range of physical processes involving  squeezed thermal baths and end with some remarks on the significance of FDRs in capturing the essence of quantum backreaction in nonequilibrium and stochastic systems.
\end{abstract}

\maketitle

\hypersetup{linktoc=all}

\tableofcontents
\clearpage
\baselineskip=18pt
\numberwithin{equation}{section}

\allowdisplaybreaks

\newpage

\section{Introduction}
Fluctuation-dissipation relations (FDR) \cite{FDT} are of fundamental significance  in many areas of physics as they describe the exact balance between the  dissipative dynamics of a system and the fluctuations in its environment. Since the system-environment interplay is at the heart of theories of open systems (TOS) \cite{qos}, the existence and nature of FDRs  are important foundational issues. The difference is, the FDRs are often framed and phrased in the setting of linear response theories (LRT) \cite{LRT} under near-equilibrium conditions, while TOS entail the fully nonequilibrium (NEq) evolution \cite{QTD1,HHPRD15,HHAoP15}  of the system, including nonMarkovian dynamics and colored noises \cite{HPZ,HalYu96,CRV}.  If one wants to access and assess the FDRs from the nonequilibrium dynamics perspective, one needs to first identify the conditions where these relations can exist, and if they do, how they emerge from a time-dependent setting, what are their physical contents and how they differ from the more familiar FDRs obtained in LRT. Understanding the attributes and functionalities of the FDRs is a foundational issue in TOS. 

Yet, one may wonder,  what does one gain doing it in this ostensibly harder way, and what difference does it make?  We shall answer this question in a simplified manner before we turn to the main subjects of this paper. 

\subsection{Advantages in connecting equilibrium conditions to  nonequilibrium processes} 
We mention two cases here, one experimental, one  theoretical, for illustration.  A more detailed description of the differences between the NEq and the LRT approaches can be found in Sec. 3 of \cite{QRad}.

1) Advances in the ability to perform \textit{real-time measurements} in high-precision  experiments in many-body quantum systems 
have added significant observational values in the study of nonequilibrium dynamics and open quantum systems. But many traditional conceptual schemes and methodologies, like the use of Gibbs ensembles, well adapted to stationary or equilibrium conditions, are no longer applicable. The study of how NEq dynamics connects with near-equilibrium phenomena, such as via the existence of FDRs, offers some way to `siphon'  useful resources plentiful in equilibrium statistical mechanics to enrich the vast and hitherto barren terrains of the NEq landscape.  

2) Expounding the connection between equilibrium and near-equilibrium states to the NEq processes which can evolve to these states in limiting conditions has  many advantages.   It is difficult to calculate  properties of  many physical quantities  defined along arbitrary trajectories in phase space or  paths in $P$-$V$ diagrams. But not so for state functions, which do not depend on the processes under arbitrary nonequilibrium dynamics, only on the end points in the phase space or in a $P$-$V$ diagram. A well-known important example is the \textit{fluctuation theorems} \cite{FlucThm} relating the nonequilibrium work function to the difference of the free energy under equilibrium conditions. Through this relation, the former quantity, which is more accessible experimentally, can be used to obtain the latter quantity. 

Granted, one may still say that being able to connect to well-established facts in equilibrium statistical mechanics is more to the benefit of nonequilibrium physics.  Thankfully so.  However, in this and our recent papers one can see the benefits go in both directions, namely,  showing that stationarity conditions exist at late times in a quantum system's nonequilibrium evolution and proving the generalized fluctuation-dissipation relations (e.g., \cite{RHA,CPR,FDRNL}) add significant substance to the limited contents of LRT while  empowering the reaches of these  relations (e.g., even to the somewhat esoteric realm of cosmological backreaction problems \cite{HuSin95,CamVer96} in semiclassical gravity~\cite{HuVer20}.

\subsection{FDRs in Nonequilibrium Open Quantum Systems}
Besides i) the NEq dynamics vis a vis the near-equilibrium conditions, two more factors enter in our considerations, namely, ii) strong coupling between the system and its environment, which LRT cannot provide as it is restricted to weak coupling; and the fact that iii) we want FDRs both for the environment -- a squeezed thermal field here, and for the system -- a quantum Brownian oscillator.  Traditional FDRs are mostly shown for the environment such as a thermal bath, but less so for the system. The two sets of relations are connected because  the system oscillator is coupled to the environment field.  Note that the existence of an FDR in the field does not automatically imply the existence of an FDR in the oscillator dynamics.  For instance, it is easy to show that  an FDR exists in an unsqueezed thermal field. It takes some extra effort to show the existence of a FDR for the system oscillator.  However, as we shall show in this paper, there is no FDR in a squeezed thermal field, and yet, somewhat surprisingly, an FDR exists in the oscillator at late times.  

 When we say it is easy to extract an FDR in a system coupled to a thermal quantum field --  two caveats are attached to `easy'.  We are dealing with a dynamical field and a system in a NEq setting, not near-equilibrium, and we want to show these relations exist under strong system-field coupling, which  many traditionally invoked perturbative methods are of no avail.  

The procedure goes schematically as follows:  Start with the two fundamental Green's functions of the field, known as the dissipation and the noise kernels. For Gaussian systems they are the causal Green's function and the Hadamard function \cite{RHK97}.  These two-point functions of free field differ in the protocol of time-ordering of the field operators, so we can straightforwardly identify their relations if they can be Fourier-transformed to the frequency domain. This is how we write down the FDR for the free thermal field, and the vacuum field as a special case.  After this,  we derive a quantum Langevin equation for the system, the Brownian oscillator, which   contains the dissipation and the noise kernels of the free field. We examine  the system's evolution dynamics and its long time behavior  to determine if the system can relax and equilibrate. We then use certain criteria like proving the power balance between the noise input and the dissipation output of the system to show that a stationarity condition is reached, and from there we establish an FDR for our system.

\subsection{Models, Methods, Issues}
Let us see in more detail the procedures in a specific yet generic model, that  of a point-like object with internal degrees of freedom described by a  quantum oscillator,  our system,  interacting with a bath made up of $M$-harmonic oscillators or a quantum field, our environment.  Our system could be a harmonic atom in the context of atomic-optical physics or an Unruh-DeWitt detector  \cite{Unr76,DeW79} in the context of relativistic quantum information \cite{RQI}.   

\subsubsection{Models and Approaches}
For the \textit{quantum Brownian motion} of a harmonic oscillator of a fixed natural frequency in a general environment of $M$-oscillators, also of fixed frequencies, an exact nonMarkovian master equation, known as the Hu-Paz-Zhang equation, has been derived \cite{HPZ} which is valid for all temperatures and spectral densities of the bath and for arbitrary coupling strengths.  It includes as subcase the Markovian Caldeira-Leggett \cite{CalLeg83} master equation valid for high temperature Ohmic baths. A Fokker-Planck equation for the Wigner function corresponding to the HPZ master equation  is derived in \cite{HalYu96}. A quantum Langevin equation \cite{QLE} governing the reduced system variables is derived in \cite{CRV} and in \cite{QTD1} (see references in these paper for other related work) by way of  the covariance matrix method.  

We now let the natural frequencies of the system oscillator and of the oscillators making up its thermal bath to be {\it time-dependent}. The master equations for the reduced density matrix of a parametric quantum oscillator in a squeezed thermal bath have  been derived by Hu and Matacz~\cite{HM94}. The theoretical framework thus established is well suited for \textit{squeezed open quantum systems}~\cite{KMH97}. We shall treat the same problem with a squeezed thermal scalar field as the bath and derive a quantum Langevin equation equivalent to the HPZ equation for parametric oscillators to investigate the conditions for possible existence of an FDR for the system oscillator. 
   
\subsubsection{Prior work on FDR via NEq methods and Issues explored}
The models we have investigated for FDRs consist of a system of one to $N$ quantum oscillator(s) linearly coupled to one or two quantum scalar field bath(s). Our methodology based on the influence functional formalism and stochastic effective actions is explained in Appendix A of~\cite{QTD1}. The procedures we formulated for obtaining the FDRs are illustrated for the one harmonic oscillator - one scalar field bath case  in Sec. III  of  \cite{QTD1}. For one \textit{anharmonic} quantum oscillator we have also provided a nonperturbative proof \cite{FDRNL}, that equilibration implies an FDR in this type of nonlinear system. 

For systems comprised of \textit{two linearly coupled oscillators} we have proven the existence of an FDR a) when they are in a common quantum field \cite{HHPRD15} and b) when each oscillator is coupled to its own thermal bath at different temperatures \cite{HHAoP15} after the system settles into a nonequilibrium steady state. We have also identified suites of FDRs and  correlation-propagation relations (CPRs) first discovered in \cite{RHA} for a system of \textit{N oscillators moving in a common field} \cite{CPR}.

The following basic issues are discussed in our earlier papers: \\
1) Limitations of \textit{FDRs obtained from LRT}  {in} \cite{FDRNL}, Sec. IA \\
2) A touch of \textit{physical insight}, an explanation of the delicate balance between  quantum fluctuations, quantum dissipation and quantum radiation as different from classical radiation and radiation reaction \cite{QRad}.\\
3) The \textit{nonMarkovian effects} in a system of N oscillators -- often referred to as Unruh-DeWitt detectors \cite{Unr76,DeW79}  in relativistic quantum information \cite{RQI} -- in a common quantum field due to the field-mediated influences of one oscillator on another is studied in detail for $N=2$ in \cite{HHPRD15} and for general $N$ in \cite{RHA,CPR}.  \\
4) The symmetry relations between the FDRs in each oscillator and the CPRs between oscillators pairwise, and the combined \textit{generalized matrix FDR} in \cite{CPR}. 
 
\subsection{FDRs for systems in a squeezed quantum  field}

With this as the background for our program of investigations, we now focus on the new issues in this paper,  namely, on the nonequilibrium dynamics of, and the FDRs in, a quantum system interacting with arbitrary strength with a \textit{squeezed thermal bath}. We introduce the three players and the four acts in the arena: The players are i) a quantum oscillator/detector, ii) a finite temperature quantum field, iii) a drive which gives time-dependence to the natural frequencies of the normal modes of the quantum field, or, in short, squeezing the field. The four acts  are a) the squeezing of a thermal field, b) the nonequilibrium dynamics of the field c) the  detector's dynamical response, and d) the FDRs for the field and for the detector. We shall not dwell on the familiar subjects which have books written about them: \textit{squeezed states} in quantum optics, e.g., \cite{Walls,LouKni,ManWol};  \textit{nonequilibrium dynamics} in dissipative and open quantum systems \cite{qos}. \textit{FDR from LRT} is also well covered by many excellent reviews since Kubo, e.g., \cite{LRT}.  The challenge here is to apply b) to find c) and d) in the setting of a).   
We shall employ the techniques in nonequilibrium quantum field theory \cite{NEqFT} for this purpose. 


Three situations we have studied where squeezing sets in are:

\begin{enumerate}[A)]

	\item The bath is initially squeezed by a {\it  mode-independent, constant squeeze} parameter.  Not only does the bath remain nonstationary and nonequilibrium in time during the entire history,   but the correlation of the bath is also not invariant under spatial translation.  An example is the squeezed thermal bath in a quantum Otto engine, as exemplified in \cite{QOtto} (see also references therein).

	\item  The bath is initially in a thermal state, but  subsequently an external drive changes the natural frequencies of the system. Such a time-dependent (parametric) process  introduces two-mode squeezing to the bath.  Well known examples are cosmological particle creation~\cite{Par69,Zel70,BirDav}  and dynamical Casimir effect~\cite{DCE}.   This is discussed in a recent cosmology paper \cite{UDWcos} (see references therein) which explores the conditions for memories in the field to be retained as a record of the expanding universe over its entire history.
Here, the {\it squeeze parameter is both time-dependent and mode-dependent}. After the parametric process stops, the squeeze parameter reduces to a mode-dependent constant. Thus the bath field remains nonstationary and nonequilibrium even at the end of the process. In contrast, the correlation of the bath field is translationally-invariant in space. It is under these rather adverse conditions that the somewhat surprising outcome from our present investigation -- that an FDR indeed exists in the system oscillator -- takes on added significance.  

\item {\it Finite coupling} between the system and the bath. Finite refers to nonvanishing, in contradistinction to the very weak coupling assumption in the definition of statistical mechanical ensembles and linear response theories.  An example of this type can be found in our recent work \cite{NEqFE}.  Here, the bath field is not initial squeezed. If the field describes a thermal bath,  the bath field is stationary both in time and space. Nonetheless the system can still acquire  squeezing due to finite system-bath coupling strength before it reaches relaxation. Distinguished from the  two following cases, this squeezing is not passed on to the system from the bath, but obtains {\it a posteriori} via the interaction, so this squeezing will not show up in the FDR of the system.

\end{enumerate}

These cases all  cause the system which the bath interacts with to acquire squeezing during the intermediate stage of the system evolution, but after the system dynamics is fully relaxed, the squeezing in the system essentially disappears except that it modifies the effective equilibrium temperature of the system.

The treatment we shall present covers all three cases.  For Case B),  without great loss of generality,  we shall assume   a statically-bounded evolution between an asymptotic  in-state and an asymptotic out-state of the field with varying frequency in between. 

Since full stationarity is a pre-condition for the existence of FDRs, we need to prove that an asymptotically constant drive allows for a stationarity condition  in the detector. Because of the time-dependent drives,  FDRs in a detector measuring a squeezed thermal quantum field is not a simple or straightforward generalization of FDRs for a thermal bath.

\subsection{Areas of Applicability}

Squeezed thermal fields have a long history and are quite commonly encountered in current research. We shall mention six areas and touch on how a study of nonequilibrium dynamics and FDRs may bear on furthering our understandings in them. They are 1) gravitation and cosmology, 2) dynamical Casimir effect and analog gravity, 3) relativistic quantum information (RQI),  4) quantum radiation and dissipation, 5) quantum friction and 6) heat engines. We shall briefly describe these processes saving the discussion of backreaction effects in 1)  2)  4) and 5) represented by FDRs to the last section.      
			

\subsubsection{Particle creation in the early universe and black holes} 

The expansion of the universe acts like a drive, parametrically amplifying the vacuum fluctuations leading to cosmological particle creation \cite{Par69,Zel70}. We can view spontaneous particle creation as the result of the vacuum being `squeezed' in the evolutionary history of the universe~\cite{GriSid}. A summary description of cosmological particle creation in terms of squeezing can be found in \cite{HKM94}. 

The Unruh effect \cite{Unr76} -- thermal radiance in an uniformly accelerated detector,  or the Hawking effect \cite{Haw74} -- thermal radiation emitted from a black hole, can be understood as the result of the event horizon present in these cases coarse-graining information from the region of space hidden by the horizon.  The Bogoliubov  transformation relating two different vacua used in the derivation of both cosmological and black hole particle creation \cite{BirDav} is a form of squeezing of the vacuum \cite{Par75,Wald75}.

\subsubsection{Dynamical Casimir effect and analog gravity}

Casimir force on two parallel conducting plates originates from the difference in the pressure exerted on the plates by the vacuum fluctuations of quantum fields in the spatial region inside and outside the plates. If one or both of the plates are moved nonadiabatically particle pairs will be created. This is known as \textit{dynamical Casimir effect} \cite{DCE}. The physical mechanism of parametric amplification applies here,  as in cosmological particle creation, the `drive' being the expansion of the universe. Thermal radiation from a moving mirror \cite{DavFul} in a specified trajectory is another famous analog of Hawking effect.  

By applying the  well established knowledge base about squeezing in quantum optics one can observe or design experiments simulating quantum processes in the early universe (e.g., \cite{CalHu04}) and in black holes (e.g., \cite{Garay}). This is the spirit of analog gravity \cite{analogG} invoking the similarity of the key physical processes and  the  commonality of the underlying issues. Perhaps the best known, the simplest and most direct analogs albeit not the easiest to implement, are  Unruh radiance in the uniformly accelerated detector for Hawking effect and the dynamical Casimir effect \cite{DCE} for cosmological particle creation.


\subsubsection{Detectors as probes of properties of fields}

The quantum Brownian oscillator, our system, studied in \cite{HPZ,HM94},  taken as a harmonic atom or an Unruh-DeWitt detector,   can be used to probe into the quantum state of the field, our environment. This is what Unruh  \cite{Unr76} did, using the response of an uniformly accelerated detector to understand the physics behind the Hawking effect. Casting it in an open quantum system theoretical framework, Hu, Koks, Matacz and Raval have applied the HPZ master equations for a parametric oscillator in a squeezed thermal bath to a range of problems from particle creation to entropy generation. In addition to the familiar cases of thermal  radiance in an accelerated detector -- the Unruh effect, and in an observer in a (static) de Sitter universe (Gibbons-Hawking) \cite{GibHaw}, the thermal radiation emitted from a 2D black hole, a moving mirror and a collapsing shell, these authors have also extended to cases emitting or experiencing near-thermal radiance in nearly-uniformly accelerated detectors or finite-time acceleration \cite{RHK97} and near-exponential cosmological expansions \cite{KHMR}. 

Changing the trajectory of a detector from uniform acceleration evokes nonequilibrium physics \cite{NEqUnr}. 
This regime is out of reach by the traditional geometric methods relying on the existence of event horizons and where the kinematical approach based on nonequilibrium quantum field theory shows its broader utility.  One can assign an arbitrary trajectory for the detector's motion and  from its response function extract useful information about the quantum field environment it interacts with.
Even with a stationary atom, its interactions with quantum fields in different states lead to interesting phenomena. For example,   the quantum radiation emitted \textsl{by} the atom, received in the far field, can be used to discern the state of the field, whether it is in a vacuum, coherent, thermal or squeezed state \cite{QRad,QRadCoh,QRadSq}.

Besides issues which concern energy of quantum systems, such as the thermality felt by a  uniformly accelerating detector,  or the nature of radiation emitted from a moving atom or the spectrum of particles created in an expanding universe,   one can also ask questions concerning the \textit{quantum information} in the system and the environment. For example, two mode squeezed states of a quantum field  have been introduced for field entanglement studies from the perspectives of noninertial observers in Hawking-Unruh effects (e.g., \cite{Ahn,Adesso}). Using many detector systems  to extract quantum information of the quantum field, such as entanglement harvesting \cite{EntHarv}, have been suggested.  These are examples of how detectors can serve as probes to extract quantum information about the field, a popular topic in the emergent field of relativistic quantum information (RQI).

This paper is organized as follows: In Section~\ref{S:etutgbdg}, we first summarize the properties of a quantum field in a squeezed thermal state, whose squeeze parameter is assumed to be a time- and mode-independent constant. We then discuss the internal dynamics of a detector coupled to this field. Since such a bath field is nonequilibrium and nonstationary in nature, we will look into how equilibration of the internal dynamics is made possible by examining the rate of energy exchange between the detector and the bath field. In Section~\ref{S:bggkdfdd}, we consider the case when the squeezing of the bath field results from a parametric process, so the squeeze parameters become time- and mode-dependent. To capture the essense without sacrifice of generality we  consider those processes wherein the field parameter of interest changes smoothly and monotonically between two constants over a  finite time interval. We will examine the relaxation and equilibration process of the internal dynamics of the detector after the parametric process of the bath field ends. After that we formulate the corresponding fluctuation-dissipation relation.  In Section~\ref{S:gbveke} we give a summary of the key findings followed by a discussion of the significance of FDRs in quantum backreaction problems.

\newpage

\section{Oscillator in a Fixed-value Squeezed Thermal Bath}\label{S:etutgbdg}

In this section we consider a detector or atom in a fixed position in space with internal degrees of freedom modeled by a harmonic oscillator interacting with a thermal bath with a fixed squeeze value.  The case of  time-dependent squeezing will be treated in the next section.

\subsection{Quantum Brownian motion in a squeezed thermal bath}\label{E:bgksbssfg}

Consider a harmonic oscillator at rest coupled to a massless scalar field initially prepared in a squeezed thermal state. Under the dipole approximation any variation of the field over the displacement of the oscillator can be ignored,  so the quantum field takes on the value at the location of the oscillator  which can be chosen to be at the origin of the coordinate system depicting its external or mechanical motion. (See, e.g., \cite{MOF} for details.)

We first address some statistical properties of the squeezed thermal (ST) bath. A  squeezed thermal state of the field is described by the density matrix operator $\hat{\rho}^{(\phi)}_{\textsc{st}}$ of the form  
\begin{equation}\label{E:ireiitre}
	\hat{\rho}^{(\phi)}_{\textsc{st}}=\hat{S}(\zeta)\hat{\rho}^{(\phi)}_{\beta}\hat{S}^{\dagger}(\zeta)\,,
\end{equation}
where $\hat{\rho}^{(\phi)}_{\beta}$ is the thermal state of the field, and $\hat{S}(\zeta)$ is the squeeze operator
\begin{equation}
	\hat{S}(\zeta)=\prod_{\mathbf{k}}\exp\Bigl[\frac{1}{2}\,\zeta^{*}\hat{a}_{\mathbf{k}}^{2}-\frac{1}{2}\,\zeta\hat{a}_{\mathbf{k}}^{\dagger2}\Bigr]\,.
\end{equation}
The squeeze parameter $\zeta\in\mathbb{C}$, assumed to be a mode-independent constant for the moment, is usually conveniently written in the polar form $\zeta=\eta\,e^{i\theta}$ with $\eta\in\mathbb{R}^{+}$ and $0\leq\theta<2\pi$. The creation and annihilation operators satisfy the standard commutation relation $[\hat{a}_{\mathbf{k}}^{\vphantom{\dagger}},\hat{a}_{\mathbf{k}}^{\dagger}]=1$. The recurring expressions are the creation and annihilation operators sandwiched by the squeeze operators,
\begin{align}
	\hat{b}_{\mathbf{k}}^{\vphantom{\dagger}}=\hat{S}^{\dagger}(\zeta)\,\hat{a}_{\mathbf{k}}^{\vphantom{\dagger}}\,\hat{S}(\zeta)&=\cosh\eta\,\hat{a}_{\mathbf{k}}^{\vphantom{\dagger}}-e^{+i\theta}\,\sinh\eta\,\hat{a}_{\mathbf{k}}^{\dagger}=\alpha\,\hat{a}_{\mathbf{k}}^{\vphantom{\dagger}}+\delta^{*}\,\hat{a}_{\mathbf{k}}^{\dagger}\,,\label{E:ngkdbr}\\
	\hat{b}_{\mathbf{k}}^{\dagger}=\hat{S}^{\dagger}(\zeta)\,\hat{a}_{\mathbf{k}}^{\dagger}\hat{S}(\zeta)&=\cosh\eta\,\hat{a}_{\mathbf{k}}^{\dagger}-e^{-i\theta}\,\sinh\eta\,\hat{a}_{\mathbf{k}}^{\vphantom{\dagger}}=\alpha^{*}\,\hat{a}_{\mathbf{k}}^{\dagger}+\delta\,\hat{a}_{\mathbf{k}}^{\vphantom{\dagger}}\,.
\end{align}
{The resulting operators can be expressed as the superpositions of the original ones. This is the so-called Bogoliubov transformation. The Bogoliubov coefficients $\alpha$, $\delta\in\mathbb{C}$ satisfies the relation $\lvert\alpha\rvert^{2}-\lvert\delta\rvert^{2}=1$.}

Since the squeezed thermal state is a Gaussian state, its statistical properties are fully described by the first two moments of $\hat{a}_{\mathbf{k}}^{\vphantom{\dagger}}$, $\hat{a}_{\mathbf{k}}^{\dagger}$. The higher moments can be obtained by a Wick expansion. Thus we need only $\langle\hat{a}_{\mathbf{k}}^{\vphantom{\dagger}}\rangle_{\textsc{st}}=0=\langle\hat{a}_{\mathbf{k}}^{\dagger}\rangle_{\textsc{st}}$ and
\begin{align}
	\langle\hat{a}_{\mathbf{k}}^{2}\rangle_{\textsc{st}}&=-e^{+i\theta}\,\sinh2\eta\,\Bigl(\langle\hat{N}_{\mathbf{k}}\rangle_{\beta}+\frac{1}{2}\Bigr)\,,\label{E:gbkf2}\\
	\langle\hat{a}_{\mathbf{k}}^{\dagger}\hat{a}_{\mathbf{k}}^{\vphantom{\dagger}}\rangle_{\textsc{st}}&=\cosh2\eta\,\Bigl(\langle\hat{N}\rangle_{\beta}+\frac{1}{2}\Bigr)-\frac{1}{2}=\cosh2\eta\,\langle\hat{N}_{\mathbf{k}}\rangle_{\beta}+\sinh^{2}\eta\,,\label{E:gbkf}
\end{align}
where $\hat{N}_{\mathbf{k}}=\hat{a}_{\mathbf{k}}^{\dagger}\hat{a}_{\mathbf{k}}^{\vphantom{\dagger}}$ is the number operator of field mode $\mathbf{k}$, and $\langle\cdots\rangle_{\beta}$ represents the expectation value taken with respect to the thermal state of the field $\hat{\rho}^{(\phi)}_{\beta}$. Note that the second term $\sinh^{2}\eta$ in \eqref{E:gbkf} is $\lvert\delta\rvert^{2}$ in the Bogoliubov transformation in \eqref{E:ngkdbr}.

A nonvanishing $\delta$ signifies the production of particles. To be more specific, for  the vacuum state $\lvert0_{\mathbf{k}}\rangle$ annihilated by $\hat{a}_{\mathbf{k}}$, that is, $\hat{a}_{\mathbf{k}}\lvert0_{\mathbf{k}}\rangle=0$, we find the number operator $\hat{b}_{\mathbf{k}}^{\dagger}\hat{b}_{\mathbf{k}}^{\vphantom{\dagger}}$, the squeeze transformation of $\hat{a}_{\mathbf{k}}^{\dagger}\hat{a}_{\mathbf{k}}^{\vphantom{\dagger}}$,  has nonzero particle content with respect to this vacuum,
\begin{equation}
	\langle0_{\mathbf{k}}^{\vphantom{\dagger}}\rvert\hat{b}_{\mathbf{k}}^{\dagger}\hat{b}_{\mathbf{k}}^{\vphantom{\dagger}}\lvert0_{\mathbf{k}}^{\vphantom{\dagger}}\rangle=\lvert\delta\rvert^{2}\,,
\end{equation}
from \eqref{E:ngkdbr}.  This describes spontaneous particle creation from the vacuum. The underlying physics will be more clearly seen when we discuss the parametrically driven bath later. Furthermore, if the  $\mathbf{k}$th mode of the bath field is in a nonzero particle number state $\lvert n_{\mathbf{k}}\rangle$, then after the squeeze transformation, the new number operator $\hat{b}_{\mathbf{k}}^{\dagger}\hat{b}_{\mathbf{k}}^{\vphantom{\dagger}}$ will see that the particle contents is amplified from $n_{\mathbf{k}}$ to
\begin{align}
	\langle n_{\mathbf{k}}^{\vphantom{\dagger}}\rvert\hat{b}_{\mathbf{k}}^{\dagger}\hat{b}_{\mathbf{k}}^{\vphantom{\dagger}}\lvert n_{\mathbf{k}}^{\vphantom{\dagger}}\rangle=n_{\mathbf{k}}+2\lvert\delta\rvert^{2}\bigl(n_{\mathbf{k}}+\frac{1}{2}\bigr)\,,\label{E:mcgfghs}
\end{align}
where we have used the relation $\lvert\alpha\rvert^{2}-\lvert\delta\rvert^{2}=1$, which is  also understood as the Wronskian condition.  The second term on the righthand side depicts stimulated production of particles. We also note that the squeeze transformation does not modify the bound in the generalized uncertainty relation for a free, linear quantum scalar field. It only distorts the quadratures in the relation.

After these preliminaries we now construct the Hadamard function of the field in a squeezed thermal state. The free field operator $\hat{\phi}_{h}(x)$ is expanded as
\begin{equation}
	\hat{\phi}_{h}(x)=\int\!\frac{d^{3}\mathbf{k}}{(2\pi)^{\frac{3}{2}}}\;\frac{1}{\sqrt{2\omega}}\,\Bigl(\hat{a}_{\mathbf{k}}^{\vphantom{\dagger}}\,e^{-i\,k\cdot x}+\hat{a}^{\dagger}_{\mathbf{k}}\,e^{+i\,k\cdot x}\Bigr)
\end{equation}
with the 4-vectors $x=(t,\mathbf{x})$ and $k=(\omega,\mathbf{k})$, and $\omega=\lvert\mathbf{k}\rvert$, $k\cdot x=\omega t-\mathbf{k}\cdot\mathbf{x}$. Thus the Hadamard function will be given by
\begin{align}
	G_{H,0}^{(\phi)}(x,x')&=\frac{1}{2}\,\operatorname{Tr}\Bigl(\hat{\rho}^{(\phi)}_{\textsc{st}}\,\bigl\{\hat{\phi}_{h}(x),\hat{\phi}_{h}(x')\bigr\}\Bigr)\notag\\
	&=-\int\!\!\frac{d^{3}\mathbf{k}}{(2\pi)^{3}}\frac{1}{4\omega}\coth\frac{\beta\omega}{2}\sinh2\eta\,e^{+i\mathbf{k}\cdot(\mathbf{x}+\mathbf{x}')}\Bigl[e^{-i\omega(t+t')+i\theta}+e^{+i\omega(t+t')-i\theta}\Bigr]\label{E:dher}\\
	&\quad+\int\!\!\frac{d^{3}\mathbf{k}}{(2\pi)^{3}}\frac{1}{4\omega}\coth\frac{\beta\omega}{2}\cosh2\eta\,e^{+i\mathbf{k}\cdot(\mathbf{x}-\mathbf{x}')}\Bigl[e^{-i\omega(t-t')}+e^{+i\omega(t-t')}\Bigr]\,.\label{E:rkjrt}
\end{align}
We will call \eqref{E:dher} the nonstationary component of the Hadamard function and \eqref{E:rkjrt} the stationary component because the latter does not change under time translation. The factor $\coth\frac{\beta\omega}{2}$ is important in the context of  thermodynamics, as often seen in the discussion of the Callen-Welton or Green-Kubo fluctuation-dissipation relation.

Accompanying the Hadamard function (noise kernel), we have the retarded Green's function (dissipation kernel)
\begin{equation}
	G_{R,0}^{(\phi)}(x,x')=i\,\theta(t-t')\,\bigl[\hat{\phi}_{h}(x),\hat{\phi}_{h}(x')\bigr]\,.
\end{equation}
This is already a $c$-number, independent of the field state. Following the previous conventions, we have
\begin{align}
	G_{R,0}^{(\phi)}(x,x')&=\frac{1}{4\pi R}\,\theta(\tau)\Bigl[\delta(\tau-R)-\delta(\tau+R)\Bigr]=\frac{1}{2\pi}\,\theta(\tau)\,\delta(\tau^{2}-R^{2})\,.\label{E:eokerhd}
\end{align}
We see the causal lightcone structure is enforced. In the special case $R\to0$ which is met for a single Brownian oscillator, we instead have
\begin{equation}\label{E:dkuerbdf}
	G_{R,0}^{(\phi)}(t,t')=-\frac{1}{2\pi}\,\theta(\tau)\lim_{R\to0}\frac{1}{2R}\Bigl[\delta(\tau+R)-\delta(\tau-R)\Bigr]=-\frac{1}{2\pi}\,\theta(\tau)\,\delta'(\tau)\,.
\end{equation}
This will be of great convenience in simplifying the equation of motion for the internal dynamics of the detector when it is coupled to a squeezed thermal bath field.

We now study the nonequilibrium evolution of the internal dynamics of a Unruh-DeWitt detector in a squeezed thermal bath.

\subsection{Internal dynamics of  an Unruh-DeWitt detector}\label{S:trvkgdfhd}
Now we consider the internal dynamics of a Unruh-DetWitt detector at rest in the squeezed thermal bath. When the internal dynamics of the detector is modeled by a quantum harmonic oscillator, its displacement $\hat{\chi}$ follows a reduced equation of motion of the form (e.g., \cite{QTD1})
\begin{equation}\label{E:thfsdf}
	\ddot{\hat{\chi}}(t)-\omega_{\textsc{b}}^{2}\,\hat{\chi}(t)-\frac{e^{2}}{m}\int_{0}^{t}\!ds\;G_{R,0}^{(\phi)}(t,s)\hat{\chi}(s)=\frac{e}{m}\,\hat{\phi}_{h}(\mathbf{0},t)\,,
\end{equation}
when all the influences from the field are taken into account. Here we assume that the coupling between the oscillator and the field is switched on at time $t=0$ and has a strength $e$, or alternatively,  represented by the damping constant $\gamma=e^{2}/8\pi m$. The oscillator depicting the internal degree of freedom of the detector is assumed to have a mass $m$ and a bare natural frequency $\omega_{\textsc{b}}$. The detector's external or mechanical degree of freedom is fixed at the spatial origin of the coordinate system.

For the retarded Green's function of the form \eqref{E:dkuerbdf}, the equation \eqref{E:thfsdf} can be simplified to a local form
\begin{equation}\label{E:kjsskw}
	\ddot{\hat{\chi}}(t)+2\gamma\,\dot{\hat{\chi}}(t)+\omega_{\textsc{r}}^{2}\,\hat{\chi}(t)=\frac{e}{m}\,\hat{\phi}_{h}(\mathbf{0},t)\,.
\end{equation}
where $m$, $\omega_{\textsc{r}}$ is the mass and the physical frequency of the internal degree of freedom of the detector. The dynamical significance has been discussed in~\cite{CPR}. Yet,  quite a few outstanding features in our problem are worthy of special notice: 1) Unlike in the thermal bath,  the driving noise, given by the free field, is not stationary. In other words, the bath from the beginning is not in equilibrium. 2) Its frequency spectrum in general is not white. 3) The internal state of the detector initially can be in any state far from equilibrium. 4) The coupling between the internal degree of freedom and the field is not required to be weak. 
5) Renormalization and the actions from the bath field need be dealt with.  6) Arguments based on linear response or perturbation theory are mostly inapplicable. The upshot is, it is by no means clear that the internal dynamics of the detector, when coupled to such a nonstationary, nonequilibrium bath, can ever come to equilibrium eventually.  This becomes the main challenge  in the present investigation.

To look into the possibility of equilibration, let us examine the internal dynamics of the detector, governed by \eqref{E:thfsdf} or \eqref{E:kjsskw}, in a squeezed thermal bath. The general solution to \eqref{E:kjsskw} is given by
\begin{equation}\label{E:dkgsrua}
	\hat{\chi}(t)=d_{1}(t)\,\hat{\chi}(0)+d_{2}(t)\,\dot{\hat{\chi}}(0)+\frac{e}{m}\int_{0}^{t}\!ds\;d_{2}(t-s)\,\hat{\phi}_{h}(\mathbf{0},s)
\end{equation}
where $d_{1}(t)$ and $d_{2}(t)$ are a special set of homogeneous solutions to \eqref{E:kjsskw},
\begin{align}
	d_{1}(0)&=1\,,&\dot{d}_{1}(0)&=0\,,&d_{2}(0)&=0\,,&\dot{d}_{2}(0)&=1\,,
\end{align}
which in the present case are given by
\begin{align}\label{E:gbkdhg}
	d_{1}(t)&=e^{-\gamma t}\Bigl[\cos\Omega t+\frac{\gamma}{\Omega}\,\sin\Omega t\Bigr]\,,&d_{2}(t)&=\frac{1}{\Omega}\,e^{-\gamma t}\sin\Omega t\,,
\end{align}
with $\Omega=\sqrt{\omega_{\textsc{r}}^{2}-\gamma^{2}}$ being the resonance frequency. Note that they decay exponentially with time, dissipation being a consequence of the interaction with the bath. 

\begin{figure}
\centering
    \scalebox{0.45}{\includegraphics{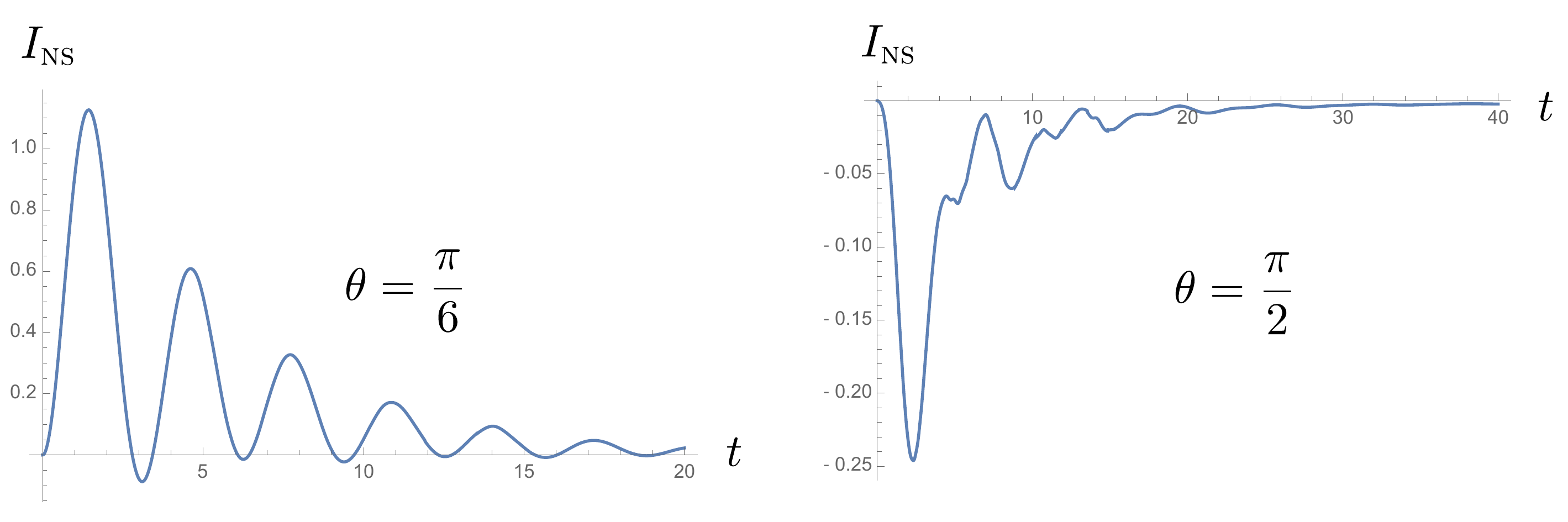}}
    \caption{The temporal behavior of $I_{\textsc{ns}}$, associated with the nonstationary component of $\langle\hat{\chi}^{2}(t)\rangle$, for different squeeze angles. We choose $m=1$, $\Omega=1$ $\gamma=0.1$, $\beta=0.3$, that is, in the high-temperature regime. we observe that $I_{\textsc{ns}}$ quickly decays to zero for time greater than the relaxation time scale $\gamma^{-1}$.}\label{Fi:004}
\end{figure}
These fundamental solutions allow us to easily construct the observables associated with the internal dynamics. For example, the building blocks of the Gaussian system, the covariance matrix elements, can be expressed as
\begin{align}
	\langle\hat{\chi}^{2}(t)\rangle&=d_{1}^{2}(t)\,\langle\hat{\chi}^{2}(0)\rangle+\frac{1}{m^{2}}\,d_{2}^{2}(t)\,\langle\hat{p}^{2}(0)\rangle\label{E:bgkeubd}\\
	&\qquad\qquad\qquad\qquad\quad+\frac{e^{2}}{m^{2}}\int_{0}^{t}\!ds\int_{0}^{t}\!ds'\;d_{2}(t-s)d_{2}(t-s')\,G_{H,0}^{(\phi)}(\mathbf{0},s;\mathbf{0},s')\,,\notag\\
	\langle\hat{p}^{2}(t)\rangle&=m^{2}\dot{d}_{1}^{2}(t)\,\langle\hat{\chi}^{2}(0)\rangle+\dot{d}_{2}^{2}(t)\,\langle\hat{p}^{2}(0)\rangle\label{E:dkjjesd}\\
	&\qquad\qquad\qquad\qquad\quad+e^{2}\int_{0}^{t}\!ds\int_{0}^{t}\!ds'\;\dot{d}_{2}(t-s)\dot{d}_{2}(t-s')\,G_{H,0}^{(\phi)}(\mathbf{0},s;\mathbf{0},s')\,,\notag\\
	\frac{1}{2}\langle\bigl\{\hat{\chi}(t),\hat{p}(t)\bigr\}\rangle&=m\,d_{1}(t)\dot{d}_{1}(t)\,\langle\hat{\chi}^{2}(0)\rangle+\frac{1}{m}\,d_{2}(t)\dot{d}_{2}(t)\,\langle\hat{p}^{2}(0)\rangle\label{E:gdkjsn}\\
	&\qquad\qquad\qquad\qquad\quad+\frac{e^{2}}{m}\int_{0}^{t}\!ds\int_{0}^{t}\!ds'\;d_{2}(t-s)\dot{d}_{2}(t-s')\,G_{H,0}^{(\phi)}(\mathbf{0},s;\mathbf{0},s')\,,\notag
\end{align}
with the conjugated momentum $\hat{p}=m\dot{\hat{\chi}}$. Here $\langle\cdots\rangle$ denotes the expectation value with respect to the initial state of the whole system in an assumed product form
\begin{equation}
	\hat{\rho}(0)=\hat{\rho}^{(\chi)}(0)\otimes\hat{\rho}^{(\phi)}_{\textsc{st}}(0)\,,
\end{equation}
with  $\hat{\rho}^{(\chi)}$ being the density matrix operator of the internal degree of freedom, and, for simplicity, it has been chosen to have the properties: $\langle\hat{\chi}(0)\rangle=0$, $\langle\hat{p}(0)\rangle=0$, and $\langle\bigl\{\hat{\chi}(0),\hat{p}(0)\bigr\}\rangle=0$. The kernel function $G_{H,0}^{(\phi)}(\mathbf{0},t;\mathbf{0},t')$, defined in \eqref{E:dher} and \eqref{E:rkjrt}, in this context, takes a simpler form
\begin{align}
	G_{H,0}^{(\phi)}(\mathbf{0},t;\mathbf{0},t')&=-\sinh2\eta\int_{0}^{\infty}\!\frac{d\omega}{2\pi}\;\frac{\omega}{4\pi}\coth\frac{\beta\omega}{2}\,\Bigl[e^{-i\omega(t+t')+i\theta}+e^{+i\omega(t+t')-i\theta}\Bigr]\\
	&\qquad\qquad\qquad\quad+\cosh2\eta\int_{0}^{\infty}\!\frac{d\omega}{2\pi}\;\frac{\omega}{4\pi}\coth\frac{\beta\omega}{2}\,\Bigl[e^{-i\omega(t-t')}+e^{+i\omega(t-t')}\Bigr]\,.\notag
\end{align}
To grasp the generic behavior of the stationary and nonstationary components of $G_{H,0}^{(\phi)}(\mathbf{0},t;\mathbf{0},t')$, and their effects on the observables of the internal dynamics, we will use $\langle\hat{\chi}^{2}(t)\rangle$ as an example. It turns out convenient to introduce the function
\begin{equation}\label{E:gkbskrjsd}
	f(t;\omega)=\int_{0}^{t}\!ds\;d_{2}(t-s)\,e^{-i\omega s}=\tilde{d}_{2}(\omega)\Bigl[e^{-i\omega t}-d_{1}(t)+i\,\omega\,d_{2}(t)\Bigr]\,,
\end{equation}
so that the integral expressions in \eqref{E:bgkeubd} can be written as
\begin{align}\label{E:skjfkjewr}
	&\quad\int_{0}^{t}\!ds\int_{0}^{t}\!ds'\;d_{2}(t-s)d_{2}(t-s')\,G_{H,0}^{(\phi)}(\mathbf{0},s;\mathbf{0},s')\\
	&=\int_{0}^{\infty}\!\frac{d\omega}{2\pi}\;\frac{\omega}{4\pi}\coth\frac{\beta\omega}{2}\biggl\{-\sinh2\eta\Bigl[f^{2}(t;\omega)\,e^{+i\theta}+f^{*2}(t;\omega)\,e^{-i\theta}\Bigr]+\cosh2\eta\times2\lvert f(t;\omega)\rvert^{2}\biggr\}\,,\notag
\end{align}
where the Fourier transform $\tilde{g}(\omega)$ of a function $g(t)$ is defined by
\begin{equation}\label{E:dkgker}
	\tilde{g}(\omega)=\int_{-\infty}^{\infty}\!dt\;g(t)\,e^{+i\omega t}\,.
\end{equation}
In this section we merely provide the figures based on numerical calculations in order to quickly glean the generic behavior of the quantities we will discuss. More detailed analysis will be given in Sec.~\ref{S:krghvjdfrg}.
\begin{figure}
\centering
    \scalebox{0.45}{\includegraphics{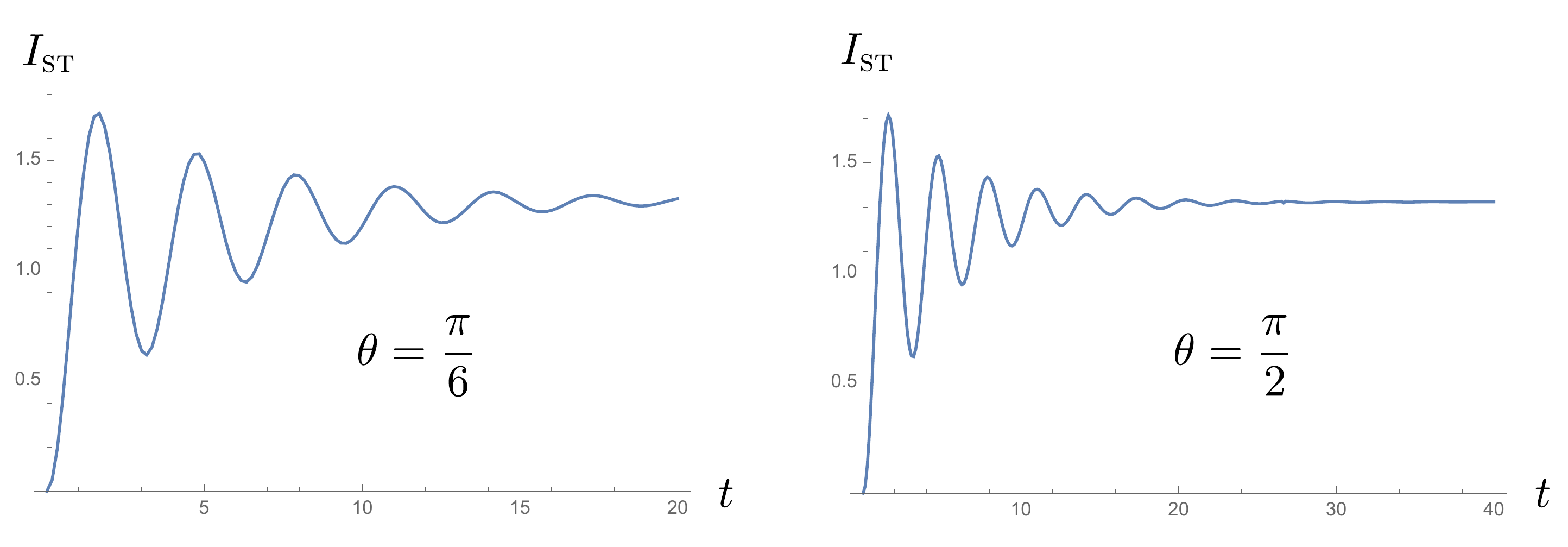}}
    \caption{The temporal behavior of $I_{\textsc{ns}}$, related to the stationary component of $\langle\hat{\chi}^{2}(t)\rangle$,  for different squeeze angles. We choose $m=1$, $\Omega=1$ $\gamma=0.1$, $\beta=0.3$. They all approach an identical constant, independent of the squeeze angle $\theta$.}\label{Fi:005}
\end{figure}

We focus on the temporal behavior of the nonstationary component in the integral expression of $\langle\hat{\chi}^{2}(t)\rangle$ in \eqref{E:bgkeubd}.  The first two terms on the righthand side of \eqref{E:bgkeubd} account for the intrinsic fluctuations of the internal dynamics are not of interest to us.  They will always decay with time to zero, and are not directly affected by the bath field fluctuations. Let us define
\begin{equation}
	I_{\textsc{ns}}=-\int_{0}^{\infty}\!\frac{d\omega}{2\pi}\;\frac{\omega}{4\pi}\coth\frac{\beta\omega}{2}\Bigl[f^{2}(t;\omega)\,e^{+i\theta}+f^{*2}(t;\omega)\,e^{-i\theta}\Bigr]\,.
\end{equation}
proportional to the nonstationary component of \eqref{E:skjfkjewr}. From Fig.~\ref{Fi:004}, we see that in general it will approach zero exponentially fast as $t$ becomes greater than the relaxation time scale $\gamma^{-1}=10$. In contrast,  {as shown in Fig.~\ref{Fi:005}}, the stationary component, proportional to
\begin{equation}
	I_{\textsc{st}}=+\int_{0}^{\infty}\!\frac{d\omega}{2\pi}\;\frac{\omega}{4\pi}\coth\frac{\beta\omega}{2}\times2\lvert f(t;\omega)\rvert^{2}\,,
\end{equation}
approaches a constant at late times. These results indicate that the contribution from the nonstationary component of the Hadamard function $G_{H,0}^{(\phi)}(\mathbf{0},t;\mathbf{0},t')$ tends to be exponentially smaller than that of the stationary component at late times in $\langle\hat{\chi}^{2}(t)\rangle$, so that the squeeze angle $\theta$ becomes irrelevant. On the other hand, since the parameter $\eta$ still appears in the contribution of the stationary component, we conclude that at late times $\langle\hat{\chi}^{2}(\infty)\rangle$ is still $\eta$ dependent. Thus measurement of $\langle\hat{\chi}^{2}(\infty)\rangle$ cannot recover the information about the angle $\theta$.

\begin{figure}
\centering
    \scalebox{0.35}{\includegraphics{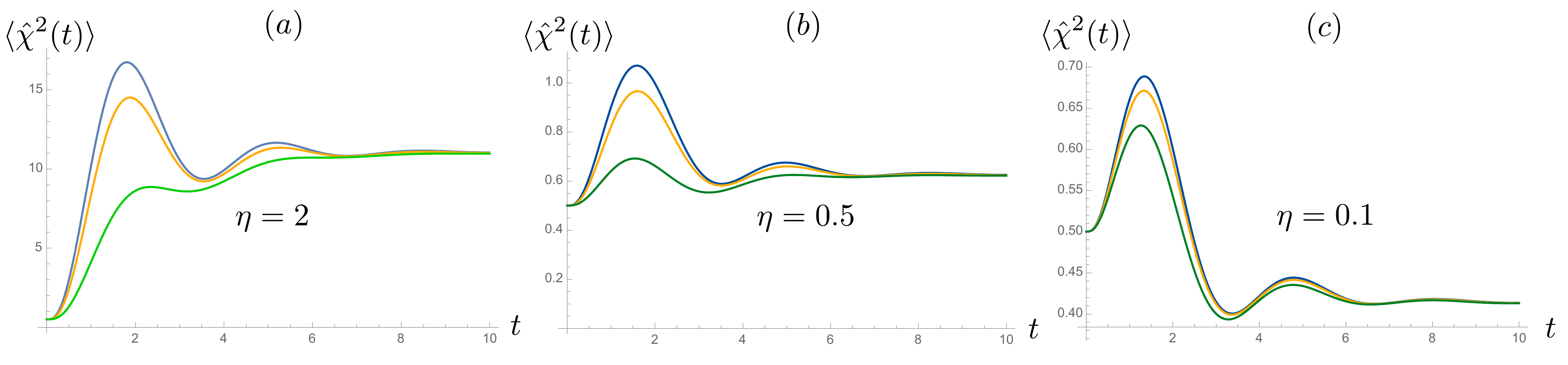}}
    \caption{(a) The temporal behavior of $\langle\hat{\chi}^{2}(t)\rangle$ on the squeeze parameter $\zeta=\eta\,e^{i\theta}$. We choose $m=1$, $\omega_{\textsc{r}}=1$ $\gamma=0.3$, $\beta=10$, a low-temperature regime. In these three plots, the blue curve corresponds to  $\theta=0$, the orange curve $\theta=\pi/6$ and the green curve $\theta=\pi/2$. The $\theta$ dependence is clearly seen during the intermediate times but this dependence is lost at times greater than $\gamma^{-1}$.}\label{Fi:006}
\end{figure}

This is more clearly seen from Fig.~\ref{Fi:006} that for sufficiently late time $\langle\hat{\chi}^{2}(t)\rangle$ is independent of $\theta$. The saturated height of $\langle\hat{\chi}^{2}(t)\rangle$ depends on $\eta$. This is expected since the excitations of the oscillators, driven by the squeezed thermal bath, is related to $\eta$. Thus as in the case of the thermal bath, certain properties of the bath can be passed on to the oscillator system coupled to the bath. In this case the information contained in $\eta$ is inherited by the oscillator but the information of $\theta$ is not.  

\begin{figure}
\centering
    \scalebox{0.5}{\includegraphics{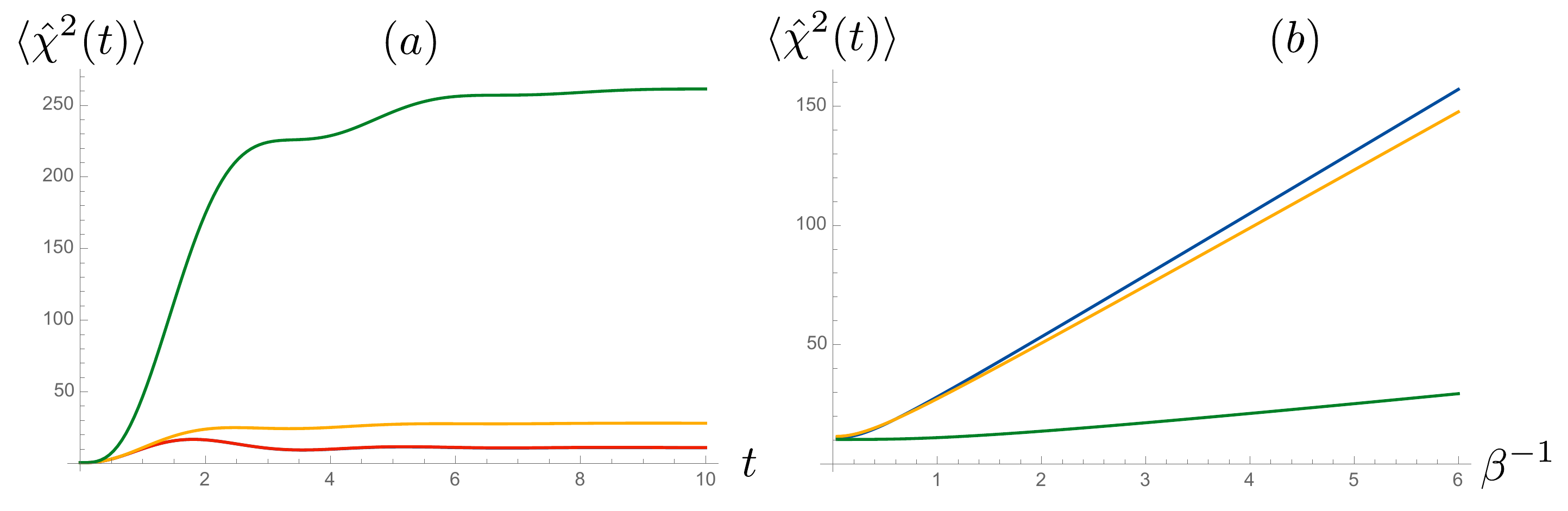}}
    \caption{The temporal behavior of $\langle\hat{\chi}^{2}(t)\rangle$ on the inverse temperature $\beta$. We choose $m=1$, $\omega_{\textsc{r}}=1$ $\gamma=0.3$, $\eta=2$, $\theta=0$. The blue, red, orange, and green curves respectively correspond to the low, intermediate and high temperature regimes by letting $\beta=100$, $\beta=10$, $\beta=1$ and $\beta=0.1$. Due to scale difference, the blue curve almost overlap with the blue curves. In (b) we sample three different times: $t=10$ for the blue curve, $t=5$ the orange curve and $t=1$ for the green curve, to present the dependence of $\langle\hat{\chi}^{2}(t)\rangle$ on the bath temperature $\beta^{-1}$ during three evolutionary stages. Note that the dimensional parameters are normalized with respect to the physical frequency $\omega_{\textsc{r}}$.}\label{Fi:007}
\end{figure}

In Fig.~\ref{Fi:007}, we show the general trend about the effect of temperature on $\langle\hat{\chi}^{2}(t)\rangle$. In particular, in Fig.~\ref{Fi:007}-(b), it has the typical feature that at the low bath temperature regime, the curves rise relatively slowly from constant values and then gradually transit to  linear growth, which is consistent with classical equipartition theorem, since $\langle\hat{\chi}^{2}(t)\rangle$ is proportional to the elastic potential energy of the oscillator. In Fig.~\ref{Fi:007}-(a),  higher temperature heightens the overall generic shape of the curve,  consistent with the monotonic behavior of the factor $\coth\frac{\beta\omega}{2}$ with respect to $\beta$.

Let us now take a closer look at the late-time result. We have
\begin{align}
	\langle\hat{\chi}^{2}(\infty)\rangle&=\cosh2\eta\,\frac{8\pi\gamma}{m}\lim_{t\to\infty}\int_{0}^{\infty}\!\frac{d\omega}{2\pi}\;\frac{\omega}{4\pi}\coth\frac{\beta\omega}{2}\times2\lvert f(t;\omega)\rvert^{2}\,.
\end{align}
Using \eqref{E:gkbskrjsd}, we arrive at
\begin{align}
	\langle\hat{\chi}^{2}(\infty)\rangle&=\cosh2\eta\,\frac{8\pi\gamma}{m}\lim_{t\to\infty}\int_{0}^{\infty}\!\frac{d\omega}{2\pi}\;\frac{\omega}{4\pi}\coth\frac{\beta\omega}{2}\times2\lvert\tilde{d}_{2}(\omega)\rvert^{2}\Bigl[1+d_{1}^{2}(t)+\omega^{2}d_{2}^{2}(t)-d_{1}(t)\,e^{-i\omega t}\Bigr.\notag\\
	&\qquad\qquad\qquad\qquad\qquad\qquad\qquad\qquad-\Bigl.d_{1}(t)\,e^{+i\omega t}-i\omega\,d_{2}(t)\,e^{-i\omega t}+i\omega\,d_{2}(t)\,e^{+i\omega t}\Bigr]\notag\\
	&=\cosh2\eta\,\langle\hat{\chi}^{2}(\infty)\rangle_{\beta}\,.\label{E:lgeure}
\end{align}
Thus $\langle\hat{\chi}^{2}(\infty)\rangle$ at late times is boosted by a factor $\cosh2\eta$, which is always greater than unity, from the corresponding value $\langle\hat{\chi}^{2}(\infty)\rangle_{\beta}$ due to the thermal bath. This factor is nothing but $2\lvert\delta\rvert^{2}+1$, related to the Bogoliubov coefficients in  \eqref{E:gbkf2}, so it is a consequence of enhanced excitation from parametric amplification.
	
Let us compare this with a free harmonic oscillator in its squeezed thermal state, denoted by a different squeeze parameter $\zeta'=\eta'e^{+i\theta'}$ but at the same temperature $1/\beta$. Since the displacement operator of the oscillator in the Heisenberg picture is
\begin{equation}
	\hat{\chi}(t)=\frac{1}{\sqrt{2m\omega_{\textsc{r}}}}\,\bigl(\hat{a}^{\dagger}\,e^{+i\omega_{\textsc{r}}t}+\hat{a}\,e^{-i\omega_{\textsc{r}}t}\bigr)\,,
\end{equation}
we find
\begin{align}
	\langle\hat{\chi}^{2}\rangle_{\textsc{st}}&=\Bigl[\cosh2\eta'-\cos(2\omega_{\textsc{r}}t-\theta')\,\sinh2\eta'\Bigr]\,\langle\hat{\chi}^{2}\rangle_{\beta}\,.\label{E:deueewowo}
\end{align}
Observe that the factor before $\langle\hat{\chi}^{2}\rangle_{\beta}$ is a positive, \textit{time-dependent} real number, so it can be smaller than unity or larger, depending on the choice of the squeeze angle $\theta$. Thus we can use the squeeze parameter to tune the coherence of the oscillator; however, in the open system scenario, it does not work the same way since the expression in \eqref{E:lgeure} is not dependent on $\theta$  after equilibration. We cannot make $\langle\hat{\chi}^{2}(\infty)\rangle$ smaller than $\langle\hat{\chi}^{2}(\infty)\rangle_{\beta}$. 

Observe that from this example we can identify two important features associated with the dynamics of the internal degrees of freedom coupled to a nonstationary, nonequilibrium bath field. The first is that even the bath has the aforementioned properties, the quantity $\langle\hat{\chi}^{2}(\infty)\rangle$ still reaches a constant on a time scale greater than the relaxation time. This signifies the existence of an asymptotic equilibrium state in the internal dynamics. Secondly, comparing with $\langle\hat{\chi}^{2}(\infty)\rangle_{\beta}$ in the weak oscillator-field coupling
\begin{equation*}
	\langle\hat{\chi}^{2}(\infty)\rangle_{\beta}=\frac{1}{2m\omega_{\textsc{r}}}\,\coth\frac{\beta\omega_{\textsc{r}}}{2}
\end{equation*}
we can write \eqref{E:lgeure}, in the same weak coupling limit, as
\begin{align}
	\langle\hat{\chi}^{2}(\infty)\rangle&=\frac{1}{2m\omega_{\textsc{r}}}\,\coth\frac{\beta\omega_{\textsc{r}}}{2}\cosh2\eta\,,&&\Rightarrow&\coth\frac{\beta_{\textsc{s}}\omega_{\textsc{r}}}{2}&=\coth\frac{\beta\omega_{\textsc{r}}}{2}\cosh2\eta\,.\label{E:gbksbkdfgs}
\end{align}
It seems to imply that the final state acts like a thermal state of a higher temperature $\beta_{\textsc{s}}^{-1}$ than $\beta^{-1}$. That is, {a detector feels hotter in a squeezed thermal bath than in a thermal bath}. This  enhancement factor due to squeezing is what gives the added  efficiency in an Otto engine \cite{QOtto} or the leverage in the quest for `hot' entanglement \cite{HotEnt}. Finally, since the result in \eqref{E:lgeure} is independent of the squeeze angle $\theta$, we cannot turn the final state $\hat{\rho}^{(\chi)}(\infty)$ back to a thermal state $\hat{\rho}_{\beta}^{(\chi)}$ of the internal dynamics by applying an unsqueezing $\hat{S}^{-1}(\zeta)$ via
\begin{equation}
	\hat{\rho}_{\beta}^{(\chi)}\stackrel{?}{=}\hat{S}(-\zeta)\hat{\rho}^{(\chi)}(\infty)\hat{S}^{\dagger}(-\zeta)\,,
\end{equation}
similar to the reversed operation of \eqref{E:ireiitre}, even in the weak coupling limit. This may invalidate proposed performance enhancement schemes  invoking unsqueezing at the end of the isothermal phase of a quantum Otto thermal engine when the harmonic oscillator is placed in contact with a squeezed thermal bath.

The momentum uncertainty $\langle\hat{p}^{2}(t)\rangle$ has similar behaviors as the displacement uncertainty $\langle\hat{\chi}^{2}(t)\rangle$ does. From \eqref{E:dkjjesd}, we see that structurally it is very similar to the displacement uncertainty except for two more time derivatives. It tends to make the integrations ill-defined because when we write \eqref{E:dkjjesd} as an integral over $\omega$, we find the expressions involving two time-integrals become
\begin{align}
	&\quad\int_{0}^{t}\!ds\int_{0}^{t}\!ds'\;\dot{d}_{2}(t-s)\dot{d}_{2}(t-s')\,G_{H,0}^{(\phi)}(\mathbf{0},s;\mathbf{0},s')\label{E:uteeriwej}\\
	&=\int_{0}^{\infty}\!\frac{d\omega}{2\pi}\;\frac{\omega}{4\pi}\coth\frac{\beta\omega}{2}\biggl\{-\sinh2\eta\Bigl[\dot{f}^{2}(t;\omega)\,e^{+i\theta}+\dot{f}^{*2}(t;\omega)\,e^{-i\theta}\Bigr]+\cosh2\eta\times2\lvert\dot{f}(t;\omega)\rvert^{2}\biggr\}\,.\notag
\end{align}
Eq.~\eqref{E:gkbskrjsd} implies that at late times $t\to\infty$, Eq.~\eqref{E:uteeriwej} reduces to a time independent constant proportional to
\begin{equation}
	\int_{-\infty}^{\infty}\!\frac{d\omega}{2\pi}\;\frac{\omega^{3}}{4\pi}\coth\frac{\beta\omega}{2}\,\lvert\tilde{d}_{2}(\omega)\rvert^{2}\,.
\end{equation}
The large $\lvert\omega\rvert$ end of the integrand grows like $\omega^{-1}$ because in that limit $\coth\frac{\beta\omega}{2}\to1$ and $\lvert\tilde{d}_{2}(\omega)\rvert^{2}\to\omega^{-4}$. The integral is logarithmically divergent. Thus regularization is needed. We may introduce a cutoff frequency $\Lambda$ to replace the limits of the integral or insert a convergent factor of the form $e^{-\lvert\omega\rvert\epsilon}$ where $\epsilon$ is a tiny positive real number and essentially plays to role of inverse cutoff frequency because its presence will highly suppress the contributions from the frequency much higher than $\epsilon^{-1}$. 

\begin{figure}
\centering
    \scalebox{0.27}{\includegraphics{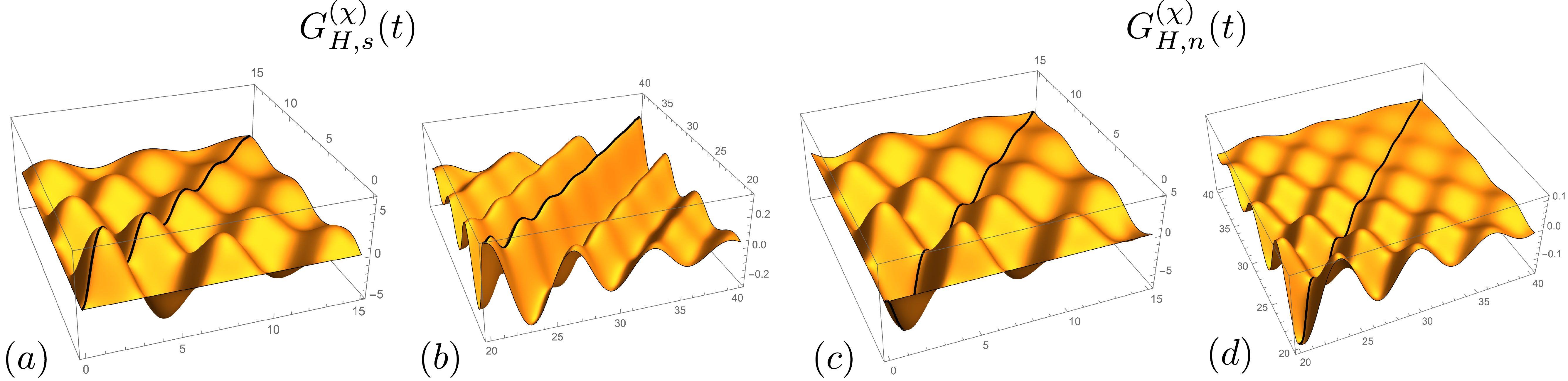}}
    \caption{The temporal behavior of $G_{H}^{(\chi)}(t,t')$ for  an oscillator in contact with a zero temperature squeezed thermal bath. In (a) and (b) the stationary component, proportional to $\cosh2\eta$, is drawn with respect to (a) $0\leq t,\,t'\leq15$ and (b) $20\leq t,\,t'\leq40$ in units of the resonance frequency $\omega_{\textsc{r}}$. The corresponding nonstationary component is shown in (c) and (d). In (a) and (c) both components are seen to exponentially decay with $t$ and $t'$; however, they differ dramatically when both $t$, $t'$ are much greater than $\gamma^{-1}$. The stationary component flattens out to a nonzero constant, but the nonstationary component continues decreasing to zero. The black curve highlights the case $t=t'$. All the parameters are chosen in units of the resonance frequency $\omega_{\textsc{r}}$ or its inverse. Here, $\omega_{\textsc{r}}=1$, the damping constant $\gamma=0.1$, the initial bath temperature $\beta^{-1}=0$ and the squeeze angle $\theta=0$. We have factored out $\cosh2\eta$ and $\sinh2\eta$ because they do not change the temporal dependence.}\label{Fi:xGrn3D}
\end{figure}

Now we turn to the two-point function of $\hat{\chi}$. This will show a different aspect of  the final state. In particular,  we are interested in the Hadamard function $G_{H}^{(\chi)}(t,t')$ associated with the operator $\hat{\chi}$, given by
\begin{align}
	G_{H}^{(\chi)}(t,t')=\frac{1}{2}\,\langle\bigl\{\hat{\chi}(t),\hat{\chi}(t')\bigr\}\rangle&=d_{1}(t)d_{1}(t')\,\langle\hat{\chi}^{2}(0)\rangle+\frac{1}{m^{2}}\,d_{2}(t)d_{2}(t')\,\langle\hat{p}(0)\rangle\label{E:fgbkfs}\\
	&\qquad\qquad\qquad+\frac{e^{2}}{m^{2}}\int_{0}^{t}\!ds\int_{0}^{t'}\!ds'\;d_{2}(t-s)d_{2}(t'-s')G_{H,0}^{(\phi)}(\mathbf{z},s;\mathbf{0},s')\,.\notag
\end{align}
This is a generalization of the covariance matrix elements we discussed earlier. We then arrive at
\begin{align}
	G_{H}^{(\chi)}(t,t')&=\cdots+\frac{e^{2}}{m^{2}}\int_{0}^{\infty}\!\frac{d\omega}{2\pi}\;\frac{\omega}{4\pi}\coth\frac{\beta\omega}{2}\biggl\{-\sinh2\eta\Bigl[f(t;\omega)f(t';\omega)\,e^{+i\theta}+f^{*}(t;\omega)f^{*}(t';\omega)\,e^{-i\theta}\Bigr]\biggr.\notag\\
	&\qquad\qquad\qquad\qquad\qquad\qquad\quad+\biggl.\cosh2\eta\Bigl[f(t;\omega)f^{*}(t';\omega)+f^{*}(t;\omega)f(t';\omega)\Bigr]\biggr\}\,,\label{E:gnhfkssd}
\end{align}
where $\dots$ represents terms that depend on the initial conditions, but become exponentially small at times greater than $\gamma^{-1}$. At first sight, similar to the Hadamard function of the free field $G_{H,0}^{(\phi)}(x,x')$, it contains a component that is not stationary in time. However we observe that the non-stationary component has {a $\tilde{d}_{2}^{2}(\omega)$ factor, rather than $\lvert\tilde{d}_{2}(\omega)\rvert^{2}$ in the integrand in \eqref{E:gnhfkssd}}. Its appearance implies a possibility that the non-stationary component may be exponentially small when both $t$ and $t'$ are greater than $\gamma^{-1}$. Indeed from the numerical calculations, we see from Fig.~\ref{Fi:xGrn3D}-(c) and (d) that generically, the nonstationary component does decay to zero or  become exponentially smaller than the stationary component when $t$, $t'$ are sufficiently large. In contrast,  the stationary components saturate to a constant in the same limit. Thus at late times the two-point function $G_{H}^{(\chi)}(t,t')$ also becomes invariant under time translation even though the oscillator is driven by a nonstationary noise at all  times
\begin{align}\label{E:gbdfkgjbs}
	G_{H}^{(\chi)}(t,t')=\cosh2\eta\,G_{H,\beta}^{(\chi)}(t-t')\,,
\end{align}
for $t$, $t'\gg\gamma^{-1}$ where $\gamma$ is the damping constant, and $\eta$ is the squeeze parameter of the bath in its initial configuration. The Hadamard function $G_{H,\beta}^{(\chi)}(t-t')$ gives the correlation of the oscillator when it is coupled to a plain thermal state. This behavior is required and is consistent if the dynamics of the oscillator can relax to an equilibrium state. This result provides additional support to identifying the final equilibrium state as a thermal state\footnote{The term ``thermal state'' is understood in the following sense: Given a fixed $\beta$ and oscillator physical frequency $\omega_{\textsc{r}}$, the oscillator will be relaxed to a state that is like a thermal state with the effective temperature given by \eqref{E:gbksbkdfgs} because the oscillator's covariance matrix elements and Hadamard function will be amplified by a common factor $\cosh2\eta$. On the other hand, strictly speaking, if we fix $\beta$ and $\eta$, but vary the physical frequency of the oscillator, then the oscillator will not have a  black-body energy spectrum at the effective temperature $\beta^{-1}_{\textsc{s}}$ because the oscillator of different physical frequencies will not see the same effective temperature.} of a Gaussian system.

On the surface this result  seems to be in conflict with our previous understanding about the properties of the squeezed state of the free  field in Sec.~\ref{E:bgksbssfg}.  For example, the Hadamard function of the free field $G_{H,0}^{(\phi)}(x,x')$ in \eqref{E:dher} gives a non-vanishing, oscillatory nonstationary component. The difference lies in the latter describing the squeezed state of a free oscillator in a closed system, while the former for the driven damped oscillator in an open system. The backactions and backreactions between the oscillator and the bath play an important role, which contribute to the distinct behaviors at late times. Simply put, the internal dynamics of the detector will respond to the bath's action in a rather complicated and delicate way:  The strength of dissipation depends on both the dissipation kernel, determined by the form of coupling and the property of the bath field, and on the state of internal motion. When the internal degree of freedom of the detector is driven by (nonstationary) quantum fluctuations of the bath, the damping will  adjust itself to match the driving force in accordance, such that in the end, equilibration is prompt  to happen. On the other hand, in this framework,  {the behavior of the internal degree of freedom} can mimic that of the free oscillator only at early times $t\ll\gamma^{-1}$, where the decaying behavior of $d_{2}(t)$ is not  yet significant. Physically speaking, the damping has not yet picked up  due to the small velocity dispersion, and the accumulative effective of the noise force is not yet notable either. Thus at this stage the oscillator  behaves quite like a free renormalized oscillator in its initial state. If the initial state of the oscillator happens to be a squeezed thermal state of the same squeeze parameter and inverse temperature, then its two point function is expected to be nonstationary, like what we have discussed earlier in Sec.~\ref{E:bgksbssfg}. 

The effects of the backactions and backreactions will be discussed in greater detail in the following. 

\subsection{Approach to Equilibrium}\label{S:etvvhret}

We have seen a few examples where the contribution  from the nonstationary component of the squeezed thermal field vanishes at late times. In fact, from the expressions of the covariance matrix elements, or the correlation function of the oscillator, we can identify two different sources of nonstationarity in the present configuration. One results from the nonequilibrium evolution of the internal degree of freedom, and the other from the nonstationarity in the bath field. The effects of the former are well investigated in~\cite{NEqFE}, so here we wish to study how the nonstationarity in the bath can affect the energy exchange between the detector and the bath field.

We first examine the power delivered by the bath field, that is,
\begin{equation}
	P_{\xi}(t)=\frac{e}{2}\,\langle\bigl\{\hat{\phi}_{h}(\mathbf{0},t),\,\dot{\hat{\chi}}(t)\bigr\}\rangle=\frac{e^{2}}{m}\int_{0}^{t}\!ds\;\dot{d}_{2}(t-s)\,G_{H,0}^{(\phi)}(s,t)\,.
\end{equation}
{This is nothing but the mechanical power if we identify $e\hat{\phi}_{h}(x)$ as a force,   similar to the Lorentz force in elemectromagnetism.} For the squeezed thermal bath, it takes the form
\begin{align}
	P_{\xi}(t)&=8\pi\gamma\int_{0}^{\infty}\!\frac{d\omega}{2\pi}\;\frac{\omega}{4\pi}\coth\frac{\beta\omega}{2}\biggl\{-\sinh2\eta\Bigl[e^{-i\omega t}e^{+i\theta}\dot{f}(t;\omega)+e^{+i\omega t}e^{-i\theta}\dot{f}^{*}(t;\omega)\Bigr]\biggr.\notag\\
	&\qquad\qquad\qquad\qquad\qquad\qquad+\biggl.\cosh2\eta\Bigl[e^{-i\omega t}\dot{f}^{*}(t;\omega)+e^{+i\omega t}\dot{f}(t;\omega)\Bigr]\biggr\}\,,\label{E:dkbgrtss}
\end{align}
where $e^{2}=8\pi\gamma m$. It is convenient to write $\dot{f}(t;\omega)$ as
\begin{align}
	\dot{f}(t;\omega)&=-i\omega\,\tilde{d}_{2}(\omega)\,e^{-i\omega t}\,g(t;\omega)\,,&g(t;\omega)&=1-i\,\frac{e^{+i\omega t}}{\omega}\,\dot{d}_{1}(t)-\dot{d}_{2}(t)\,e^{+i\omega t}\,,
\end{align}
so that Eq.~\eqref{E:dkbgrtss} becomes
\begin{align}
	P_{\xi}(t)&=8\pi\gamma\int_{0}^{\infty}\!\frac{d\omega}{2\pi}\;\frac{\omega}{4\pi}\coth\frac{\beta\omega}{2}\biggl\{-\sinh2\eta\Bigl[-i\omega\,\tilde{d}_{2}^{\vphantom{*}}(\omega)\,e^{-i2\omega t}e^{+i\theta}g(t;\omega)+i\omega\,\tilde{d}^{*}_{2}(\omega)\,e^{+i2\omega t}e^{-i\theta}g^{*}(t;\omega)\Bigr]\biggr.\notag\\
	&\qquad\qquad\qquad\qquad\qquad\qquad+\biggl.\cosh2\eta\Bigl[+i\omega\,\tilde{d}^{*}_{2}(\omega)\,g^{*}(t;\omega)-i\omega\,\tilde{d}^{\vphantom{*}}_{2}(\omega)\,g(t;\omega)\Bigr]\biggr\}\,.
\end{align}
In the limit $t\to\infty$, we find $g(t;\omega)\to1$, and then we find
\begin{align}
	P_{\xi}(\infty)&=8\pi\gamma\int_{0}^{\infty}\!\frac{d\omega}{2\pi}\;\frac{\omega}{4\pi}\coth\frac{\beta\omega}{2}\biggl\{-\sinh2\eta\Bigl[-i\omega\,\tilde{d}_{2}^{\vphantom{*}}(\omega)\,e^{-i2\omega t}e^{+i\theta}+i\omega\,\tilde{d}^{*}_{2}(\omega)\,e^{+i2\omega t}e^{-i\theta}\Bigr]\biggr.\notag\\
	&\qquad\qquad\qquad\qquad\qquad\qquad+\biggl.\cosh2\eta\Bigl[+i\omega\,\tilde{d}^{*}_{2}(\omega)-i\omega\,\tilde{d}^{\vphantom{*}}_{2}(\omega)\Bigr]\biggr\}\,.\label{E:gkrjtfd}
\end{align}
Noticeably the stationary component approaches a constant, but the nonstationary component still has a sinusoidal factor $e^{\pm i2\omega t}$, so it is not immediately clear  whether the nonstationary component will vanish at late times or approach a constant as well.  Looking more closely, we see the nonstationary  term in \eqref{E:gkrjtfd} contains the integral
\begin{align}
	J_{n}&=\int_{0}^{\infty}\!\frac{d\omega}{2\pi}\;\frac{\omega}{4\pi}\,e^{-n\beta\omega}\,\Bigl[-i\omega\,\tilde{d}_{2}^{\vphantom{*}}(\omega)\,e^{-i2\omega t}\Bigr]\notag\\
	&=-\frac{i}{16\pi^{2}}\biggl[\frac{2i}{2t-in\beta}+\frac{1}{\Omega}\Bigl\{e^{(\gamma-i\Omega)(2t-in\beta)}\bigl(\gamma-i\Omega\bigr)^{2}\operatorname{E}_{1}[(\gamma-i\Omega)(2t-in\beta)]\Bigr.\biggr.\notag\\
	&\qquad\qquad\qquad\qquad\qquad\qquad-\biggl.\Bigl.e^{(\gamma+i\Omega)(2t-in\beta)}\bigl(\gamma+i\Omega\bigr)^{2}\operatorname{E}_{1}[(\gamma-i\Omega)(2t+in\beta)]\Bigr\}\biggr]\,,
\end{align}
where the factor $e^{-n\beta\omega}$ results from
\begin{equation}\label{E:ljfer}
	\coth\frac{\beta\omega}{2}=1+2\sum_{n=1}^{\infty}e^{-n\beta\omega}\,,
\end{equation}
and $\operatorname{E}_{n}(z)$ is the exponential integral function defined by
\begin{equation}
	\operatorname{E}_{n}(z)=\int_{1}^{\infty}\!dt\;\frac{e^{-zt}}{t^{n}}\,,
\end{equation}
which has a branch cut along $-\infty$ to $0$ in the complex $z$ plane. Since we are interested in the large time $t\to\infty$ and the low temperature $\beta\to\infty$ limits,  and in \eqref{E:ljfer} the summation index $n$  runs from 1 to $\infty$, so we can introduce a ``large'' complex parameter $z=2t-in\beta$ and then carry out a large $z$ expansion. Doing so, we obtain
\begin{align}
	J_{n}\simeq\frac{1}{4\pi^{2}\omega_{\textsc{r}}^{2}z^{3}}-\frac{3\gamma}{2\pi^{2}\omega_{\textsc{r}}^{4}z^{4}}+\cdots\,,
\end{align}
with $z=2t-in\beta$. We will need to evaluate the following summations
\begin{align}
	\sum_{n=1}^{\infty}\frac{1}{(2t-in\beta)^{3}}&=\frac{i}{2\beta^{3}}\,\psi_{2}(1+i\,\frac{2t}{\beta})\simeq\frac{1}{8\beta t^{2}}+\cdots\,,\\
	\sum_{n=1}^{\infty}\frac{1}{(2t-in\beta)^{4}}&=\frac{1}{6\beta^{4}}\,\psi_{3}(1+i\,\frac{2t}{\beta})\simeq\frac{1}{24\beta t^{3}}+\cdots\,,
\end{align}
where we have assumed $t\gg\beta$ for any finitely large $\beta$. We do not need to worry about the $\beta\to\infty$ limit for the moment. Thus we see the finite temperature contributions in the nonstationary component of \eqref{E:gkrjtfd} falls off to zero at least like $t^{-2}$, not exponentially fast. The vacuum contribution in \eqref{E:ljfer} can be found by replacing $n\beta$ in $J_{n}$ by $\epsilon$, and we obtain
\begin{equation}
	J_{0}\simeq\frac{1}{4\pi^{2}\omega_{\textsc{r}}^{2}(2t-i\epsilon)^{3}}-\frac{3\gamma}{2\pi^{2}\omega_{\textsc{r}}^{4}(2t-i\epsilon)^{4}}+\cdots\,,
\end{equation}
so this contribution falls off a little faster like $t^{-3}$. Thus we conclude that in the limit $t\to\infty$, the nonstationary contribution in $P_{\xi}$ approaches zero. Note that these do not tell completely the temporal behavior of $P_{\xi}(\infty)$; it is just to show that the nonstationary contribution of the power $P_{\xi}$ does vanish in the limit $t\to\infty$. Thus we find
\begin{align}
	P_{\xi}(\infty)&=8\pi\gamma\int_{0}^{\infty}\!\frac{d\omega}{2\pi}\;\cosh2\eta\,\frac{\omega}{4\pi}\coth\frac{\beta\omega}{2}\times2\omega\,\operatorname{Im}\tilde{d}_{2}(\omega)\notag\\
	&=e^{2}\int_{-\infty}^{\infty}\!\frac{d\omega}{2\pi}\;\cosh2\eta\,\frac{\omega^{2}}{4\pi}\coth\frac{\beta\omega}{2}\,\operatorname{Im}\tilde{G}_{R}^{(\chi)}(\omega)\,,
\end{align}
where we recalled that 
\begin{equation}
	\tilde{G}_{R}^{(\chi)}(\omega)=\frac{\tilde{d}_{2}(\omega)}{m}\,.
\end{equation}
The power delivered by the bath field into the internal degree of freedom of the detector approaches a time-independent constant at late times, even though the bath is nonstationary and nonequilibrium.

On the other hand, the power trickling back to the bath at late times is given by
\begin{align}
	P_{\gamma}(\infty)&=-2m\gamma\,\langle\dot{\hat{\chi}}^{2}(\infty)\rangle=-\frac{2\gamma}{m}\,\langle\hat{p}^{2}(\infty)\rangle\,.
\end{align}
From \eqref{E:lgeure} and the definition $\hat{p}=m\dot{\hat{\chi}}$, we arrive at
\begin{align}\label{E:kgkerue}
	P_{\gamma}(\infty)&=-\cosh2\eta\,\frac{2\gamma}{m}\,e^{2}\int_{-\infty}^{\infty}\!\frac{d\omega}{2\pi}\;\frac{\omega^{3}}{4\pi}\coth\frac{\beta\omega}{2}\,\lvert\tilde{d}_{2}(\omega)\rvert^{2}\,.
\end{align}
It is straightforward to show that
\begin{equation}
	2\gamma\omega\,\lvert\tilde{d}_{2}(\omega)\rvert^{2}=\operatorname{Im}\tilde{d}_{2}(\omega)\,,
\end{equation}
and this enables us to write \eqref{E:kgkerue} as
\begin{align}
	P_{\gamma}(\infty)&=-e^{2}\int_{-\infty}^{\infty}\!\frac{d\omega}{2\pi}\;\cosh2\eta\frac{\omega^{2}}{4\pi}\coth\frac{\beta\omega}{2}\,\operatorname{Im}\tilde{G}_{R}^{(\chi)}(\omega)\,.
\end{align}
It is also independent of time, and therefore we indeed have
\begin{equation}
	P_{\xi}(\infty)+P_{\gamma}(\infty)=0\,.
\end{equation}
That is, the dissipative energy flux in the end matches with the power delivered by the nonstationary quantum fluctuations of the bath. Therefore we have a balanced energy exchange between the internal degree of freedom of the detector and the nonstationary, nonequilibrium squeezed bath field. This tells us that the dynamics of the internal degrees of freedom of our detector or atom, the Brownian oscillator,  does reach equilibration. In fact, the above arguments also offer a stronger statement that not only does the sum of two powers vanish at late times, both of them are time-independent constants of the same magnitude even when the squeezed thermal bath is time dependent.

Finally, due to the vanishing contribution of the nonstationary component in the late-time limit, we recognize that
\begin{equation}
	\langle\hat{\chi}^{2}(\infty)\rangle=\int_{-\infty}^{\infty}\!\frac{d\omega}{2\pi}\;\tilde{G}_{H,s}^{(\chi)}(\omega)\,,
\end{equation}
such that from \eqref{E:lgeure} we have
\begin{align}
	\tilde{G}_{H,s}^{(\chi)}(\omega)=\frac{2\gamma\omega}{m}\,\cosh2\eta\,\coth\frac{\beta\omega}{2}\,\lvert\tilde{d}_{2}(\omega)\rvert^{2}&=\cosh2\eta\,\coth\frac{\beta\omega}{2}\,\operatorname{Im}\tilde{G}_{R}^{(\chi)}(\omega)\notag\\
	&=\coth\frac{\beta_{\textsc{s}}\omega}{2}\,\operatorname{Im}\tilde{G}_{R}^{(\chi)}(\omega)
\end{align}
 This is a fluctuation-dissipation relation for the internal degree of freedom of the detector coupled to the squeezed thermal bath. It has two features distinct  from the plain thermal bath cases we have studied before: 1) only the stationary component\footnote{There is actually no need to emphasize the ``stationary component'' because at late times, the whole noise kernel of the internal degrees of  freedom  contains only the stationary part.} of the noise kernel of the internal degree of freedom of the detector/atom is involved, and 2) the proportionality constant has an additional factor due to squeezing. This factor is always equal to or grater than unity in the open system configuration. The proportionality constant can be re-written in terms of the effective temperature $\beta_{\textsc{s}}^{-1}$ with the help of \eqref{E:gbksbkdfgs}, so that  it assumes the conventional form.

{A remark is in place here on the nature of  the effective temperature  introduced in \eqref{E:gbksbkdfgs} and the somewhat intriguing role it plays in the final equilibrium state of the internal dynamics.}
 Since $\langle\hat{\chi}^{2}(\infty)\rangle$ and $\langle\hat{p}^{2}(\infty)\rangle$ are proportional to $\cosh2\eta$ but independent of $\theta$, the effective temperature will depend on $\eta$ only in this equilibrium state. The effective temperature defined  in \eqref{E:gbksbkdfgs} is consistent with the nonequilibrium effective temperature we introduced in~\cite{NEqFE}
\begin{equation}
	\beta_{\textsc{eff}}=\frac{2}{\omega_{\textsc{r}}}\,\ln\frac{1+\sqrt{1+4\mathfrak{S}}}{2\sqrt{\mathfrak{S}}}\,.
\end{equation}
and is a special case of the latter when the reduced system reaches its equilibrium state. The uncertainty function 
\begin{align}
	\mathfrak{S}&=\langle\hat{\chi}^{2}\rangle\langle\hat{p}^{2}\rangle-\frac{1}{4}\langle\bigl\{\hat{\chi}(t),\hat{p}(t)\bigr\}\rangle^{2}-\frac{1}{4}\,,
\end{align}
is related to the Robertson-Schr\"odinger uncertainty relation. Thus it will be a monotonic function of $\eta$. Owing to the factor $\cosh2\eta$, the system may appear to have a much higher effective temperature than the bath's temperature $\beta^{-1}$, but since we have shown that $\cosh2\eta$ is an overall factor multiplied to the covariance matrix elements of the oscillator coupled to a thermal bath,
\begin{align}
	\langle\hat{\chi}^{2}(\infty)\rangle&=\cosh2\eta\,\langle\hat{\chi}^{2}(\infty)\rangle_{\beta}\,,&\langle\hat{p}^{2}(\infty)\rangle&=\cosh2\eta\,\langle\hat{p}^{2}(\infty)\rangle_{\beta}\,,
\end{align}
the quantum nature of the system does not seem to be affected by the squeeze parameter $\eta$. We may on the surface claim that the quantumness survives at higher system temperatures, but this could be illusory\footnote{{Although the detector feels hotter in the squeezed bath than the thermal bath of the same $\beta$,  since the detector's observables scale up by a factor $\cosh2\eta$,  but the statistics does not change accordingly. Take an extreme example with a very large $\eta$,   the effective temperature hikes high up, and at such a high temperature one would expect the thermal fluctuations to become random. But in fact it is not. The system  still behaves like a low-temperature one. A squeezed state remains quantum in nature.  See Footnote 1 for additional comments. }}.

At this point, we have investigated the dynamics and equilibration of the internal degree of freedom of the detector, bilinearly coupled to a single Gaussian squeezed thermal bath. In the next section we will consider a more general case, when the bath is driven by some external agent which results in different modes in the bath acquiring a different time-dependent squeezing.   Oscillators  with time-dependent natural frequencies are called parametric oscillators, so we shall call a bath made up of parametric oscillators a parametric bath for short, likewise for fields whose normal modes are represented by oscillators with time-dependent frequencies.  A system coupled to such a parametric bath  in general is not expected to have an equilibrium state, which makes it  more interesting and challenging for our purpose.

\section{Oscillator in a thermal bath with time-dependent squeezing}\label{S:bggkdfdd}
 In contrast to the case  in the previous section  where the squeeze parameter of the bath is fixed, here we treat a bath whose squeezing changes with time {in a parametric process}.  For simplicity and without loss of generality in bringing out the key physics, we will consider the  {so-called `statically-bounded'} situation where the parameter of the bath field that accounts for the parametric process begins with a constant value, {and  smoothly transits to} a different constant value according to some specified functional form over a finite time interval.  The rate of change can be arbitrary, and not restricted to be gradual. In fact,  nonadiabatic changes produce qualitatively different effects,  such as particle creation in dynamical Casimir effort or cosmological particle creation.

We first consider the dynamics of a quantum scalar field undergoing the aforementioned parametric process.  {Then we derive the dynamics of the internal degrees of freedom (idf) of  a harmonic atom or a Unruh-DeWitt detector coupled to such a field.  From there we analyze the late time behavior of the idf and ask if  {equilibration of the internal dynamics is possible and if} a FDR exists for the idf of the detector/atom.  
   
Consider a real bath field described by the action
\begin{align}\label{E:dgksjfbrt}
	S&=\frac{1}{2}\int\!dtd^{3}\mathbf{x}\;\biggl\{\Bigl[\partial_{t}\phi(\mathbf{x},t)\Bigr]^{2}-\Bigl[\nabla\phi(\mathbf{x},t)\Bigr]^{2}-\mathfrak{m}^{2}(t)\,\phi^{2}(\mathbf{x},t)\biggr\}\,.
\end{align}
The time dependent function $\mathfrak{m}(t)$ acting like a mass accounts for the effects of the parametric process of interest. Suppose it changes from one constant value $\mathfrak{m}_{i}$ at time $t=\mathfrak{t}_{i}\geq0$ monotonically and sufficiently continuously to another constant value $\mathfrak{m}_{f}$ at time $t=\mathfrak{t}_{f}$.  Thus before $\mathfrak{t}_{i}$ and after $\mathfrak{t}_{f}$, the field $\chi(\mathbf{x},t)$ behaves like a free, real massive (with fixed value) scalar field.

Let us expand $\phi(\mathbf{x},t)$ by
\begin{equation}
	\phi(\mathbf{x},t)=\int\!\!\frac{d^{3}\mathbf{k}}{(2\pi)^{\frac{3}{2}}}\;e^{+i\mathbf{k}\cdot\mathbf{x}}\,\varphi_{\mathbf{k}}(t)\,,
\end{equation}
whence the action \eqref{E:dgksjfbrt} becomes
\begin{equation}
	S=\int\!dt\;\frac{1}{2}\int\!d^{3}\mathbf{k}\,\Bigl\{\dot{\varphi}_{\mathbf{k}}^{\vphantom{*}}(t)\dot{\varphi}_{\mathbf{k}}^{*}(t)-\omega^{2}(t)\,\varphi_{\mathbf{k}}^{\vphantom{*}}(t)\varphi_{\mathbf{k}}^{*}(t)\Bigr\}\,,
\end{equation}
with $\omega^{2}(t)=\mathbf{k}^{2}+\mathfrak{m}^{2}(t)$. The mode amplitude function  $\varphi_{\mathbf{k}}(t)$ then satisfies an equation of motion of a parametric oscillator
\begin{equation}\label{E:rirtyrd}
	\ddot{\varphi}_{\mathbf{k}}(t)+\omega^{2}(t)\,\varphi_{\mathbf{k}}(t)=0\,.
\end{equation}
The solutions to the corresponding Heisenberg equation are formally given by
\begin{align}
	\hat{\varphi}_{\mathbf{k}}(t)&=d^{(1)}_{+\mathbf{k}}(t)\,\hat{\varphi}_{+\mathbf{k}}(0)+d^{(2)}_{+\mathbf{k}}(t)\,\hat{\pi}_{-\mathbf{k}}(0)\,,\label{E:dgbksbgsd1}\\
	\hat{\pi}_{\mathbf{k}}(t)&=\dot{d}^{(1)}_{-\mathbf{k}}(t)\,\hat{\varphi}_{-\mathbf{k}}(0)+\dot{d}^{(2)}_{-\mathbf{k}}(t)\,\hat{\pi}_{+\mathbf{k}}(0)\,,
\end{align}
where $\hat{\pi}_{\mathbf{k}}^{\vphantom{*}}(t)=\dot{\varphi}_{\mathbf{k}}^{*}(t)=\dot{\varphi}_{-\mathbf{k}}(t)$ is the canonical momentum conjugated to $\hat{\varphi}_{\mathbf{k}}(t)$. In the context of dynamical evolution, it proves convenient to introduce  a special set of homogeneous solutions to \eqref{E:rirtyrd}, $d^{(1)}_{\mathbf{k}}(t)$, $d^{(2)}_{\mathbf{k}}(t)$ which satisfy the initial conditions
\begin{align}
	d^{(1)}_{\mathbf{k}}(0)&=1\,,&\dot{d}^{(1)}_{\mathbf{k}}(0)&=0\,,&d^{(2)}_{\mathbf{k}}(0)&=0\,,&\dot{d}^{(2)}_{\mathbf{k}}(0)&=1\,,
\end{align}
for each mode $\mathbf{k}$. The canonical commutation relation $[\hat{\varphi}_{\mathbf{k}}(t),\hat{\pi}_{\mathbf{k}}(t)]=i$ gives the Wronskian or the normalization condition,
\begin{equation}\label{E:fgksbrt}
	d^{(1)}_{\mathbf{k}}(t)\dot{d}^{(2)}_{\mathbf{k}}(t)-d^{(2)}_{\mathbf{k}}(t)\dot{d}^{(1)}_{\mathbf{k}}(t)=1\,.
\end{equation}
From \eqref{E:rirtyrd}, we observe that $d^{(i)}_{\mathbf{k}}(t)$ in fact depends on $k=\lvert\mathbf{k}\rvert$.

Now suppose at the initial time $t=0$, we expand $\hat{\varphi}_{\mathbf{k}}$, $\hat{\pi}_{\mathbf{k}}$ in term of the creation and annihilation operators $\hat{a}^{\dagger}_{\mathbf{k}}$, $\hat{a}^{\vphantom{\dagger}}_{\mathbf{k}}$
\begin{align}
	\hat{\varphi}_{+\mathbf{k}}&=\frac{1}{\sqrt{2\omega_{i}}}\,\bigl(\hat{a}^{\dagger}_{-\mathbf{k}}+\hat{a}^{\vphantom{\dagger}}_{\mathbf{k}}\bigr)\,,&\hat{\pi}_{-\mathbf{k}}=i\sqrt{\frac{\omega_{i}}{2}}\,\bigl(\hat{a}^{\dagger}_{-\mathbf{k}}-\hat{a}^{\vphantom{\dagger}}_{\mathbf{k}}\bigr)\,,
\end{align}
which satisfies $[\hat{a}^{\vphantom{\dagger}}_{\mathbf{k}},\hat{a}^{\dagger}_{\mathbf{k}}]=1$. Then by \eqref{E:dgbksbgsd1}, we can express $\hat{\chi}(t)$ in terms of $\hat{a}^{\vphantom{\dagger}}_{\mathbf{k}}(0)$ and $\hat{a}^{\dagger}_{\mathbf{k}}(0)$ 
\begin{align}
	\hat{\varphi}_{\mathbf{k}}(t)=\frac{1}{\sqrt{2\omega_{i}}}\Bigl\{\Bigl[d^{(1)}_{\mathbf{k}}(t)-i\,\omega_{i}\,d^{(2)}_{\mathbf{k}}(t)\Bigr]\,\hat{a}^{\vphantom{\dagger}}_{\mathbf{k}}(0)+\Bigl[d^{(1)}_{\mathbf{k}}(t)+i\,\omega_{i}\,d^{(2)}_{\mathbf{k}}(t)\Bigr]\,\hat{a}^{\dagger}_{\mathbf{k}}(0)\Bigr\}\,,
\end{align}
with the shorthand notation $\omega(t)=\omega_{i}$ when $t\leq\mathfrak{t}_{i}$. Thus the field operator $\hat{\phi}(\mathbf{x},t)$ has a plane-wave expansion
\begin{align}\label{E:riyhtiygh}
	\hat{\phi}(\mathbf{x},t)=\int\!\!\frac{d^{3}\mathbf{k}}{(2\pi)^{\frac{3}{2}}}\;\frac{1}{\sqrt{2\omega_{i}}}\Bigl\{\hat{a}^{\vphantom{\dagger}}_{\mathbf{k}}(0)\,e^{+i\mathbf{k}\cdot\mathbf{x}}\,f_{\mathbf{k}}(t)+\hat{a}^{\dagger}_{\mathbf{k}}(0)\,e^{-i\mathbf{k}\cdot\mathbf{x}}\,f_{\mathbf{k}}^{*}(t)\Bigr\}\,,
\end{align}
and the corresponding momentum operator $\hat{\pi}(\mathbf{x},t)$ is given by $\hat{\pi}(\mathbf{x},t)=\dot{\hat{\chi}}(\mathbf{x},t)$, with $f_{\mathbf{k}}(t)=d^{(1)}_{\mathbf{k}}(t)-i\,\omega_{i}\,d^{(2)}_{\mathbf{k}}(t)$. We can verify that they satisfy the standard commutation relation $\bigl[\hat{\phi}(\mathbf{x},t),\hat{\pi}(\mathbf{x}',t)\bigr]=i\,\delta^{(3)}(\mathbf{x}-\mathbf{x}')$, with the help of \eqref{E:fgksbrt}.

\subsection{Retarded Green's function}\label{S:gbsgs}
With the field expansion \eqref{E:riyhtiygh}, the retarded Green's function $G_{R,0}^{(\phi)}(\mathbf{x},t;\mathbf{x}',t')$ of the free field $\hat{\phi}(\mathbf{x},t)$ is given by
\begin{align}
	G_{R,0}^{(\phi)}(\mathbf{x},t;\mathbf{x}',t')&=i\,\theta(t-t')\,\bigl[\hat{\phi}(\mathbf{x},t),\hat{\phi}(\mathbf{x}',t')\bigr]\notag\\
	&=-\theta(t-t')\int\!\!\frac{d^{3}\mathbf{k}}{(2\pi)^{3}}\;e^{+i\mathbf{k}\cdot(\mathbf{x}-\mathbf{x}')}\Bigl\{d^{(1)}_{\mathbf{k}}(t)d^{(2)}_{\mathbf{k}}(t')-d^{(2)}_{\mathbf{k}}(t)d^{(1)}_{\mathbf{k}}(t')\Bigr\}\,.\label{E:ptyiort}
\end{align}
When the field undergoes a parametric process, its retarded Green's function in general is not stationary, so the integrand in \eqref{E:ptyiort} will not reduce to $d^{(2)}_{\mathbf{k}}(t-t')$. However, in the regimes of either $t$, $t'<\mathfrak{t}_{i}$ or $t$, $t'>\mathfrak{t}_{f}$, the Green's function behaves like the standard two-point function of the massive field, but with different mass $\mathfrak{m}$. The retarded Green's function and the Hadamard function of the free massive field of mass $m$ take on the form
\begin{align}
	G_{R,0}^{(\phi)}(x,x')&=\frac{\theta(T)}{2\pi}\,\Bigl[\delta(\sigma^{2})-\theta(\sigma^{2})\,\frac{m}{2\sqrt{\sigma^{2}}}\,J_{1}(m\sqrt{\sigma^{2}})\Bigr]\,,\label{E:ierjie}\\
	G_{H,0}^{(\phi)}(x,x')&=\frac{1}{4\pi}\Bigl[\theta(+\sigma^{2})\,\frac{m}{2\sqrt{+\sigma^{2}}}\,Y_{1}(m\sqrt{+\sigma^{2}})+\theta(-\sigma^{2})\,\frac{m}{\pi\sqrt{-\sigma^{2}}}\,K_{1}(m\sqrt{-\sigma^{2}})\Bigr]\,,
\end{align}
where $T=t-t'$ and $\sigma^{2}=T^{2}-\mathbf{R}^{2}$ with $\mathbf{R}=\mathbf{x}-\mathbf{x}'$. In our case $\mathbf{R}=0$, so $\sigma^{2}>0$ always, and the retarded Green's function \eqref{E:ptyiort} is simplified to
\begin{equation}
	G_{R,0}^{(\phi)}(\sigma;\mathbf{z})=-\frac{\theta(\sigma)}{4\pi}\Bigl\{2\delta'(\sigma)+\operatorname{sgn}(\sigma)\,\frac{m}{\sqrt{\sigma^{2}}}\,J_{1}(m\sqrt{\sigma^{2}})\Bigr\}\,,\label{E:turrnfjgdd}
\end{equation}
where $\theta(\sigma)\operatorname{sgn}(\sigma)=\theta(\sigma)$ and now $\sigma=t-t'$. Eq.~\eqref{E:ierjie}   {implies that the internal dynamics of the detector} can be influenced by its own radiation of this massive field emitted at earlier times.  {Therefore the internal dynamics} depends on  its past evolutionary history. The duration of the memory is quantified by $m^{-1}$. {For a very light field, it has a very long memory span, but since} the second term in \eqref{E:ierjie} or \eqref{E:turrnfjgdd} scales with $m^{2}$, it will have a negligible contribution, so that the influence of the field is essentially confined by the first term on the lightcone of the detector.

\subsection{Hadamard function}\label{S:rhgjfgfgdf}
The Hadamard function $G_{H,0}^{(\phi)}(\mathbf{x},t;\mathbf{x}',t')$, defined by
\begin{align}\label{E:gesmfs}
	G_{H,0}^{(\phi)}(\mathbf{x},t;\mathbf{x}',t')&=\frac{1}{2}\,\langle\bigl\{\hat{\phi}(\mathbf{x},t),\hat{\phi}(\mathbf{x}',t')\bigr\}\rangle
\end{align}
plays a special role in the nonequilibrium dynamics. It describes the correlation of the fluctuating force of the bath that imparts a stochastic component in our system's dynamics. That is why it is often called the noise kernel, in correspondence to the retarded Green's function, the dissipation kernel. Their dynamical significance has been discussed  in the earlier case when the internal dynamics of an Unruh-DeWitt detector is coupled to the fixed-value squeezed thermal bath field. Here the bath field, subjected to time-dependent squeezing in a parametric process, has much more complicated correlations and time dependence.

From \eqref{E:gesmfs} and the field expansion \eqref{E:riyhtiygh}, the Hadamard function of the parametric bath field takes the form 
\begin{align}
	G_{H,0}^{(\phi)}(\mathbf{x},t;\mathbf{x}',t')&=\int\!\!\frac{d^{3}\mathbf{k}}{(2\pi)^{3}}\frac{1}{2\omega_{i}}\biggl\{\,e^{+i\mathbf{k}(\mathbf{x}-\mathbf{x}')}\Bigl(\langle\hat{N}_{\mathbf{k}}(0)\rangle+\frac{1}{2}\Bigr)\Bigl[2d^{(1)}_{\mathbf{k}}(t)d^{(1)}_{\mathbf{k}}(t')+2\omega_{i}^{2}d^{(2)}_{\mathbf{k}}(t)d^{(2)}_{\mathbf{k}}(t')\Bigr]\biggr.\notag\\
	&\qquad\qquad\qquad\;+e^{+i\mathbf{k}(\mathbf{x}+\mathbf{x}')}\,\langle\hat{a}_{\mathbf{k}}^{\hphantom{\dagger}2}(0)\rangle\Bigl[d^{(1)}_{\mathbf{k}}(t)d^{(1)}_{\mathbf{k}}(t')-\omega_{i}^{2}\,d^{(2)}_{\mathbf{k}}(t)d^{(2)}_{\mathbf{k}}(t')\Bigr.\notag\\
	&\qquad\qquad\qquad\qquad\qquad\qquad\qquad\qquad-\Bigl.i\,\omega_{i}\,d^{(1)}_{\mathbf{k}}(t)d^{(2)}_{\mathbf{k}}(t')-i\,\omega_{i}\,d^{(2)}_{\mathbf{k}}(t)d^{(1)}_{\mathbf{k}}(t')\Bigr]\notag\\
	&\qquad\qquad\qquad\;+\biggl.e^{-i\mathbf{k}(\mathbf{x}+\mathbf{x}')}\,\langle\hat{a}_{\mathbf{k}}^{\dagger2}(0)\rangle\Bigl[d^{(1)}_{\mathbf{k}}(t)d^{(1)}_{\mathbf{k}}(t')-\omega_{i}^{2}\,d^{(2)}_{\mathbf{k}}(t)d^{(2)}_{\mathbf{k}}(t')\Bigr.\label{E:brtyufb}\\
	&\qquad\qquad\qquad\qquad\qquad\qquad\qquad\qquad+\Bigl.i\,\omega_{i}\,d^{(1)}_{\mathbf{k}}(t)d^{(2)}_{\mathbf{k}}(t')+i\,\omega_{i}\,d^{(2)}_{\mathbf{k}}(t)d^{(1)}_{\mathbf{k}}(t')\Bigr]\biggr\}\,.\notag
\end{align}
Here $\langle\cdots\rangle$ is the expectation value taken with respect to the initial state of the field at $t=0$. We note that there are two types of nonstationary in $G_{H,0}^{(\phi)}(\mathbf{x},t;\mathbf{x}',t')$ according to \eqref{E:brtyufb}. One results from the parametric process of the field and is contained in the fundamental solutions $d^{(i)}_{\mathbf{k}}(t)$; the other is due to the presence of nonvanishing $\langle\hat{a}_{\mathbf{k}}^{2}(0)\rangle$, much like the squeezed thermal state we discussed in the previous section. Thus, here we only focus on the first type of nonstationarity and let the initial state be a stationary state. The last four lines of \eqref{E:brtyufb} will then not be considered. However, it is interesting to emphasize that even under this consideration, the state of the parametric field will still tend to a squeezed state, a characteristic of the parametric process of a Gaussian system. This can be understood by the fact that the most general Gaussian state of a Gaussian system is the squeezed (thermal) state~\cite{OL12,NEqFE}. We also note that the operators inside the expectation values are evaluated at the initial time.

The nonstationarity from squeezing encapsulated in the evolution of the field operator will be made manifest if the field operator at any moment can be mapped from the ``in''-field, 
\begin{equation}\label{E:ornlnksbd}
	\hat{\phi}_{\textsc{in}}(\mathbf{x},t)=\int\!\!\frac{d^{3}\mathbf{k}}{(2\pi)^{\frac{3}{2}}}\;\frac{1}{\sqrt{2\omega_{i}}}\Bigl[e^{-i\omega_{i}t}\,\hat{a}^{\vphantom{\dagger}}_{\mathbf{k}}(0)\,e^{+i\mathbf{k}\cdot\mathbf{x}}+e^{+i\omega_{i}t}\,\hat{a}^{\dagger}_{\mathbf{k}}(0)\,e^{-i\mathbf{k}\cdot\mathbf{x}}\Bigr]\,,
\end{equation}
i.e., the free-field operator before the parametric process begins, by a suitable two-mode squeeze transformation (see Appendix~\ref{S:eotie} for some essential materials about two-mode squeezing.) 
\begin{equation}
	\hat{S}_{2}^{\dagger}(\zeta)\,\hat{\phi}_{\textsc{in}}(\mathbf{x},t)\,\hat{S}_{2}(\zeta)\,,
\end{equation}
where $\zeta=\{\zeta_{\mathbf{k}}\}$, and each $\zeta_{\mathbf{k}}$ has a polar decomposition of the form $\zeta_{\mathbf{k}}=\eta_{\mathbf{k}}\,e^{i\theta_{\mathbf{k}}}$
\begin{equation}\label{E:povghey}
	\hat{S}_{2}^{\vphantom{\dagger}}(\zeta_{\mathbf{k}}^{\vphantom{\dagger}})=\exp\Bigl[\zeta_{\mathbf{k}}^{*\vphantom{\dagger}}\,\hat{a}_{\mathbf{k}}^{\vphantom{\dagger}}\hat{a}_{-\mathbf{k}}^{\vphantom{\dagger}}-\zeta_{\mathbf{k}}^{\vphantom{\dagger}}\,\hat{a}_{\mathbf{k}}^{\dagger}\hat{a}_{-\mathbf{k}}^{\dagger}\Bigr]\,.
\end{equation}
Note that due to \eqref{E:rirtyrd} and $\omega^{2}(t)=\mathbf{k}^{2}+\mathfrak{m}^{2}(t)$, the squeezing in general is mode-dependent and,  for this, the squeeze parameter will carry a $\mathbf{k}$ subscript. Eq.~\eqref{E:povghey} implies that the squeeze parameter in fact is a function of $k=\lvert\mathbf{k}\rvert$.

Implementing squeezing by the Bogoliubov transformation
\begin{equation}
	\hat{S}_{2}^{\dagger}(\zeta_{\mathbf{k}}^{\vphantom{\dagger}})\,\hat{a}_{\mathbf{k}}^{\vphantom{\dagger}}\,\hat{S}_{2}(\zeta_{\mathbf{k}}^{\vphantom{\dagger}})=\alpha_{\mathbf{k}}^{\vphantom{\dagger}}\,\hat{a}_{\mathbf{k}}^{\vphantom{\dagger}}+\beta_{-\mathbf{k}}^{\vphantom{\dagger}*}\,\hat{a}_{-\mathbf{k}}^{\dagger}=\cosh\eta_{\mathbf{k}}^{\vphantom{\dagger}}\,\hat{a}_{\mathbf{k}}^{\vphantom{\dagger}}-e^{+i\theta_{\mathbf{k}}}\sinh\eta_{\mathbf{k}}^{\vphantom{\dagger}}\,\hat{a}_{-\mathbf{k}}^{\dagger}\,, 
\end{equation}	
then we have
\begin{align}\label{E:oeifbdvd}
	&\quad\hat{S}_{2}^{\dagger}(\zeta)\,\hat{\phi}_{\textsc{in}}(\mathbf{x},t)\,\hat{S}_{2}(\zeta)\\
	&=\int\!\!\frac{d^{3}\mathbf{k}}{(2\pi)^{\frac{3}{2}}}\;\frac{1}{\sqrt{2\omega_{i}}}\,\Bigl\{e^{+i\mathbf{k}\cdot\mathbf{x}}e^{-i\omega_{i}t}\,\Bigl[\alpha_{\mathbf{k}}^{\vphantom{\dagger}}\,\hat{a}^{\vphantom{\dagger}}_{\mathbf{k}}(0)+\beta_{-\mathbf{k}}^{*\vphantom{\dagger}}\,\hat{a}^{\dagger}_{-\mathbf{k}}(0)\Bigr]+e^{-i\mathbf{k}\cdot\mathbf{x}}e^{+i\omega_{i}t}\,\Bigl[\alpha_{\mathbf{k}}^{*\vphantom{\dagger}}\,\hat{a}^{\dagger}_{\mathbf{k}}(0)+\beta_{-\mathbf{k}}^{\vphantom{\dagger}}\,\hat{a}^{\vphantom{\dagger}}_{-\mathbf{k}}(0)\Bigr]\Bigr\}\notag\\
	&=\int\!\!\frac{d^{3}\mathbf{k}}{(2\pi)^{\frac{3}{2}}}\;\frac{1}{\sqrt{2\omega_{i}}}\,\Bigl\{\Bigl[d^{(1)}_{\mathbf{k}}(t)-i\,\omega_{i}\,d^{(2)}_{\mathbf{k}}(t)\Bigr]\,\hat{a}^{\vphantom{\dagger}}_{\mathbf{k}}(0)\,e^{+i\mathbf{k}\cdot\mathbf{x}}+\Bigl[d^{(1)}_{\mathbf{k}}(t)+i\,\omega_{i}\,d^{(2)}_{\mathbf{k}}(t)\Bigr]\,\hat{a}^{\dagger}_{\mathbf{k}}(0)\,e^{-i\mathbf{k}\cdot\mathbf{x}}\Bigr\}\,,\notag
\end{align}
where a change of variables $\mathbf{k}\to-\mathbf{k}$ is carried out as needed. We thus obtain
\begin{equation}\label{E:fnljsb1}
	d^{(1)}_{\mathbf{k}}(t)-i\,\omega_{i}\,d^{(2)}_{\mathbf{k}}(t)=e^{-i\omega_{i}t}\,\alpha_{\mathbf{k}}^{\vphantom{*}}(t)+e^{+i\omega_{i}t}\,\beta_{\mathbf{k}}^{\vphantom{*}}(t)\,,
\end{equation}
for $t>0$. Similarly, for the conjugate momentum $\hat{\pi}(\mathbf{x},t)$, we have
\begin{align}
	&\quad\hat{S}_{2}^{\dagger}(\zeta)\,\hat{\pi}_{\textsc{in}}(\mathbf{x},t)\,\hat{S}_{2}(\zeta)\\
	&=\int\!\!\frac{d^{3}\mathbf{k}}{(2\pi)^{\frac{3}{2}}}\;\frac{1}{\sqrt{2\omega_{i}}}\,\Bigl\{\Bigl[\dot{d}^{(1)}_{\mathbf{k}}(t)-i\,\omega_{i}\,\dot{d}^{(2)}_{\mathbf{k}}(t)\Bigr]\,\hat{a}^{\vphantom{\dagger}}_{\mathbf{k}}(0)\,e^{+i\mathbf{k}\cdot\mathbf{x}}+\Bigl[\dot{d}^{(1)}_{\mathbf{k}}(t)+i\,\omega_{i}\,\dot{d}^{(2)}_{\mathbf{k}}(t)\Bigr]\,\hat{a}^{\dagger}_{\mathbf{k}}(0)\,e^{-i\mathbf{k}\cdot\mathbf{x}}\Bigr\}\,,\notag
\end{align}
and find
\begin{equation}\label{E:fnljsb2}
	\dot{d}^{(1)}_{\mathbf{k}}(t)-i\,\omega_{i}\,\dot{d}^{(2)}_{\mathbf{k}}(t)=-i\,\omega_{i}\,e^{-i\omega_{i}t}\,\alpha_{\mathbf{k}}(t)+i\,\omega_{i}\,e^{+i\omega_{i}t}\,\beta_{\mathbf{k}}(t)\,.
\end{equation}
Eqs.~\eqref{E:fnljsb1} and \eqref{E:fnljsb2} lead to
\begin{align}
	\alpha_{\mathbf{k}}(t)&=\frac{1}{2\omega_{i}}\,e^{+i\omega_{i}t}\,\Bigl[\omega_{i}\,d^{(1)}_{\mathbf{k}}(t)+i\,\dot{d}^{(1)}_{\mathbf{k}}(t)-i\,\omega_{i}^{2}d^{(2)}_{\mathbf{k}}(t)+\omega_{i}\,\dot{d}^{(2)}_{\mathbf{k}}(t)\Bigr]\,,\label{E:gbkrjbkdg1}\\
	\beta_{\mathbf{k}}(t)&=\frac{1}{2\omega_{i}}\,e^{-i\omega_{i}t}\,\Bigl[\omega_{i}\,d^{(1)}_{\mathbf{k}}(t)-i\,\dot{d}^{(1)}_{\mathbf{k}}(t)-i\,\omega_{i}^{2}d^{(2)}_{\mathbf{k}}(t)-\omega_{i}\,\dot{d}^{(2)}_{\mathbf{k}}(t)\Bigr]\,,\label{E:gbkrjbkdg2}
\end{align}
and we can verify that 
\begin{equation}
	\lvert\alpha_{\mathbf{k}}\rvert^{2}-\lvert\beta_{\mathbf{k}}\rvert^{2}=d^{(1)}_{\mathbf{k}}(t)\dot{d}^{(2)}_{\mathbf{k}}(t)-\dot{d}^{(1)}_{\mathbf{k}}(t)d^{(2)}_{\mathbf{k}}(t)=1\,,
\end{equation}
and
\begin{equation}
	\lvert\alpha_{\mathbf{k}}\rvert^{2}+\lvert\beta_{\mathbf{k}}\rvert^{2}=\frac{1}{2\omega_{i}^{2}}\Big[\omega_{i}^{2}d^{(1)2}_{\mathbf{k}}(t)+\dot{d}^{(1)2}_{\mathbf{k}}(t)+\omega_{i}^{4}d^{(2)2}_{\mathbf{k}}(t)+\omega_{i}^{2}\dot{d}^{(2)2}_{\mathbf{k}}(t)\Bigr]=\cosh2\eta_{\mathbf{k}}\,.
\end{equation}
In this way, we can relate the evolution of the field, expressed in terms of the fundamental solutions, to the suitable squeeze parameter or the coefficients of the Bogoliubov transformation. During the parametric process, these parameters and coefficients are in general time-dependent. However, once the process ends at $t=\mathfrak{t}_{f}$, they turn into time-independent but still mode-dependent constants. Thus in the regime $t>\mathfrak{t}_{f}$, we can apply the results in Sec.~\ref{E:bgksbssfg} for a fixed-value squeezed thermal field to this parametrically driven field.

Suppose the initial state of the field before the parametric process is in a thermal state, then for $t$, $t'>\mathfrak{t}_{f}$, we can write the two-point function of the scalar field $\hat{\phi}(\mathbf{x},t)$ as
\begin{equation}
	\operatorname{Tr}\Bigl\{\hat{\rho}_{\beta}^{(\phi)}(0)\hat{\phi}(\mathbf{x},t)\hat{\phi}(\mathbf{x}',t')\Bigr\}=\operatorname{Tr}\Bigl\{\hat{\rho}_{\textsc{st}}^{(\phi)}(0)\,\hat{\phi}_{\textsc{in}}(\mathbf{x},t)\hat{\phi}_{\textsc{in}}(\mathbf{x}',t')\Bigr\}\,,
\end{equation}
where we have defined the squeezed thermal state $\hat{\rho}_{\textsc{st}}^{(\phi)}$ by
\begin{equation}
	\hat{\rho}_{\textsc{st}}^{(\phi)}=\hat{S}_{2}(\zeta)\hat{\rho}_{\beta}^{(\phi)}\hat{S}_{2}^{\dagger}(\zeta)\,.
\end{equation} 
Since the retarded Green's function is independent of the field state, here we focus on the Hadamard function of the scalar field
\begin{align}
	G_{H,0}^{(\phi)}(\mathbf{x},t;\mathbf{x}',t')&=\frac{1}{2}\operatorname{Tr}\Bigl[\hat{\rho}_{\beta}^{(\phi)}(0)\bigl\{\hat{\phi}(\mathbf{x},t),\,\hat{\phi}(\mathbf{x}',t')\bigr\}\Bigr]\,.
\end{align}	
Before writing down the explicit expression for the Hadamard function, it is convenient to first spell out the field operator expansion in terms of the Bogoliubov coefficients and the in-modes,
\begin{align}
	\hat{\phi}(\mathbf{x},t)=\int\!\frac{d^{2}\mathbf{k}}{(2\pi)^{\frac{3}{2}}}\frac{1}{\sqrt{2\omega_{i}}}\;\Bigl[\alpha_{\mathbf{k}}^{\vphantom{\dagger}}\,\hat{a}_{\mathbf{k}}^{\vphantom{\dagger}}(0)\,e^{+i\mathbf{k}\cdot\mathbf{x}-i\omega_{i}t}+\beta_{\mathbf{k}}^{*\vphantom{\dagger}}\hat{a}_{\mathbf{k}}^{\dagger}(0)\,e^{-i\mathbf{k}\cdot\mathbf{x}-i\omega_{i}t}+\text{H.C.}\Bigr]\,.
\end{align}
Therefore we obtain
\begin{align}
	&\quad G_{H,0}^{(\phi)}(\mathbf{x},t;\mathbf{x}',t')\notag\\
	&=\int\!\frac{d^{3}\mathbf{k}}{(2\pi)^{3}}\frac{1}{4\omega_{i}}\,\coth\frac{\beta\omega_{i}}{2}\Bigl[\alpha_{\mathbf{k}}^{\vphantom{\dagger}}\beta_{\mathbf{k}}^{*\vphantom{\dagger}}\,e^{+i\mathbf{k}\cdot(\mathbf{x}-\mathbf{x}')-i\omega_{i}(t+t')}+\alpha_{\mathbf{k}}^{*\vphantom{\dagger}}\beta_{\mathbf{k}}^{\vphantom{\dagger}}\,e^{-i\mathbf{k}\cdot(\mathbf{x}-\mathbf{x}')+i\omega_{i}(t+t')}\Bigr.\notag\\
	&\qquad\qquad\qquad\qquad\qquad+\Bigl.\lvert\alpha_{\mathbf{k}}^{\vphantom{\dagger}}\rvert^{2}e^{+i\mathbf{k}\cdot(\mathbf{x}-\mathbf{x}')-i\omega_{i}(t-t')}+\lvert\beta_{\mathbf{k}}^{\vphantom{\dagger}}\rvert^{2}e^{-i\mathbf{k}\cdot(\mathbf{x}-\mathbf{x}')-i\omega_{i}(t-t')}+\text{C.C.}\Bigr]\notag\\
	&=\int\!\frac{d^{3}\mathbf{k}}{(2\pi)^{3}}\frac{1}{4\omega_{i}}\,\coth\frac{\beta\omega_{i}}{2}\Bigl[-e^{+i\theta_{-\mathbf{k}}^{\vphantom{\dagger}}}\cosh\eta_{\mathbf{k}}^{\vphantom{\dagger}}\sinh\eta_{-\mathbf{k}}^{\vphantom{\dagger}}\,e^{+i\mathbf{k}\cdot(\mathbf{x}-\mathbf{x}')-i\omega_{i}(t+t')}\Bigr.\notag\\
	&\qquad\qquad\qquad\qquad\qquad\quad-e^{-i\theta_{-\mathbf{k}}^{\vphantom{\dagger}}}\cosh\eta_{\mathbf{k}}^{\vphantom{\dagger}}\sinh\eta_{-\mathbf{k}}^{\vphantom{\dagger}}\,e^{-i\mathbf{k}\cdot(\mathbf{x}-\mathbf{x}')+i\omega_{i}(t+t')}\notag\\
	&\qquad\qquad\qquad\quad+\Bigl.\cosh^{2}\eta_{\mathbf{k}}^{\vphantom{\dagger}}\,e^{+i\mathbf{k}\cdot(\mathbf{x}-\mathbf{x}')-i\omega_{i}(t-t')}+\sinh^{2}\eta_{-\mathbf{k}}^{\vphantom{\dagger}}e^{-i\mathbf{k}\cdot(\mathbf{x}-\mathbf{x}')-i\omega_{i}(t-t')}+\text{C.C.}\Bigr]\label{E:gnfkgfkd}
\end{align}
with $\alpha_{\mathbf{k}}^{\vphantom{\dagger}}=\cosh\eta_{\mathbf{k}}^{\vphantom{\dagger}}$, $\beta_{\mathbf{k}}^{\vphantom{\dagger}}=-e^{-i\theta_{-\mathbf{k}}^{\vphantom{\dagger}}}\sinh\eta_{-\mathbf{k}}^{\vphantom{\dagger}}$,
\begin{align}
	\langle\hat{a}_{\mathbf{k}}^{\dagger}(0)\hat{a}_{\mathbf{k}}^{\vphantom{\dagger}}(0)\rangle_{\beta}+\frac{1}{2}&=\frac{1}{2}\,\coth\frac{\beta\omega_{i}}{2}\,.
\end{align}
Note that $\eta_{\mathbf{k}}$ in fact is a function of $\lvert\mathbf{k}\rvert$, so \eqref{E:gnfkgfkd} reduces to
\begin{align}
	G_{H,0}^{(\phi)}(\mathbf{x},t;\mathbf{x}',t')&=-\int\!\frac{d^{3}\mathbf{k}}{(2\pi)^{3}}\frac{1}{4\omega_{i}}\,\coth\frac{\beta\omega_{i}}{2}\,e^{+i\mathbf{k}\cdot(\mathbf{x}-\mathbf{x}')}\biggl\{\sinh2\eta_{k}\Big[e^{-i\omega_{i}(t+t')+i\theta_{k}}+e^{+i\omega_{i}(t+t')-i\theta_{k}}\Bigr]\biggr.\notag\\
&\qquad\qquad\qquad\qquad+\biggl.\cosh2\eta_{k}\,e^{+i\mathbf{k}\cdot(\mathbf{x}-\mathbf{x}')}\Big[e^{-i\omega_{i}(t-t')}+e^{+i\omega_{i}(t-t')}\Bigr]\biggr\}\,.\label{E:gbkdgsjt}
\end{align}
It is interesting to compare this with \eqref{E:rkjrt}. They look almost identical except that 1) the former now has mode-dependent parameters, and 2) they have different spatial dependence. Since in the current case, squeezing results from the global parametric process of the scalar field which is initially in a state that respects translational invariance in space, the factor $e^{+i\mathbf{k}\cdot(\mathbf{x}-\mathbf{x}')}$ will be preserved by the parametric process.

Therefore the following discussions about the internal dynamics of the detector coupled to such a parametrically drive bath field will be in close parallel to what we presented in the last section with minor modifications to account for the mode-dependent parameters. Again, we assume that the detector is fixed at the origin of the spatial coordinates, so that the difference in point 2) becomes moot.

\subsection{Detector dynamics in the parametric bath}

Suppose that the internal degrees of freedom of the detector is coupled to such a bath at the end of the parametric process of the bath field, then the internal dynamics will follow the same equation of motion as \eqref{E:thfsdf}. Now for convenience we will shift the origin of time coordinate to $\mathfrak{t}_{f}$ so that the parametric process of the field occurs at $t\leq 0$. In contrast to the case discussed in Sec.~\ref{S:trvkgdfhd}, since the bath field has acquired an effective mass from the parametric process, the discussions in Sec.~\ref{S:gbsgs} tell us that the internal dynamics now will be history-dependent, and additional nonlocal terms that account  for this memory effect will emerge in the equation of motion
\begin{equation}\label{E:bdfhdd}
	\ddot{\hat{\chi}}(t)+\omega^{2}_{\textsc{r}}\,\hat{\chi}(t)+2\gamma\,\dot{\chi}(t)+2\gamma\int_{0}^{t}\!ds\;\frac{\mathfrak{m}_{f}}{t-s}\,J_{1}(\mathfrak{m}_{f}(t-s))\,\hat{\chi}(s)=\frac{e}{m}\,\hat{\phi}_{h}(\mathbf{z},t)\,,
\end{equation}
for $t>0$, in comparison with \eqref{E:kjsskw}. Note that when the mass $\mathfrak{m}_{f}$ of the field quantum goes to zero, the fourth term on the lefthand side vanishes. On the other hand in general it does not vanish for $t>s$, that is, a timelike interval. Therefore the massive bath field can induce a non-Markovian effect on the internal dynamics of the detector from the same detector at earlier moments.

As before, we can construct a special set of homogeneous solution to \eqref{E:bdfhdd}, $d_{1}^{(\chi)}(t)$ and $d_{2}^{(\chi)}(t)$,   taking on the forms
\begin{align}
	d_{1}^{(\chi)}(t)&=e^{-\scriptstyle{\Upsilon} t}\Bigl[\cos\varpi t+\frac{\scriptstyle{\Upsilon}}{\varpi}\,\sin\varpi t\Bigr]\,,&d_{2}^{(\chi)}(t)&=\frac{e^{-\scriptstyle{\Upsilon} t}}{\varpi}\,\sin\varpi t\,,
\end{align}
where $\mathfrak{z}=-\sqrt{\smash[b]{\gamma}^{2}-\Omega^{2}\mp2\smash[b]{\gamma}\sqrt{\smash[b]{\mathfrak{m}_{f}^{2}}-\Omega^{2}}}$, and $\Upsilon=+\operatorname{Re}\mathfrak{z}$, $\varpi=-\operatorname{Im}\mathfrak{z}$. In the limit of small mass $\mathfrak{m}_{f}$ of the bath field particle, we note
\begin{equation}
	\mathfrak{z}\simeq-\gamma\pm i\Omega+\Bigl[\frac{\gamma}{2(\Omega^{2}+\gamma^{2})}\mp i\,\frac{\gamma^{2}}{2\Omega(\Omega^{2}+\gamma^{2})}\Bigr]\,\mathfrak{m}_{f}^{2}\,,
\end{equation}
so $d_{1,2}^{(\chi)}(t)$ will be similar to their counterparts \eqref{E:gbkdhg} for the detector coupled to a massless bath field.

\subsection{Energy balance and the FDR}\label{S:krghvjdfrg}
Now we are ready to examine the lat-time internal dynamics of the detector coupled to the parametrically driven bath field. The power delivered by the bath field is given by
\begin{align}
	P_{\xi}(t)=\frac{e^{2}}{m}\int_{0}^{t}\!ds\;\dot{d}_{2}^{(Q)}(t-s)\,G_{H,0}^{(\chi)}(t,s;\mathbf{z})\,,
\end{align}
which, according to \eqref{E:gbkdgsjt}, is
\begin{align}
	P_{\xi}(t)&=-8\pi\gamma\int\!\frac{d^{3}\mathbf{k}}{(2\pi)^{3}}\frac{1}{4\omega_{i}}\coth\frac{\beta\omega_{i}}{2}\,\sinh2\eta_{k}\,\int_{0}^{t}\!ds\;\dot{d}_{2}^{(\chi)}(t-s)\Bigl[e^{-i\omega_{i}(t+s)+i\theta_{k}}+e^{+i\omega_{i}(t+s)-i\theta_{k}}\Bigr]\notag\\
	&\quad+8\pi\gamma\int\!\frac{d^{3}\mathbf{k}}{(2\pi)^{3}}\frac{1}{4\omega_{i}}\,\coth\frac{\beta\omega_{i}}{2}\,\cosh2\eta_{k}\,\int_{0}^{t}\!ds\;\dot{d}_{2}^{(\chi)}(t-s)\Bigl[e^{-i\omega_{i}(t-s)}+e^{+i\omega_{i}(t-s)}\Bigr]\,.\label{E:nbyruiezc}
\end{align}
Like before,  we may introduce the same auxiliary functions $f(t;\omega)$ and $g(t;\omega)$,   now  with $\Upsilon$, $\varpi$ replacing $\gamma$, $\Omega$ respectively. However, since now the squeeze parameter is mode-dependent, we may use this characteristic to simplify the arguments to show energy balance.

To do this we only need to examine the temporal behavior of the integral $J_{n}$, associated with the nonstationary term in \eqref{E:nbyruiezc}
\begin{equation}
	J_{n}=\int_{0}^{\infty}\!\frac{dk}{2\pi}\;\frac{k^{2}}{4\pi}\,e^{-n\beta\omega_{i}}\,\sinh2\eta_{k}\,\tilde{d}_{2}^{(\chi)}(\omega_{i})\,e^{-i2\omega_{i}t+i\theta_{k}}\,.
\end{equation}
with $n=0$, 1, $\cdots$, $k=\lvert\mathbf{k}\rvert$, $\omega_{i}^{2}=k^{2}+\mathfrak{m}_{i}^{2}$ and the assumption of spatial isotropy. Observe that the integrand now is an even function of $k$, so we write the integral as
\begin{equation}
	J_{n}=\frac{1}{2}\int_{-\infty}^{\infty}\!\frac{dk}{2\pi}\;\frac{k^{2}}{4\pi}\,e^{-n\beta\omega_{i}}\,\sinh2\eta_{k}\,\tilde{d}_{2}^{(\chi)}(\omega_{i})\,e^{-i2\omega_{i}t+i\theta_{k}}\,.
\end{equation}
The exponential factor $e^{-n\beta\omega_{i}}$ facilitates the convergence of the integral when $n\neq0$.  Furthermore we expect that the factor $\sinh2\eta_{k}$ should approach zero rather fast because if $\sinh^{2}\eta_{\mathbf{k}}$ is related to the production of the massive bath quanta during the parametric process, then for sufficiently high modes the production rate should be extremely low due to the finite energy involved in the process. Thus in a physically realistic configuration, this integral should be better defined than its counterpart in Sec.~\ref{S:etvvhret} when the squeeze parameter is a mode-independent constant. Finally we note that the integrand has poles in $\tilde{d}_{2}^{(\chi)}(\omega_{i})$, but their imaginary parts lie on the lower half of the complex $\omega_{i}$ or $k$ plane because we have required $\mathfrak{m}_{f}$ to be small compared with $\varpi$. Thus, the integral is well defined for large $k$, and then according to the residue theorem, we expect that the integral will decay with time to zero. It implies that at late times, the power pumped by the bath field contains only the stationary component, and thus is constant in time and signifies a steady state of the internal dynamics of the detector,
\begin{align}\label{E:hiridfnnf}
	\lim_{t\gg\Upsilon^{-1}}P_{\xi}(t)&=8\pi\gamma\int_{-\infty}^{\infty}\!\frac{dk}{2\pi}\;\frac{k^{2}}{4\pi}\,\coth\frac{\beta \omega_{i}}{2}\,\cosh2\eta_{k}\,\operatorname{Im}\tilde{d}_{2}^{(\chi)}(\omega_{i})\,,
\end{align}
even though at this moment the bath remains nonstationary and nonequilibrium.

To disclose the underlying physics of these two ostensibly paradoxical statements, we further investigate the power associated with the nonlocal expression like the one in \eqref{E:thfsdf} with the dissipation kernel of the bath field given by \eqref{E:turrnfjgdd}. We first rewrite \eqref{E:thfsdf} to isolate the contribution in \eqref{E:turrnfjgdd} to the frequency renormalization, i.e., the term proportional to $\delta'(\sigma)$. Let us introduce a kernel function $\Gamma^{(\phi)}(t,s)$
\begin{equation}
	G_{R,0}^{(\phi)}(t-s)=-\frac{\partial}{\partial t}\Gamma^{(\phi)}(t-s)\,,
\end{equation}
such that \eqref{E:turrnfjgdd} now has the form
\begin{align}
	&&\ddot{\hat{\chi}}(t)+\omega^{2}_{\textsc{b}}\,\hat{\chi}(t)-\frac{e^{2}}{m}\int_{0}^{t}\!ds\;\Bigl[\frac{\partial}{\partial s}\Gamma^{(\phi)}(t-s)\Bigr]\,\hat{\chi}(s)&=\frac{e}{m}\,\hat{\phi}_{h}(\mathbf{z},t)\,,\notag\\
	&\Rightarrow&\ddot{\hat{\chi}}(t)+\Bigl[\omega^{2}_{\textsc{b}}-\frac{e^{2}}{m}\Gamma^{(\phi)}(0)\Bigr]\,\hat{\chi}(t)+\frac{e^{2}}{m}\Gamma^{(\phi)}(t)\,\hat{\chi}(0)+\frac{e^{2}}{m}\int_{0}^{t}\!ds\;\Gamma^{(\phi)}(t-s)\dot{\hat{\chi}}(s)&=\frac{e}{m}\,\hat{\phi}_{h}(\mathbf{z},t)\,.\label{E:gkkjeds}
\end{align}
However, from \eqref{E:turrnfjgdd}, we note that the kernel function $\Gamma^{(\phi)}(t)$ will contain more than the delta function $\delta(t)$, as for the case in Sec.~\ref{S:etutgbdg}, say Eq.~\eqref{E:dkuerbdf}, so we may wonder whether we should include the full $\Gamma^{(\phi)}(0)$ into the frequency renormalization or only the part related to $\delta(0)$? It turns out that the additional terms does not have any contribution at $t=0$. In the current setup, this can be explicitly seen by evaluating the integral
\begin{equation}
	\int\!ds\;\frac{m}{\sqrt{(t-s)^{2}}}\,J_{1}(m\sqrt{(t-s)^{2}})=-\frac{m^{2}}{2}\bigl(t-s\bigr)\,{}_{p}F_{q}(\{\frac{1}{2}\};\{\frac{3}{2},2\};-\frac{m^{2}(t-s)^{2}}{4})\,,
\end{equation}
for $t>s$, where ${}_{p}F_{q}(a;b;z)$ is the generalized hypergeometric function. In the limit $s\to t$, the integral accounts for the contribution from the additional term in $\Gamma^{(\phi)}(0)$ other than $\delta(0)$, but  we find it is zero. Thus it means that the additional term in $\Gamma^{(\phi)}(0)$ will not contribute to the frequency renormalization.

We also note that in \eqref{E:gkkjeds}, we have a new term proportional to $\Gamma^{(\phi)}(t)\,\hat{\chi}(0)$. This will not contribute to the calculation of the energy exchange between the internal degree of freedom of the detector and the bath field at late times. For example, the associated energy flux will be proportional to $\langle\bigl\{\hat{\chi}(0),\dot{\hat{\chi}}(t)\bigr\}\rangle$, given by
\begin{equation}
	\langle\bigl\{\hat{\chi}(0),\dot{\hat{\chi}}(t)\bigr\}\rangle=\dot{d}_{1}^{(\chi)}(t)\,\langle\bigl\{\hat{\chi}(0),\hat{\chi}(0)\bigr\}\rangle+\dot{d}_{2}^{(\chi)}(t)\,\langle\bigl\{\hat{\chi}(0),\dot{\hat{\chi}}(0)\bigr\}\rangle\,.
\end{equation}
When $t\gg\Upsilon^{-1}$, it is exponentially small due to the damping behavior of $d_{i}^{(\chi)}(t)$.

The remaining nonlocal expression in \eqref{E:gkkjeds} can be shown to reduce to a local damping term and a history-dependent term. At this stage we do not need their explicit expressions. We define the power associated with the dissipation kernel $G_{R,0}^{(\phi)}$ by
\begin{align}\label{E:ruturabd}
	P_{\gamma}(t)&=-\frac{e^{2}}{2}\int_{0}^{t}\!ds\;\Gamma^{(\phi)}(t-s)\,\langle\bigl\{\dot{\hat{\chi}}(s),\dot{\hat{\chi}}(t)\bigr\}\rangle=-e^{2}\int_{0}^{t}\!ds\;\Gamma^{(\phi)}(t-s)\,\frac{\partial^{2}}{\partial t\partial s}G_{H}^{(\chi)}(t,s)\,.
\end{align}
To proceed further, we need to examine the late-time properties of the Hadamard function of the internal degree of freedom of the detector,
\begin{align}
	G_{H}^{(\chi)}(t,t')=\frac{1}{2}\,\langle\big\{\hat{\chi}(t),\hat{\chi}(t')\bigr\}\rangle\simeq\frac{e^{2}}{m^{2}}\int_{0}^{t}\!ds\int_{0}^{t'}\!ds'\;d_{2}^{(\chi)}(t-s)d_{2}^{(\chi)}(t'-s')\,G_{H,0}^{(\phi)}(s,s';\mathbf{z})\,,\label{E:fgbkrhtsd}
\end{align}
if $t$, $t'\gg\Upsilon^{-1}$. Let us take a closer look at the righthand side of \eqref{E:fgbkrhtsd}. Plugging in the explicit expression of $G_{H,0}^{(\phi)}(s,s';\mathbf{z})$, we have
\begin{align}
	&=\frac{e^{2}}{m^{2}}\int_{0}^{t}\!ds\int_{0}^{t'}\!ds'\;d_{2}^{(\chi)}(t-s)d_{2}^{(\chi)}(t'-s')\notag\\
	&\qquad\qquad\qquad\qquad\times\biggl\{-\int\!\frac{d^{3}\mathbf{k}}{(2\pi)^{3}}\frac{1}{4\omega_{i}}\coth\frac{\beta\omega_{i}}{2}\,\sinh2\eta_{k}\,\Bigl[e^{-i\omega_{i}(s+s')+i\theta_{k}}+e^{+i\omega_{i}(s+s')-i\theta_{k}}\Bigr]\biggr.\notag\\
	&\qquad\qquad\qquad\qquad\quad\;\,+\biggl.\int\!\frac{d^{3}\mathbf{k}}{(2\pi)^{3}}\frac{1}{4\omega_{i}}\,\coth\frac{\beta\omega_{i}}{2}\,\cosh2\eta_{k}\,\Bigl[e^{-i\omega_{i}(s-s')}+e^{+i\omega_{i}(s-s')}\Bigr]\biggr\}\,.\label{E:ghjkdsdff}
\end{align}
Now we will use $f(t;\omega)$ defined in \eqref{E:gkbskrjsd},
\begin{align}
	f(t;\omega)=\int_{0}^{t}\!ds\;d_{2}^{(\chi)}(t-s)\,e^{-i\omega s}=\tilde{d}_{2}^{(\chi)}(\omega)\,e^{-i\omega t}\Bigl[1-e^{+i\omega t}\,d_{1}^{(\chi)}(t)+i\omega\,e^{+i\omega t}\,d_{2}^{(\chi)}(t)\Bigr]\,,
\end{align}
where we note that $d_{1}^{(\chi)}(t)$, $d_{2}^{(\chi)}(t)$ are decaying functions of time, so they are exponentially small at late times $t$, $t'\gg\Upsilon^{-1}$. Thus, Eq.~\eqref{E:ghjkdsdff} reduces to
\begin{align}
	&\quad G_{H}^{(Q)}(t,t')\notag\\
	&=\frac{e^{2}}{m^{2}}\int_{0}^{\infty}\!\frac{dk}{2\pi}\frac{k^{2}}{4\pi\omega_{i}}\coth\frac{\beta\omega_{i}}{2}\biggl[-\sinh2\eta_{k}\Bigl\{\tilde{d}_{2}^{(\chi)2}(\omega_{i})\,e^{-i\omega_{i}(t+t')+i\theta_{k}}+\tilde{d}_{2}^{(\chi)*2}(\omega)\,e^{+i\omega_{i}(t+t')-i\theta_{k}}\Bigr\}\biggr.\notag\\
	&\qquad\qquad\qquad\qquad\qquad\qquad\qquad+\biggl.\cosh2\eta_{k}\,\lvert\tilde{d}_{2}^{(\chi)}(\omega_{i})\rvert^{2}\,\Bigl\{e^{-i\omega_{i}(t-t')}+e^{+i\omega_{i}(t-t')}\Bigr\}\biggr]\,.
\end{align}
The presence of $(t+t')$ in the exponential terms of the nonstationary component, in contrast to $(t-t')$ in the corresponding exponentials\footnote{This can be roughly seen by substituting the poles $\omega=i\,\Upsilon\pm \varpi$ of $\tilde{d}_{2}^{(\chi)}(\omega)$ in the nonstationary component. On the other hand, for the stationary component, we also need to take into account the sign of $t-t'$ because it will determine which half of the complex $\omega$ plane will be used to evaluate the integral.} of the stationary components, implies that,  generically speaking,  the nonstationary component of $G_{H}^{(Q)}(t,t')$ will fall exponentially faster with increasing $t$ and $t'$ than the stationary component does, when both $t$ and $t'$ are much greater than the relaxation time $\Upsilon^{-1}$, which is roughly a mode-independent constant in the current setting.

Thus we may conclude that at late times the Hadamard function of the internal degree of freedom of the detector will become stationary, and it has the form
\begin{equation*}
	\lim_{t,t'\gg\Upsilon^{-1}}G_{H}^{(Q)}(t,t')=\frac{e^{2}}{m^{2}}\int_{0}^{\infty}\!\frac{dk}{2\pi}\;\frac{k^{2}}{4\pi\omega_{i}}\coth\frac{\beta\omega_{i}}{2}\,\cosh2\eta_{k}\,\lvert\tilde{d}_{2}^{(\chi)}(\omega_{i})\rvert^{2}\,\Bigl[e^{-i\omega_{i}(t-t')}+e^{+i\omega_{i}(t-t')}\Bigr]\,.
\end{equation*}
In this regime, the Fourier transformation of this Hadamard function is given by
\begin{align}\label{E:kfhdvsjf}
	\tilde{G}_{H}^{(\chi)}(\omega)&=\int_{-\infty}^{\infty}\!d\tau\;G_{H}^{(\chi)}(\tau)\,e^{+i\omega\tau}\notag\\
	&=\frac{e^{2}}{m^{2}}\int_{0}^{\infty}\!dk\;\frac{k^{2}}{4\pi\omega_{i}}\coth\frac{\beta\omega_{i}}{2}\,\cosh2\eta_{k}\,\lvert\tilde{d}_{2}^{(\chi)}(\omega_{i})\rvert^{2}\Bigl[\delta(\omega_{i}-\omega)+\delta(\omega_{i}+\omega)\Bigr]\,,
\end{align}
with $\omega_{i}^{2}=k^{2}+\mathfrak{m}_{i}^{2}$.

The late-time stationarity of $G_{H}^{(\chi)}(t,t')$ allows us to rewrite Eq.~\eqref{E:ruturabd} into a simpler form
\begin{align}
	\lim_{t\gg\Upsilon^{-1}}P_{\gamma}(t)&=-e^{2}\int_{-\infty}^{\infty}\!\frac{d\omega}{2\pi}\;\omega^{2}\,\tilde{\Gamma}^{(\phi)}(\omega)\tilde{G}_{H}^{(\chi)}(\omega)\notag\\
	&=-\frac{e^{4}}{m^{2}}\int_{0}^{\infty}\!dk\;\frac{k^{2}}{4\pi\omega_{i}}\coth\frac{\beta\omega_{i}}{2}\,\cosh2\eta_{k}\,\lvert\tilde{d}_{2}^{(\chi)}(\omega_{f})\rvert^{2}\notag\\
	&\qquad\qquad\qquad\qquad\qquad\times\int_{-\infty}^{\infty}\!\frac{d\omega}{2\pi}\;\omega^{2}\,\tilde{\Gamma}^{(\phi)}(\omega)\,\Bigl[\delta(\omega_{i}-\omega)+\delta(\omega_{i}+\omega)\Bigr]\,.\label{E:trjtbgd}
\end{align}
Since the kernel function $\Gamma^{(\phi)}(t-s)$ is related to the original retarded Green's function of the bath field $G_{R,0}^{(\phi)}(t-s)$ by
\begin{equation}
	G_{R,0}^{(\phi)}(\tau)=-\frac{\partial}{\partial\tau}\Gamma^{(\phi)}(\tau)\,,
\end{equation}
we find $\tilde{G}_{R,0}^{(\phi)}(\omega)=i\,\omega\,\tilde{\Gamma}^{(\phi)}(\omega)$. Thus, Eq.~\eqref{E:trjtbgd} becomes
\begin{align}
	P_{\gamma}(\infty)&=i\,\frac{e^{4}}{m^{2}}\int_{0}^{\infty}\!\frac{dk}{2\pi}\;\frac{k^{2}}{4\pi}\coth\frac{\beta\omega_{i}}{2}\,\cosh2\eta_{k}\,\lvert\tilde{d}_{2}^{(\chi)}(\omega_{i})\rvert^{2}\,\Bigl[\tilde{G}_{R,0}^{(\phi)}(\omega_{i})-\tilde{G}_{R,0}^{(\phi)*}(\omega_{i})\Bigr]\,.
\end{align}
Since
\begin{equation}
	\tilde{d}_{2}^{(\chi)}(\omega_{i})-\tilde{d}_{2}^{(\chi)*}(\omega_{i})=\frac{e^{2}}{m}\,\lvert\tilde{d}_{2}^{(\chi)}(\omega_{i})\rvert^{2}\Bigl[\tilde{G}_{R,0}^{(\phi)}(\omega_{i})-\tilde{G}_{R,0}^{(\phi)*}(\omega_{i})\Bigr]\,,
\end{equation}
we arrive at
\begin{align}
	P_{\gamma}(\infty)&=-8\pi\gamma\int_{-\infty}^{\infty}\!\frac{dk}{2\pi}\;\frac{k^{2}}{4\pi}\coth\frac{\beta\omega_{i}}{2}\,\cosh2\eta_{k}\,\operatorname{Im}\tilde{d}_{2}^{(\chi)}(\omega_{i})\,.\label{E:rtgjfddf}
\end{align}
Thus at late time the net energy exchange between the detector and the bath field does vanish:
\begin{equation}
	P_{\xi}(\infty)+P_{\gamma}(\infty)=0\,,
\end{equation}
by \eqref{E:hiridfnnf} and \eqref{E:rtgjfddf}. This means that  the detector's internal degree of freedom does reach equilibrium even though it is driven by the nonstationary, nonequilibrium bath noise, originated from the parametric process of the bath field.

Although we have shown the energy balance without explicitly referencing  the fluctuation-dissipation relation associated with the internal dynamics, this relation in fact is a paraphrasing of the energy balance~\cite{QTD1}.  Now, instead of showing their close connection, we will directly construct the FDR for the internal degree of freedom of the detector.

We start with the two-point functions of the free bath field. The Fourier transformation of the retarded Green's function of the bath field is given by
\begin{align}
	\tilde{G}_{R,0}^{(\phi)}(\omega)&=i\int\!\frac{d^{3}\mathbf{k}}{(2\pi)^{3}}\;\frac{1}{2\omega_{i}}\int_{0}^{\infty}\!d\tau\;e^{+i\omega\tau}\Bigl[e^{-i\omega_{f}\tau}-e^{+i\omega_{f}\tau}\Bigr]\\
	&=\int_{0}^{\infty}\!\frac{dk}{2\pi}\;\frac{k^{2}}{2\pi\omega_{i}}\Bigl\{\Bigl[P(\frac{1}{(\omega_{i}-\omega)})+P(\frac{1}{(\omega_{i}+\omega)})\Bigr]+i\,\pi\,\Bigl[\delta(\omega_{i}-\omega)-\delta(\omega_{i}+\omega)\Bigr]\Bigr\}\,,\notag
\end{align}
where a factor $e^{-\tau\epsilon}$ with $\epsilon>0$ has been inserted in the integrand to ensure the convergence of the integral. We thus have
\begin{equation}
	\operatorname{Im}\tilde{G}_{R,0}^{(\phi)}(\omega)=\int_{0}^{\infty}\!\frac{dk}{2\pi}\;\frac{k^{2}}{2\omega_{i}}\,\Bigl[\delta(\omega_{i}-\omega)-\delta(\omega_{i}+\omega)\Bigr]\,.\label{E:itortiudf}
\end{equation}
Knowing  that the Hadamard function of the bath field is nonstationary in general, we only keep its stationary component, and carry out the Fourier transformation,
\begin{align}
	\tilde{G}_{H,\textsc{s}}^{(\phi)}(\omega)&=\int\!\frac{d^{3}\mathbf{k}}{(2\pi)^{3}}\;\frac{1}{4\omega_{i}}\,\coth\frac{\beta\omega_{i}}{2}\,\cosh2\eta_{k}\int_{-\infty}^{\infty}\!d\tau\;e^{+i\omega\tau}\Bigl(e^{-i\omega_{i}\tau}+e^{+i\omega_{i}\tau}\Bigr)\notag\\
	&=\int_{0}^{\infty}\!\frac{dk}{2\pi}\;\frac{k^{2}}{2\omega_{i}}\,\coth\frac{\beta\omega_{i}}{2}\,\cosh2\eta_{k}\,\Bigl[\delta(\omega_{i}-\omega)+\delta(\omega_{i}+\omega)\Bigr]\,.\label{E:gbfjsds}
\end{align}
We observe that the integrands of~\eqref{E:itortiudf} and \eqref{E:gbfjsds} look quite alike, so we attempt to put them together into a form similar to the conventional FDR.

We first try to carry out the $k$ integrals in \eqref{E:itortiudf} and \eqref{E:gbfjsds}. For later convenience, we introduce $\omega^{2}=\kappa^{2}+\mathfrak{m}_{i}^{2}$, in comparison with $\omega_{i}=\sqrt{k^{2}+\smash{\mathfrak{m}_{i}^{2}}}>0$. Suppose $\omega>0$. Then we have
\begin{align}
	\operatorname{Im}\tilde{G}_{R,0}^{(\phi)}(\omega)&=\int_{0}^{\infty}\!\frac{dk}{2\pi}\;\frac{k^{2}}{2\omega_{i}}\,\delta(\omega_{i}-\omega)=\int_{0}^{\infty}\!\frac{dk}{2\pi}\;\frac{k}{2}\,\delta(k-\kappa)=\frac{\kappa}{4\pi}\,,&\kappa&=\sqrt{\smash[b]{\omega^{2}_{\vphantom{f}}}-\smash[b]{\mathfrak{m}_{i}^{2}}}\,,\label{E:kgjb}
\end{align}
and
\begin{align}
	\tilde{G}_{H,\textsc{s}}^{(\phi)}(\omega)=\int_{0}^{\infty}\!\frac{dk}{2\pi}\;\frac{k^{2}}{2\omega_{i}}\,\coth\frac{\beta\omega_{i}}{2}\,\cosh2\eta_{k}\,\delta(\omega_{i}-\omega)&=\int_{0}^{\infty}\!\frac{dk}{2\pi}\;\frac{k}{2}\,\coth\frac{\beta \omega_{i}}{2}\,\cosh2\eta_{k}\,\delta(k-\kappa)\notag\\
	&=\frac{\kappa}{4\pi}\,\coth\frac{\beta\omega}{2}\,\cosh2\eta_{\kappa}\,,\label{E:trugbs}
\end{align}
where we have re-written the delta function as
\begin{equation}\label{E:bgvrute}
	\delta(\omega_{i}-\omega)=\frac{\delta(k-\kappa)}{\lvert\frac{\partial\omega_{i}}{\partial k}\rvert}=\frac{\omega_{i}}{k}\,\delta(k-\kappa)\,.
\end{equation}
Then a formal relation similar to the FDR can be set up between Eqs.~\eqref{E:kgjb} and \eqref{E:trugbs}
\begin{equation}
	\tilde{G}_{H,\textsc{s}}^{(\phi)}(\omega)=\coth\frac{\beta\omega}{2}\,\cosh2\eta_{\kappa}\,\operatorname{Im}\tilde{G}_{R,0}^{(\phi)}(\omega)\,,
\end{equation}
for $\omega>0$. After taking $\omega<0$ into account by the same arguments, we arrive at
\begin{equation}\label{E:rtbbjswere}
	\tilde{G}_{H,\textsc{s}}^{(\phi)}(\omega)=\operatorname{sgn}(\omega)\,\coth\frac{\beta\omega}{2}\,\cosh2\eta_{\kappa}\,\operatorname{Im}\tilde{G}_{R,0}^{(\phi)}(\omega)\,.
\end{equation}
This FDR is slightly different the conventional FDR in two aspects:  First,  only the stationary component of the Hadamard function of the free bath field is involved due to the nonstationary, nonequilibrium nature of the bath field considered here.  Second,  the proportionality factor between two kernels has an additional element $\cosh2\eta_{\kappa}$, which accounts for particle production due to the parametric process
\begin{equation}
	\cosh2\eta_{\kappa}=2\Bigl(\sinh^{2}\eta_{\kappa}+\frac{1}{2}\Bigr)\,,
\end{equation}
in which $\sinh^{2}\eta_{\kappa}$ is essentially $\lvert\delta_{\kappa}\rvert^{2}$, as is discussed in Sec.~\ref{E:bgksbssfg}.

Now we turn to the kernel functions of the internal degree of freedom of the detector  at late times. From \eqref{E:kfhdvsjf}, we have
\begin{align}
	\tilde{G}_{R}^{(\chi)}(\omega)&=\frac{1}{m}\frac{1}{-\omega^{2}+\omega_{\textsc{b}}^{2}-\frac{e^{2}}{m}\,\tilde{G}_{R,0}^{(\phi)}(\omega)}\,,\notag\\
	\tilde{G}_{H}^{(\chi)}(\omega)&=e^{2}\int_{0}^{\infty}\!\frac{dk}{2\pi}\;\frac{k^{2}}{2\omega_{i}}\coth\frac{\beta\omega_{i}}{2}\,\cosh2\eta_{k}\,\lvert\tilde{G}_{R}^{(\chi)}(\omega_{i})\rvert^{2}\Bigl[\delta(\omega_{i}-\omega)+\delta(\omega_{i}+\omega)\Bigr]\,.
\end{align}
In the case $\omega>0$, we find 
\begin{align}\label{E:rotidnfg}
	\operatorname{Im}\tilde{G}_{R}^{(\chi)}(\omega)&=e^{2}\,\lvert\tilde{G}_{R}^{(\chi)}(\omega)\rvert^{2}\,\operatorname{Im}\tilde{G}_{R,0}^{(\phi)}(\omega)\,,
\end{align}
and with the help of \eqref{E:bgvrute},
\begin{align}
	\tilde{G}_{H}^{(\chi)}(\omega)&=e^{2}\int_{0}^{\infty}\!\frac{dk}{2\pi}\;\frac{k^{2}}{2\omega_{i}}\,\coth\frac{\beta\omega_{i}}{2}\,\cosh2\eta_{k}\,\lvert\tilde{G}_{R}^{(\chi)}(\omega_{i})\rvert^{2}\,\delta(\omega_{i}-\omega)\notag\\
	&=e^{2}\,\coth\frac{\beta\omega}{2}\,\cosh2\eta_{\kappa}\,\lvert\tilde{G}_{R}^{(\chi)}(\omega)\rvert^{2}\,\operatorname{Im}\tilde{G}_{R,0}^{(\phi)}(\omega)\,.
\end{align}
By \eqref{E:rtbbjswere} and\eqref{E:rotidnfg}, we then have
\begin{equation}
	\tilde{G}_{H}^{(\chi)}(\omega)=\coth\frac{\beta\omega}{2}\,\cosh2\eta_{\kappa}\,\operatorname{Im}\tilde{G}_{R}^{(\chi)}(\omega)\,.
\end{equation}
Likewise after incorporating the $\omega<0$ case, we finally obtain
\begin{equation}
	\tilde{G}_{H}^{(\chi)}(\omega)=\operatorname{sgn}(\omega)\,\coth\frac{\beta\omega}{2}\,\cosh2\eta_{\kappa}\,\operatorname{Im}\tilde{G}_{R}^{(\chi)}(\omega)\,.
\end{equation}
Thus it inherits the common proportionality factor from the FDR of the bath field \eqref{E:rtbbjswere}. However, the  physical settings are quite different in the two cases. For the bath field at the moment the detector-field interaction is switched on, the field is nonstationary and nonequilibrium due to the parametric process, so its noise kernel cannot be reduced to a function of two-time difference and we do not have a FDR in the usual sense. Nonetheless its stationary component can be used to formulate a relation similar to FDR. In comparison, after the interaction is switched on, the internal dynamics of the detector is also nonstationary due to its own nonequilibrium evolution. Although the bath field may remain nonstationary over the entire history because the backaction from the detector has negligible effect on it due to the gigantic difference in the numbers of degree of freedom, the dissipation force on the internal degree of freedom of the detector will adjust itself to match up with the effects from the nonstationary bath noise so that the equilibration of the internal dynamics of the detector is made possible. When the dynamics is fully relaxed, the internal dynamics becomes stationary, thus enabling  the existence of an FDR for the internal degree of freedom.

\begin{figure}
\centering
    \scalebox{0.35}{\includegraphics{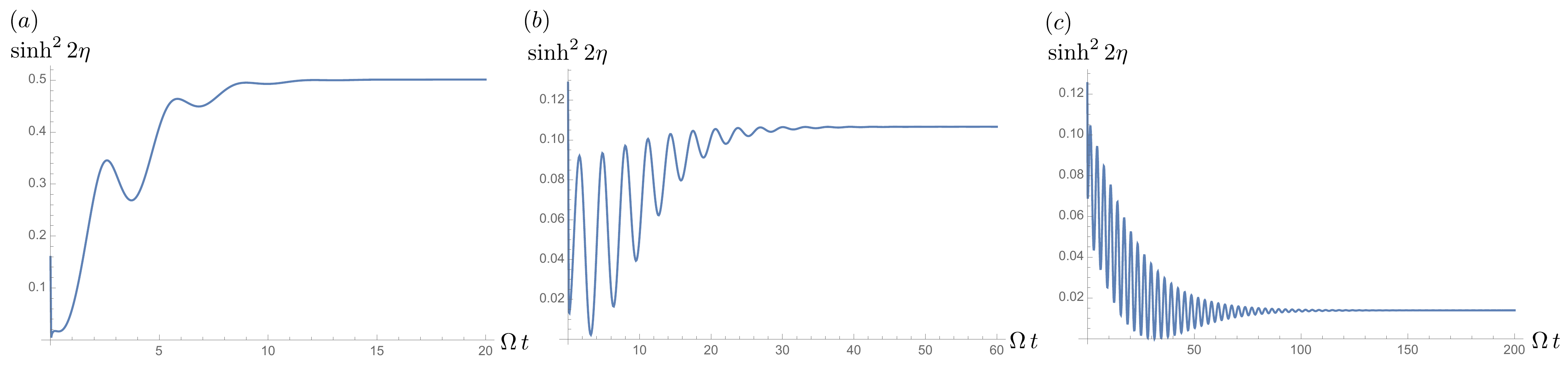}}
    \caption{The time evolution of the squeeze parameter $\eta$ in terms of $\sinh^{2}2\eta$ where the parameters are normalized with respect to the resonance frequency $\Omega=\sqrt{\omega_{\textsc{r}}^{2}-\gamma^{2}}$ such that $m=1\,\Omega$, $\beta=10\,\Omega^{-1}$, and the cutoff frequency $\Lambda=1000\,\Omega$. We choose the initial displacement dispersion $\langle\hat{\chi}^{2}(0)\rangle=2$ and the momentum dispersion $\langle\hat{p}^{2}(0)\rangle=1$. In (a) $\gamma=0.3\,\Omega$, (b) $\gamma=0.1\,\Omega$,  and (c) $\gamma=0.03\,\Omega$. This parameter relaxes, at late times, to constants proportional to $\gamma$, as can be read out from the vertical scales, so it will approach zero when the oscillator-bath  coupling  becomes vanishingly weak.}\label{Fi:tan2eta}
\end{figure}

In this and the previous sections we have treated the two cases of squeezed baths in detail, corresponding to Case A) and B) described in the Introduction.  We now wish to make some comments on Case C), namely, squeezing corresponding to finite coupling strength between the oscillator and the field.   

\subsection{Squeezing due to finite coupling}
In all our derivations of energy balance and FDRs we made no perturbative arguments, because we don't need to make the assumption of vanishingly weak  coupling between the bath field and the internal degree of freedom of the detector. {This} finite coupling strength will introduce an additional degree of time-dependent squeezing to the internal dynamics during its course of evolution~\cite{NEqFE}, even if initially neither the detector nor the bath is in a squeezed state.  It is most clearly seen from the covariance elements of the internal dynamics at any moment
\begin{align}
	\langle\hat{\chi}^{2}(t)\rangle&=\frac{1}{2m\omega_{\textsc{r}}}\,\Xi(t)\bigl[\cosh2\eta(t)-\sinh2\eta(t)\,\cos\theta(t)\bigr]\,,\label{E:fkbgdkf1}\\
	\langle\hat{p}^{2}(t)\rangle&=\frac{m\omega_{\textsc{r}}}{2}\,\Xi(t)\bigl[\cosh2\eta(t)+\sinh2\eta(t)\,\cos\theta(t)\bigr]\,,\\
	\frac{1}{2}\langle\bigl\{\hat{\chi}(t),\hat{p}(t)\bigr\}\rangle&=-\frac{1}{2}\,\Xi(t)\sinh2\eta(t)\,\sin\theta(t)\,,\label{E:fkbgdkf3}
\end{align}
because a Gaussian system will remain Gaussian during the evolution. Here $\eta$ and $\theta$ together give the squeeze parameter $\zeta(t)=\eta(t)\,e^{+i\theta(t)}$, while $\Xi(t)=\coth\vartheta(t)/2$, with $\vartheta(t)$ being related to the inverse effective temperature of the internal dynamics. At finite coupling, for example, $\langle\hat{\chi}^{2}(t)\rangle$, $\langle\hat{p}^{2}(t)\rangle$ are different from the values
\begin{align}
	\langle\hat{\chi}^{2}(t)\rangle&\neq\frac{1}{2m\omega_{\textsc{r}}}\,\coth\frac{\beta\omega_{\textsc{r}}}{2}\,,&\langle\hat{p}^{2}(t)\rangle&\neq\frac{m\omega_{\textsc{r}}}{2}\,\coth\frac{\beta\omega_{\textsc{r}}}{2}\,,&\frac{1}{2}\langle\bigl\{\hat{\chi}(t),\hat{p}(t)\bigr\}\rangle&\neq0\,,
\end{align}
as given by the internal degree of freedom in a thermal state of temperature $\beta^{-1}$. These differences amount to squeezing -- the corresponding squeeze parameter and the effective temperature can be found~\cite{NEqFE} by inverting \eqref{E:fkbgdkf1}--\eqref{E:fkbgdkf3}.

Fig.~\ref{Fi:tan2eta} shows the time dependence of $\sinh^{2}2\eta$ for three different choices of damping constants $\gamma$, a manifestation of the oscillator-bath coupling strength. We observe that at late times, the parameter $\eta$ approaches a constant that depends on $\gamma$, and the   late-time saturated values   goes to zero in the vanishing $\gamma$ limit. Furthermore, the plots in Fig.~\ref{Fi:tanphi} show that $\sin\theta$ oscillates with time but the amplitude of oscillations decays with time regardless of the oscillator-bath coupling constants, so at late times $\sin\theta$ asymptotically goes to zero, so will $\theta$.

\begin{figure}
\centering
    \scalebox{0.35}{\includegraphics{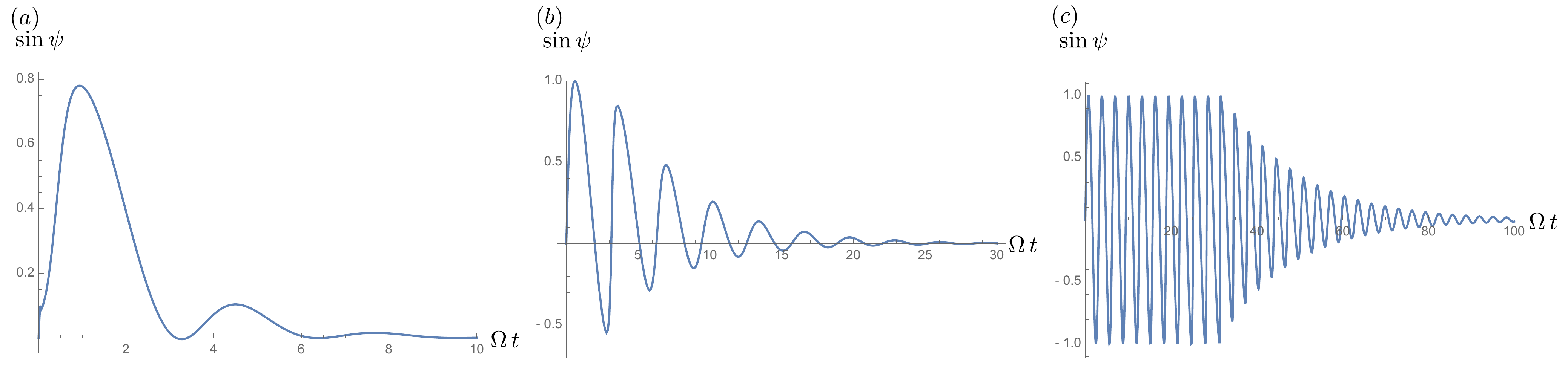}}
    \caption{The time evolution of the squeeze parameter $\theta$ in terms of $\sin\theta$, where the parameters are normalized with respect to the resonance frequency $\Omega$ such that $m=1\,\Omega$, $\beta=10\,\Omega^{-1}$, and the cutoff frequency $\Lambda=1000\,\Omega$. We choose the initial displacement dispersion $\langle\hat{\chi}^{2}(0)\rangle=2$ and the momentum dispersion $\langle\hat{p}^{2}(0)\rangle=1$. In (a) $\gamma=0.3\,\Omega$, (b) $\gamma=0.1\,\Omega$,  and (c) $\gamma=0.03\,\Omega$. This parameter oscillates with decreasing amplitude in time. At late times, it asymptotically approaches zero independent of the oscillator-bath coupling strength.}\label{Fi:tanphi}
\end{figure}

 These results imply that at late time the state of the internal degree of freedom will be described by the covariance matrix elements of the form
\begin{align}
	\langle\hat{\chi}^{2}\rangle&=\frac{1}{2m\omega_{\textsc{r}}}\,\Xi\,\cosh2\eta\,,&\langle\hat{p}^{2}\rangle&=\frac{m\omega_{\textsc{r}}}{2}\,\Xi\,\cosh2\eta\,,&\frac{1}{2}\langle\bigl\{\hat{\chi},\,\hat{p}\bigr\}\rangle&=0\,,\label{E:gbfbhs6}
\end{align}
for $t\gg\gamma^{-1}$. They are exactly like what we had derived in Sec.~\ref{S:trvkgdfhd}, namely, the final equilibrium state the internal system  relaxes to is a thermal state at a temperature different from the initial bath temperature. The effective temperature  is given by \eqref{E:gbksbkdfgs}. Only when $\gamma/\Omega\to0$ will the squeeze parameter $\eta$ also approach zero such that in this limit, the equilibrium temperature is reduced to the initial bath temperature.

If we assume that neither the internal degree of freedom of the detector nor the bath field is initially squeezed, then in the end the internal degree of freedom will acquire an FDR of the form
\begin{equation}
	\tilde{G}_{H}^{(\chi)}(\omega)=\operatorname{sgn}(\omega)\,\coth\frac{\beta\omega}{2}\,\operatorname{Im}\tilde{G}_{R}^{(\chi)}(\omega)\,,
\end{equation}
if it is coupled to a plain thermal bath field of temperature $\beta^{-1}$ at finite coupling strength. The squeezing and the effective temperature due to finite coupling strength will not enter in the FDR~\cite{NEqFE}. This is a characteristic of the nonequilibrium dynamics of  Gaussian systems where some initial traits of the bath will imprint on the final state of the system in contact with it.

\section{Summary and Discussions}\label{S:gbveke}

\subsection{Summary of Key Findings}
{In this paper we address the nonequilibrium evolution and the approach to equilibration of the internal degree of freedom of a Unruh-DeWitt detector when it is coupled to a nonequilibrium and nonstationary bath field. We discuss two different basic mechanisms which makes the bath field nonstationary: 
	\begin{enumerate}[A)]
		\item The bath field is initially in a squeezed thermal state, whose squeeze parameter is a mode- and time-independent constant. The setting is often encountered in quantum optics and quantum thermodynamics. Since the bath field has a much greater number of degrees of freedom than the detector has, the bath field will essentially remain in the same initial squeezed state through out the equilibration process of the internal dynamics of the detector, except for the modes that are in resonance with the internal motion of the detector.
	\item The bath field is initially in a thermal state, but undergoes a parametric process such that the parameter of the field changes monotonically and smoothly from one constant to another. In this case, the bath field will acquire mode- and time-dependent squeezing during the parametric process, but once the process stops, the squeeze parameters reduce to mode-dependent constants. This scenario is more often encountered in cosmology and dynamical Casimir effects, where the time-dependent backgrounds cause the field's parameters to change with time.  
\end{enumerate}
The squeezing in the bath in these two types of processes will be passed on to the detector because the internal degree of freedom of the detector is driven by the nonequilibrium and nonstationary  squeezed field noise. During the nonequilibrium evolution of the internal dynamics  the internal degree of freedom  acquires  a time-dependent squeezing. However, once the internal degree of freedom approaches equilibration (stationarity), the trait of squeezing is imbued in its effective equilibrium temperature. 

In these two cases, since the bath field is not stationary and not in equilibrium, we cannot write down a fluctuation-dissipation relation as we can do for a thermal field, but as was shown in Sec. II and III,  there does exist such a relation for the internal dynamics of the detector after it reaches equilibrium. The proportionality factor of the FDR  {for the internal degree of freedom of the detector} is only related to the stationary component of the noise kernel of the bath field. As is seen more clearly in the parametric field case, this is a consequence of   particle creation during the parametric process of the field. Thus because of the nonequilibrium nature of both the detector and the field dynamics, through their interaction,  information of the field can be transferred to and measured by the detector.
	\begin{enumerate}[C)]
		\item There is another subtler way the internal dynamics of the detector can obtain squeezing. The finite system-bath coupling strength can bring the internal dynamics to a squeezed state \cite{NEqFE}  even though the bath field is stationary and does not engage in any parametric process. The squeezing of the system in this case is in general time-dependent but becomes constant when the internal dynamics is fully relaxed. Since the squeezing is related to the coupling strength, it is typically very mild, not as strong as the previous two cases. 
\end{enumerate}

\subsection{FDR and Backreaction}

Historically, a primary motivation for one of the present authors to investigate into FDRs in NEq  quantum processes  was to find a way to elucidate the balance of fluctuations in a quantum field with the dissipation in a dynamical system interacting with it.  In fact, the dynamical system was the early universe and the quantum process was cosmological particle creation.   Particle creation in the early universe prevalent  at the Planck time originated from the parametric amplification of vacuum fluctuations of  quantum matter field in an expanding spacetime.  This process, as we saw earlier, can be understood simply as the squeezing of the vacuum.  Particles created   backreacts   on the  background spacetime resulting in its isotropization and homogenization \cite{CalHu87,CamVer94}.  It has been shown that this kind of backreaction can be represented as a FDR  \cite{HuSin95, CamVer96} in the framework of semiclassical gravity \cite{HuVer20}: fluctuations or noise in the quantum field related to dissipative geometrodynamics. (See also \cite{CanSci,Mottola} for black holes).  

Note, however,  the difference from the FDR discussed in this work.  Here we show, by solving for the nonequilibrium dynamics of a quantum harmonic oscillator in a squeezed thermal scalar field bath, that at late times a FDR governing the system exists, which depends on the squeezing of the bath.   But for cosmological backreaction there is no detector or atom involved. Only the field and its driver are present.  Thus the FDR in early universe cosmology is of a different type, that which connects the quantum field activities and the external driver dynamics.  

Belonging to the same class of problems is dynamical Casimir effect,  with a moving mirror playing the role of the expanding universe.  We speculate that a FDR in the field may exist which shows the power input from the drive balancing the power output of the field. It requires the stipulation of how the external drive acts on the field over the evolutionary time span, and how the backreaction from the  particles created changes the drive input.  Therefore, to begin with, one needs to treat the displacement of the mirror as a dynamical variable obeying some equation of motion, then solve it together with the field equation to get the particle creation rate, its backreaction on the drive and the balance between the powers delivered by the noise forces and the  dissipative drive dynamics.

A slightly more involved problem is quantum friction \cite{QFric}.  There, a neutral atom moving at constant speed near a dielectric plane experiences a reactive drag force. Here  there are four players -- atom, field, dielectric, and the external drive and three acts -- the medium  with certain dynamical susceptibility modifies the quantum field in the space between the atom and the dielectric,  the medium-modified quantum field acting on the moving atom generates a reactive drag force on the atom  and the external drive pumping energy into the atom.  Therefore there should be a FDR between fluctuations in this medium-modified field and the dissipation of the moving atom.  Now,  between the  drive and the atom: The drive delivers the extra amount of energy to the atom which replenishes the dissipated energy due to quantum friction, thus keeping the atom moving at a constant speed.   The atom would backreact on the drive also, and that depends on the mechanism of how the atom is driven. So there should be a balance relation between the dissipative power in the moving atom and the rate of energy output of the drive.  It would be very interesting to see how the  FDR and power balance conditions  come about and how they connect all the players in their coordinated acts to maintain a nonequilibrium steady state \cite{QFricNESS} in the whole system.

To end this discussion we return to the detector-field system studied in this paper and mention that  fluctuation-dissipation relations in an $N$ atom/detector-quantum field system, and the lesser known but equally important correlation-propagation relations  amongst the $N$ detectors, have been shown to exist~\cite{RHA,CPR}.   Just like the backreaction problems in atom/detector-quantum field systems,  in cosmological particle creation, dynamical Casimir effect and quantum friction, FDRs provide a very important self-consistency condition between the dynamics of a system and the environment it interacts with. \\
 


\noindent\textbf{Acknowledgment}  J.-T. Hsiang is supported by the Ministry of Science and Technology of Taiwan under Grant No.~MOST 110-2811-M-008-522.

\newpage 
\appendix
\section{Two-mode squeezing and Bogoliubov transformation}\label{S:eotie}
Here we summarize the connection of the Bogoliubov transformations and the two-mode squeezing. Suppose the field is expanded by the in-mode $\{u^{\textsc{in}}_{\mathbf{k}}\}$,
\begin{equation}\label{E:gnller}
	\hat{\phi}(x)=\sum_{\mathbf{k}}\hat{a}_{\mathbf{k}}^{\vphantom{\dagger}}u^{\textsc{in}\vphantom{\dagger}}_{\mathbf{k}}(x)+\hat{a}_{\mathbf{k}}^{\dagger}u^{\textsc{in}*\vphantom{\dagger}}_{\mathbf{k}}(x)\,,
\end{equation}
and $\hat{a}_{\mathbf{k}}$ is the annihilation operator corresponding to the in-mode. The field operator can equally well be expanded an alternative complete set of mode functions, called out-mode $\{u^{\textsc{out}}_{\mathbf{k}}\}$
\begin{equation}
	\hat{\phi}(x)=\sum_{\mathbf{k}}\hat{b}_{\mathbf{k}}^{\vphantom{\dagger}}u^{\textsc{out}\vphantom{\dagger}}_{\mathbf{k}}(x)+\hat{b}_{\mathbf{k}}^{\dagger}u^{\textsc{out}*\vphantom{\dagger}}_{\mathbf{k}}(x)\,,
\end{equation}
and the annihilation operator $\hat{b}_{\mathbf{k}}$ is associated with the out modes. Since both set of modes are supposed to be complete, the mode function $u^{\textsc{in}}_{\mathbf{k}}(x)$ can be expressed as a superposition of the out mode $u^{\textsc{out}}_{\mathbf{k}}(x)$ by
\begin{equation}\label{E:othoue}
	u^{\textsc{in}}_{\mathbf{k}}(x)=\alpha_{\mathbf{k}}^{\vphantom{\dagger}}\,u^{\textsc{out}\vphantom{\dagger}}_{\mathbf{k}}(x)+\beta_{\mathbf{k}}^{\vphantom{\dagger}}\,u^{\textsc{out}*\vphantom{\dagger}}_{-\mathbf{k}}(x)\,.
\end{equation}
This enables us to write the expansion of the field operator $\hat{\phi}(x)$ in \eqref{E:gnller} by the out mode $u^{\textsc{out}}_{\mathbf{k}}(x)$
\begin{align}
	\hat{\phi}(x)&=\sum_{\mathbf{k}}\Bigl[\alpha_{\mathbf{k}}^{\vphantom{\dagger}}\,\hat{a}_{\mathbf{k}}^{\vphantom{\dagger}}+\beta_{-\mathbf{k}}^{\vphantom{\dagger}*}\,\hat{a}_{-\mathbf{k}}^{\dagger}\Bigr]\,u^{\textsc{out}\vphantom{\dagger}}_{\mathbf{k}}(x)+\Bigl[\alpha_{\mathbf{k}}^{*\vphantom{\dagger}}\,\hat{a}_{\mathbf{k}}^{\dagger}+\beta_{-\mathbf{k}}^{\vphantom{\dagger}}\,\hat{a}_{-\mathbf{k}}^{\vphantom{\dagger}}\Bigr]\,u^{\textsc{out}*\vphantom{\dagger}}_{\mathbf{k}}(x)\,,
\end{align}
so that the annihilation operators $\hat{b}_{\mathbf{k}}^{\vphantom{\dagger}}$ is expanded by $\hat{a}_{\mathbf{k}}^{\vphantom{\dagger}}$ and $\hat{a}_{\mathbf{k}}^{\dagger}$,
\begin{equation}\label{E:ghjrbdfdf}
	\hat{b}_{\mathbf{k}}^{\vphantom{\dagger}}=\alpha_{\mathbf{k}}^{\vphantom{\dagger}}\,\hat{a}_{\mathbf{k}}^{\vphantom{\dagger}}+\beta_{-\mathbf{k}}^{\vphantom{\dagger}*}\,\hat{a}_{-\mathbf{k}}^{\dagger}\,.
\end{equation}
We have a transformation that mixes the $\pm\mathbf{k}$ modes, so in this case $\hat{b}_{\mathbf{k}}^{\vphantom{\dagger}}$ can be related to $\hat{a}_{\mathbf{k}}^{\vphantom{\dagger}}$ by a two-mode squeezing
\begin{align}
	\hat{b}_{\mathbf{k}}^{\vphantom{\dagger}}&=\hat{S}_{2}^{\dagger}(\zeta_{\mathbf{k}}^{\vphantom{\dagger}})\,\hat{a}_{\mathbf{k}}^{\vphantom{\dagger}}\,\hat{S}_{2}^{\vphantom{\dagger}}(\zeta_{\mathbf{k}}^{\vphantom{\dagger}})\,,&&\text{such that} &\hat{b}_{\mathbf{k}}^{\vphantom{\dagger}}&=\cosh\eta_{\mathbf{k}}^{\vphantom{\dagger}}\,\hat{a}_{\mathbf{k}}^{\vphantom{\dagger}}-e^{+i\theta_{\mathbf{k}}}\sinh\eta_{\mathbf{k}}^{\vphantom{\dagger}}\,\hat{a}_{-\mathbf{k}}^{\dagger}\,,
\end{align}
where the two-mode squeeze operator $\hat{S}_{2}^{\vphantom{\dagger}}(\zeta_{\mathbf{k}}^{\vphantom{\dagger}})$ takes the form
\begin{equation}
	\hat{S}_{2}^{\vphantom{\dagger}}(\zeta_{\mathbf{k}}^{\vphantom{\dagger}})=\exp\Bigl[\zeta_{\mathbf{k}}^{*\vphantom{\dagger}}\,\hat{a}_{\mathbf{k}}^{\vphantom{\dagger}}\hat{a}_{-\mathbf{k}}^{\vphantom{\dagger}}-\zeta_{\mathbf{k}}^{\vphantom{\dagger}}\,\hat{a}_{\mathbf{k}}^{\dagger}\hat{a}_{-\mathbf{k}}^{\dagger}\Bigr]\,.
\end{equation}
It has a useful factorization
\begin{align}
	\hat{S}_{2}^{\vphantom{\dagger}}(\zeta_{\mathbf{k}}^{\vphantom{\dagger}})&=\exp\Bigl[-\tanh\eta_{\mathbf{k}}\,e^{+i\theta_{\mathbf{k}}}\,\hat{a}_{\mathbf{k}}^{\dagger}\hat{a}_{-\mathbf{k}}^{\dagger}\Bigr]\exp\Bigl[-\ln\cosh\eta_{\mathbf{k}}\bigl(\hat{a}_{\mathbf{k}}^{\dagger}\hat{a}_{-\mathbf{k}}^{\vphantom{\dagger}}+\hat{a}_{\mathbf{k}}^{\dagger}\hat{a}_{-\mathbf{k}}^{\vphantom{\dagger}}+1\bigr)\Bigr]\notag\\
	&\qquad\qquad\qquad\qquad\qquad\qquad\times\exp\Bigl[+\tanh\eta_{\mathbf{k}}\,e^{-i\theta_{\mathbf{k}}}\,\hat{a}_{\mathbf{k}}^{\vphantom{\dagger}}\hat{a}_{-\mathbf{k}}^{\vphantom{\dagger}}\Bigr]\,,
\end{align}
so that
\begin{equation}
	\hat{S}_{2}^{\vphantom{\dagger}}(\zeta_{\mathbf{k}}^{\vphantom{\dagger}})\,\lvert0_{+\mathbf{k}},0_{-\mathbf{k}}\rangle=\frac{1}{\cosh\eta_{\mathbf{k}}}\sum_{n=0}^{\infty}\bigl(-\tanh\eta_{\mathbf{k}}\,e^{+i\theta_{\mathbf{k}}}\bigr)^{n}\lvert n_{+\mathbf{k}},n_{-\mathbf{k}}\rangle\,.
\end{equation}
It creates particles in pair, which have the opposite momenta.

Consider a two-point functions $\langle\hat{\chi}(x)\hat{\chi}(x')\rangle_{\textsc{in}}$ in the in-state, which takes the form
\begin{align}\label{E:gbkreddfd}
	\langle\hat{\chi}(x)\hat{\chi}(x')\rangle_{\textsc{in}}=\sum_{\mathbf{k}}\langle\hat{a}_{\mathbf{k}}^{2\vphantom{\dagger}}\rangle_{\textsc{in}}\,u^{\textsc{in}\vphantom{\dagger}}_{\mathbf{k}}(x)u^{\textsc{in}\vphantom{\dagger}}_{\mathbf{k}}(x')+\langle\hat{a}_{\mathbf{k}}^{\vphantom{\dagger}}\hat{a}_{\mathbf{k}}^{\dagger}\rangle_{\textsc{in}}\,u^{\textsc{in}\vphantom{\dagger}}_{\mathbf{k}}(x)u^{\textsc{in}*\vphantom{\dagger}}_{\mathbf{k}}(x')+\text{C.C.}\,.
\end{align}
Suppose that this in-state is stationary such that $\langle\hat{a}_{\mathbf{k}}^{2\vphantom{\dagger}}\rangle_{\textsc{in}}=0$. Let us take a look at $u^{\textsc{in}\vphantom{\dagger}}_{\mathbf{k}}(x)u^{\textsc{in}*\vphantom{\dagger}}_{\mathbf{k}}(x')$ and express it by the out-modes with the help of \eqref{E:othoue}, the term $u^{\textsc{in}\vphantom{\dagger}}_{\mathbf{k}}(x)u^{\textsc{in}*\vphantom{\dagger}}_{\mathbf{k}}(x')$ in the two-point function \eqref{E:gbkreddfd} becomes
\begin{align}\label{E:bgfkjgsd}
	u^{\textsc{in}\vphantom{\dagger}}_{\mathbf{k}}(x)u^{\textsc{in}*\vphantom{\dagger}}_{\mathbf{k}}(x')&=\lvert\alpha_{\mathbf{k}}^{\vphantom{\dagger}}\rvert^{2}\,u^{\textsc{out}\vphantom{\dagger}}_{\mathbf{k}}(x)u^{\textsc{out}*\vphantom{\dagger}}_{\mathbf{k}}(x')+\alpha_{\mathbf{k}}^{\vphantom{\dagger}}\beta_{\mathbf{k}}^{*\vphantom{\dagger}}\,u^{\textsc{out}\vphantom{\dagger}}_{\mathbf{k}}(x)u^{\textsc{out}\vphantom{\dagger}}_{-\mathbf{k}}(x')\notag\\
	&\qquad\qquad\qquad\qquad+\beta_{\mathbf{k}}^{\vphantom{\dagger}}\alpha_{\mathbf{k}}^{*\vphantom{\dagger}}\,u^{\textsc{out}*\vphantom{\dagger}}_{-\mathbf{k}}(x)u^{\textsc{out}*\vphantom{\dagger}}_{\mathbf{k}}(x')+\lvert\beta_{\mathbf{k}}^{\vphantom{\dagger}}\rvert^{2}\,u^{\textsc{out}*\vphantom{\dagger}}_{-\mathbf{k}}(x)u^{\textsc{out}\vphantom{\dagger}}_{-\mathbf{k}}(x')\,.
\end{align}
If we suppose that the in- and out-mode takes the form
\begin{align}
	u^{\textsc{in}\vphantom{\dagger}}_{\mathbf{k}}(x)&=\frac{1}{\sqrt{2\omega_{\textsc{in}}}}\,e^{+i\mathbf{k}\cdot\mathbf{x}-i\omega_{\textsc{in}}t}\,,&u^{\textsc{out}\vphantom{\dagger}}_{\mathbf{k}}(x)&=\frac{1}{\sqrt{2\omega_{\textsc{out}}}}\,e^{+i\mathbf{k}\cdot\mathbf{x}-i\omega_{\textsc{out}}t}\,,
\end{align}
then we have
\begin{align*}
	u^{\textsc{out}\vphantom{\dagger}}_{\mathbf{k}}(x)u^{\textsc{out}*\vphantom{\dagger}}_{\mathbf{k}}(x')&=\frac{1}{2\omega_{\textsc{out}}}\,e^{+i\mathbf{k}\cdot(\mathbf{x}-\mathbf{x}')-i\omega_{\textsc{out}}(t-t')}\,,&u^{\textsc{out}\vphantom{\dagger}}_{\mathbf{k}}(x)u^{\textsc{out}\vphantom{\dagger}}_{-\mathbf{k}}(x')&=\frac{1}{2\omega_{\textsc{out}}}\,e^{+i\mathbf{k}\cdot(\mathbf{x}-\mathbf{x}')-i\omega_{\textsc{out}}(t+t')}\,,\\
	u^{\textsc{out}*\vphantom{\dagger}}_{-\mathbf{k}}(x)u^{\textsc{out}*\vphantom{\dagger}}_{\mathbf{k}}(x')&=\frac{1}{2\omega_{\textsc{out}}}\,e^{+i\mathbf{k}\cdot(\mathbf{x}-\mathbf{x}')+i\omega_{\textsc{out}}(t+t')}\,,&u^{\textsc{out}*\vphantom{\dagger}}_{-\mathbf{k}}(x)u^{\textsc{out}\vphantom{\dagger}}_{-\mathbf{k}}(x')&=\frac{1}{2\omega_{\textsc{out}}}\,e^{+i\mathbf{k}\cdot(\mathbf{x}-\mathbf{x}')+i\omega_{\textsc{out}}(t-t')}\,.
\end{align*}
We see that with the choice of the expansion \eqref{E:othoue}, both sides of \eqref{E:bgfkjgsd} will preserve the same spatial dependence $\displaystyle e^{+i\mathbf{k}\cdot(\mathbf{x}-\mathbf{x}')}$.

The density of particle production in this implementation is then
\begin{align}
	\langle\hat{b}_{\mathbf{k}}^{\dagger}\hat{b}_{\mathbf{k}}^{\vphantom{\dagger}}\rangle_{\textsc{in}}&=\langle\Bigl[\alpha_{\mathbf{k}}^{*\vphantom{\dagger}}\,\hat{a}_{\mathbf{k}}^{\dagger}+\beta_{-\mathbf{k}}^{\vphantom{\dagger}}\,\hat{a}_{-\mathbf{k}}^{\vphantom{\dagger}}\Bigr]\Bigl[\alpha_{\mathbf{k}}^{\vphantom{\dagger}}\,\hat{a}_{\mathbf{k}}^{\vphantom{\dagger}}+\beta_{-\mathbf{k}}^{\vphantom{\dagger}*}\,\hat{a}_{-\mathbf{k}}^{\dagger}\Bigr]\rangle_{\textsc{in}}\notag\\
	&=\lvert\alpha_{\mathbf{k}}^{\vphantom{\dagger}}\rvert^{2}\,\langle\hat{a}_{\mathbf{k}}^{\dagger}\hat{a}_{\mathbf{k}}^{\vphantom{\dagger}}\rangle_{\textsc{in}}+\beta_{-\mathbf{k}}^{\vphantom{\dagger}}\alpha_{\mathbf{k}}^{\vphantom{\dagger}}\,\langle\hat{a}_{-\mathbf{k}}^{\vphantom{\dagger}}\hat{a}_{\mathbf{k}}^{\vphantom{\dagger}}\rangle_{\textsc{in}}+\alpha_{\mathbf{k}}^{*\vphantom{\dagger}}\beta_{-\mathbf{k}}^{\vphantom{\dagger}*}\,\langle\hat{a}_{\mathbf{k}}^{\dagger}\hat{a}_{-\mathbf{k}}^{\dagger}\rangle_{\textsc{in}}+\lvert\beta_{-\mathbf{k}}^{\vphantom{\dagger}}\rvert^{2}\,\langle\hat{a}_{-\mathbf{k}}^{\vphantom{\dagger}}\hat{a}_{-\mathbf{k}}^{\dagger}\rangle_{\textsc{in}}\,.
\end{align}
When $\langle\hat{a}_{-\mathbf{k}}^{\vphantom{\dagger}}\hat{a}_{\mathbf{k}}^{\vphantom{\dagger}}\rangle_{\textsc{in}}=0$, we obtain
\begin{align}
	\langle\hat{b}_{\mathbf{k}}^{\dagger}\hat{b}_{\mathbf{k}}^{\vphantom{\dagger}}\rangle_{\textsc{in}}&=\lvert\alpha_{\mathbf{k}}^{\vphantom{\dagger}}\rvert^{2}\,\langle\hat{a}_{\mathbf{k}}^{\dagger}\hat{a}_{\mathbf{k}}^{\vphantom{\dagger}}\rangle_{\textsc{in}}+\lvert\beta_{-\mathbf{k}}^{\vphantom{\dagger}}\rvert^{2}\,\langle\hat{a}_{-\mathbf{k}}^{\vphantom{\dagger}}\hat{a}_{-\mathbf{k}}^{\dagger}\rangle_{\textsc{in}}\notag\\
	&=\Bigl(\lvert\beta_{\mathbf{k}}^{\vphantom{\dagger}}\rvert^{2}+1\Bigr)\,\langle\hat{N}_{\mathbf{k}}^{a\vphantom{\dagger}}\rangle_{\textsc{in}}+\lvert\beta_{-\mathbf{k}}^{\vphantom{\dagger}}\rvert^{2}\,\Bigl(\langle\hat{N}_{-\mathbf{k}}^{a\vphantom{\dagger}}\rangle_{\textsc{in}}+1\Bigr)\,.
\end{align}
A similar result applies to $\langle\hat{b}_{-\mathbf{k}}^{\dagger}\hat{b}_{-\mathbf{k}}^{\vphantom{\dagger}}\rangle_{\textsc{in}}$. Therefore we sum over all modes, we find
\begin{align}
	\sum_{\mathbf{k}>0}\langle\hat{b}_{\mathbf{k}}^{\dagger}\hat{b}_{\mathbf{k}}^{\vphantom{\dagger}}\rangle_{\textsc{in}}+\langle\hat{b}_{-\mathbf{k}}^{\dagger}\hat{b}_{-\mathbf{k}}^{\vphantom{\dagger}}\rangle_{\textsc{in}}=\sum_{\mathbf{k}}2\Bigl(\lvert\beta_{\mathbf{k}}^{\vphantom{\dagger}}\rvert^{2}+\frac{1}{2}\Bigr)\Bigl(\langle\hat{a}_{\mathbf{k}}^{\dagger}\hat{a}_{\mathbf{k}}^{\vphantom{\dagger}}\rangle_{\textsc{in}}+\frac{1}{2}\Bigr)-\frac{1}{2}\,,
\end{align}
so we obtain the usual expression for the total number of particles of the in-state, quantified by the out-number operator.



\begin{thebibliography}{99}
\bibitem{FDT}
	H. B. Callen, and T. A. Welton, {\it Irreversibility and generalized noise}, Phys, Rev. {\bf83}, 34 (1951).

	M. S. Green, {\it Markoff random processes and the statistical mechanics of time‐dependent phenomena, II. Irreversible processes in fluids}, J. Chem. Phys. {\bf22}, 398, (1954).

	R. Kubo, {\it The fluctuation-dissipation theorem}, Rep. Prog. Phys. {\bf 29}, 255 (1966). 

	G. W. Ford, {\it The fluctuation-dissipation theorem}, Contemp. Phys. {\bf 58}, 244 (2017).

\bibitem{qos}
		U. Weiss, 
		{\textsl{Quantum Dissipative Systems, 4th Edition}} (World Scientific, Singapore, 2012).

 		H. P. Breuer, and F. Petruccione, 
		{\textsl{The Theory of Open Quantum Systems, 2nd Edition}} (Oxford University Press, Oxford, 2007).	
		
		A. Rivas, and  S. F. Huelga, 
		{\textsl{Open quantum systems: An Introduction}} (Springer, Berlin, Heidelberg, 2012).
	
\bibitem{LRT}
	A. L. Fetter, and J. D. Walecka, {\sl Quantum Theory of Many-particle Systems}  (Courier Corporation, Dover, 2003).

	S. W. Lovesey, {\sl Condensed Matter Physics: Dynamic Correlations} (Addison-Wesley, Reading, 1986).

\bibitem{QTD1}
	J.-T. Hsiang, C. H. Chou, Y. Suba{\c s}{\i}, and B. L. Hu, {\it Quantum thermodynamics from the nonequilibrium dynamics of open systems: Energy, heat capacity, and the third law},  Phys. Rev. E \textbf{97}, 0125135 (2018).

\bibitem{HHPRD15}  
	J.-T. Hsiang and B. L. Hu, {\it Distance and coupling dependence of entanglement in the presence of a quantum field}, Phys. Rev. D {\bf 92}, 125026 (2015).

\bibitem{HHAoP15}
	J.-T. Hsiang, and B. L. Hu, {\it Nonequilibrium steady state in open quantum systems: influence action, stochastic equation and power balance}, Ann. Phys. \textbf{362}, 139 (2015).

\bibitem{HPZ} 
	B. L. Hu, J. P. Paz, and Y. Zhang, {\it Quantum Brownian motion in a general environment: exact master equation with nonlocal dissipation and colored noise}, Phys. Rev. D {\bf45}, 2843 (1992). 
	
	B. L. Hu, J. P. Paz, and Y. Zhang, \textit{Quantum Brownian motion in a general environment II. Nonlinear coupling and perturbative approach}, Phys. Rev. D \textbf{47}, 1576 (1993).

\bibitem{HalYu96} 
	J. J. Halliwell, and T. Yu. {\it Alternative derivation of the Hu-Paz-Zhang master equation of quantum Brownian motion}, Phys. Rev. D {\bf53}, 2012 (1996).

\bibitem{CRV}
	E. Calzetta, A. Roura, and E. Verdaguer, {\it Stochastic description for open quantum systems}, Physica A {\bf 319}, 188 (2003).

\bibitem{QRad}
	J.-T. Hsiang, and B. L. Hu, {\it Atom-field interaction: From vacuum fluctuations to quantum radiation and quantum dissipation or radiation reaction}, Physics \textbf{1}, 430 (2019).

\bibitem{FlucThm}
	C. Jarzynski, {\it Nonequilibrium equality for free energy differences}, Phys. Rev. Lett. {\bf78}, 2690 (1997). 	
	
	C. Jarzynski, {\it Equilibrium free-energy differences from nonequilibrium measurements: a master-equation approach}, Phys. Rev. E {\bf56}, 5018 (1997). 	
	
	C. Jarzynski, {\it Equalities and inequalities: irreversibility and the second law of thermodynamics at the nanoscale}, Ann. Rev. Cond. Mat. Phys. {\bf2}, 329 (2011).
	
	G. E. Crooks, {\it Entropy production fluctuation theorem and the nonequilibrium work relation for free energy
differences}, Phys. Rev. E {\bf60}, 2721 (1999).

\bibitem{CPR}
	J.-T. Hsiang, B. L. Hu, and S.-Y. Lin, {\it Fluctuation-dissipation and correlation-propagation relations from the nonequilibrium dynamics of detector-quantum field systems},  Phys. Rev. D \textbf{100}, 025019 (2019).
 	
	J.-T. Hsiang, B. L. Hu, S.-Y. Lin, and K. Yamamoto, {\it Fluctuation-dissipation and correlation-propagation relations in (1+3)D moving detector-quantum field systems}, Phys. Lett. B \textbf{795}, 694 (2019).

\bibitem{FDRNL}
	J.-T. Hsiang and B. L. Hu, {\it Fluctuation-dissipation relation from the nonequilibrium dynamics of a nonlinear open quantum system},  Phys. Rev. D {\bf101}, 125003 (2020). 

\bibitem{RHA}
	A. Raval, B. L. Hu, and J. Anglin, \textit{Stochastic theory of accelerated detectors in a quantum field}, Phys. Rev. D \textbf{53}, 7003 (1996).

\bibitem{HuSin95}  
	B. L. Hu, and S. Sinha, {\it Fluctuation-dissipation relation for semiclassical cosmology}, Phys. Rev. D {\bf 51}, 1587 (1995).

\bibitem{CamVer96}  
	A. Campos, and E. Verdaguer, {\it Stochastic semiclassical equations for weakly inhomogeneous cosmologies}, Phys. Rev. D \textbf{53}, 1927 (1996).

\bibitem{HuVer20} 
	E. Calzetta, and B. L. Hu, \textsl{ Nonequilibrium Quantum Field Theory} (Cambridge University Press, Cambridge, 2008).

\bibitem{RHK97}  
	A. Raval, B. L. Hu, and D. Koks, {\it Near-thermal radiation in detectors, mirrors, and black holes: A stochastic approach}, Phys. Rev.  D {\bf55}, 4795 (1997).

\bibitem{Unr76} 
	W. G. Unruh, \textit{Notes on black-hole evaporation}, Phys. Rev. D \textbf{14}, 870 (1976). 

\bibitem{DeW79} 
	B. S. DeWitt,  {\it Quantum gravity : the new synthesis}, in \textsl{General Relativity: An Einstein Centenary Survey}, edited by S. W. Hawking and W. Israel (Cambridge University Press, Cambridge, UK, 1979). 

\bibitem{RQI} 
	http://www.isrqi.net/

\bibitem{CalLeg83}  
	A. O. Caldeira, and A. J. Leggett, {\it Path integral approach to quantum Brownian motion}, Phys. A (Amsterdam) {\bf121}, 587 (1983).

\bibitem{QLE}  
	G. W. Ford, J. T. Lewis, and R. F. O'Connell, {\it Quantum Langevin equation}, Phys. Rev, A, {\bf37}, 4419 (1988).

\bibitem{HM94} 
	B. L. Hu, and A. Matacz, \textit{Quantum Brownian motion in a bath of parametric oscillators: A model for system-field interactions},  Phys. Rev. D \textbf{49}, 6612 (1994).

\bibitem{KMH97}	
	D. Koks, A. Matacz, and B. L. Hu,  {\it Entropy and uncertainty of squeezed quantum open systems}, Phys. Rev. D {\bf55}, 5917 (1997).

\bibitem{Walls} 
	D. F. Walls, {\it Squeezed states of light}, Nature {\bf306}, 141 (1981).

\bibitem{LouKni} 
	R. Loudon, and P. L. Knight. {\it Squeezed light}, J. of Mod. Opt. {\bf34}, 709 (1987).

\bibitem{ManWol} 
	L. Mandel, and E. Wolf, {\sl Optical Coherence and Quantum Optics} (Cambridge University Press, Cambridge, 1995).

\bibitem{NEqFT}
	L. Kadanoff, and G. Baym, {\sl Quantum Statistical Mechanics} (Benjamin, New York, 1962). 

		E. Calzetta, and B. L. Hu, 
		{\textit{Nonequilibrium quantum fields: Closed-time-path effective action, Wigner function and Boltzmann equation}}, Phys. Rev. D \textbf{37}, 2878 (1988).
		
		E. Calzetta, and B. L. Hu, 
		{\textsl{Nonequilibrium Quantum Field Theory}} (Cambridge University Press, Cambridge, 2008). 
	
		J. Rammer, 
		{\textsl{Quantum Field Theory of Non-equilibrium States}} (Cambridge University Press, Cambridge, 2009). 
		
		A. Kamenev, 
		{\textsl{Field Theory of Non-Equilibrium Systems}} (Cambridge University Press, Cambridge, 2011).
		
		J. Berges, 
		{\textit{Nonequilibrium Quantum fields: From cold atoms to cosmology}}, in Lecture Notes of the Les Houches Summer School, Vol. 99, \textsl{Strongly Interacting Quantum Systems out of Equilibrium} (Oxford University Press, Oxford, 2016); 
		{[arxiv:1503.02907]}. 

\bibitem{QOtto}
 	O. Ar{\i}soy, J.-T. Hsiang, and B. L. Hu,  {\it Quantum parametric oscillator heat engine in squeezed thermal baths:  Foundational theoretical issues}, submitted to Phys. Rev. E;  [arXiv:2106.12325]. 

\bibitem{BirDav}  
	N. D. Birrell, and P. C. W. Davies, \textsl{Quantum Fields in Curved Space} (Cambridge University Press, Cambridge, UK, 1982). 


\bibitem{Par69}
	L. Parker, {\it Quantized fields and particle creation in expanding universes. I}, Phys. Rev. {\bf183}, 1057 (1969).

\bibitem{Zel70}  
	Y. B. Zel'dovich, {\it Particle production in cosmology}, Pis'ma Zh. Eksp. Teor. Fiz, \textbf{12}, 443 (1970); [JETP Lett. \textbf{12}, 307 (1970)].
	
	Y. B. Zel'dovich, and A. A. Starobinsky, {\it Particle production and vacuum polarization in an anisotropic gravitational field},
Sov. Phys. JETP {\bf 34}, 1159 (1972).
%
\bibitem{DCE}
	V. V. Dodonov, {\it Fifty years of the dynamical Casimir effect}, Physics {\bf2}, 67 (2020).

\bibitem{UDWcos}
	J.-T. Hsiang, and B. L. Hu,  {NonMarkovianity in cosmology: Memories kept in a quantum field}, Ann. Phys. {\bf434}, 168656 (2021); [arXiv:2107.04862].

\bibitem{NEqFE}
  	J. T. Hsiang, and B. L. Hu,  {\it Nonequilibrium quantum free energy and effective temperature, generating functional and influence action}, Phys. Rev. D {\bf103}, 065001 (2021).
  
\bibitem{GriSid} 
	L. P. Grishchuk, and Y. V. Sidorov, \textit{Squeezed quantum states of relic gravitons and primordial density fluctuations}, Phys. Rev. D \textbf{42}, 3413 (1990).

\bibitem{HKM94} 
	B. L. Hu,  G. Kang, and  A.  Matacz, \textit{Squeezed vacua and the quantum statistics of cosmological particle creation},  Int. J. Mod. Phys. \textbf{A9}, 991 (1994).

\bibitem{Haw74}  
	S. W. Hawking, \textit{Black hole explosions}, Nature \textbf{248}, 30 (1974). 
	
	S. W. Hawking, {\it Particle creation by black holes}, Comm, Math. Phys. {\bf43}, 199 (1975).

\bibitem{Par75}
	L. Parker, {\it Probability distribution of particles created by a black hole}, Phys. Rev. D {\bf12}, 1519 (1975).

\bibitem{Wald75}
	R. M. Wald, {\it On particle creation by black holes}, Comm. Math. Phys., {\bf45}, 9 (1975).

\bibitem{DavFul}
 	S. A. Fulling, and P. C. W.  Davies, \textit{Radiation from a moving mirror in two dimensional space-time: conformal anomaly}, Proc. Roy. Soc.  London \textbf{348}A, 393 (1976).

	P. C. W. Davies, and S. A. Fulling, \textit{Radiation from moving mirrors and from black holes}, Proc. Roy. Soc. London \textbf{356}A, 237 (1977).

\bibitem{CalHu04} 
	E. Calzetta, and B. L. Hu, \textit{Bose-Einstein condensate collapse and dynamical squeezing of vacuum fluctuations}, Phys. Rev. A \textbf{68},  043625 (2003); \textit{Early universe quantum processes in BEC collapse experiments},  Int. J. Theor. Phys. \textbf{44}, 1691  (2005). 

\bibitem{Garay} 
	L. J. Garay, J. R. Anglin, J. I. Cirac, and P. Zoller, \textit{Sonic analog of gravitational black holes in Bose-Einstein condensates}, Phys. Rev. Lett. \textbf{85}, 4643 (2000).

\bibitem{analogG} 
	C. Barcel\'o, S. Liberati, and M. Visser, \textit{Analogue gravity}, Living Rev. Rel. \textbf{14}, 1 (2011).

\bibitem{GibHaw}  	
	G. W. Gibbons, and S. W. Hawking, \textit{Cosmological event horizons, thermodynamics, and particle creation}, Phys. Rev. D \textbf{15}, 2738 (1977).

\bibitem{KHMR} 
	D. Koks, B. L. Hu, A. Matacz, and A. Raval,  {\it Thermal particle creation in cosmological spacetimes: A stochastic approach}, Phys. Rev. D \textbf{56}, 4905 (1997).

\bibitem{NEqUnr}
 	B. L. Hu, and P. R. Johnson, {\it Beyond Unruh effect: Nonequilibrium quantum dynamics of moving charges} in {\sl Quantum Aspects of Beam Physics}, edited by P. Chen. (World-Scientific, Singapore, 2001);  [arXiv:quant-ph/0012132].

\bibitem{QRadCoh}
	J.-T. Hsiang, and B. L. Hu, {\it Quantum radiation and dissipation in relation to classical radiation and radiation reaction}, in preparation for  {\sl Symmetry}, special issue on Accelerated Radiation, edited by S. A. Fulling (2021).  

\bibitem{QRadSq}
	M. Bravo, J.-T. Hsiang, and B. L. Hu, {\it Quantum radiation and dissipation from an atom in a squeezed quantum field}, in preparation for {\sl Atom} (2021).

\bibitem{Ahn}
	D. Ahn, and  M. S. Kim, {\it Hawking–Unruh effect and the entanglement of two-mode squeezed states in Riemannian space–time}, Phys. Lett. A {\bf366}, 202  (2007).

\bibitem{Adesso}
	G. Adesso, and I. Fuentes-Schuller, and M. Ericsson, {\it Continuous-variable entanglement sharing in noninertial frames}, Phys. Rev. A {\bf76}, 062112 (2007).

\bibitem{EntHarv}
	E. Mart\'in-Mart\'inez, and N. C. Menicucci, {\it Cosmological quantum entanglement}, Class. Quantum Grav. {\bf29} 224003 (2012);  {\it Entanglement in curved spacetimes and cosmology}, Class. Quantum Grav. {\bf31}, 214001 (2014). 

 	N. Stritzelberger, L. J. Henderson, V. Baccetti, N. C. Menicucci, and A. Kempf, {\it Entanglement harvesting with coherently delocalized matter}, Phys. Rev. D {\bf103}, 016007 (2021).

\bibitem{MOF}
	C. R. Galley, R. Behunin, and B. L. Hu, {\it Oscillator-field models of  moving mirrors in quantum optomechanics}, Phys. Rev. A {\bf87}, 043832 (2013).  
	
	K. Sinha, S.-Y. Lin, and B. L. Hu,  {\it Mirror-Field entanglement in a microscopic model for quantum optomechanics}, Phys. Rev. A {\bf92}, 023852 (2015).  

\bibitem{HotEnt}
	F. Galve, L. A. Pach\'on, and D. Zueco, {\it Bringing entanglement to the high temperature limit}, Phys. Rev. Lett. {\bf 105}, 180501 (2010).
	
	V. Vedral, {\it Quantum physics: hot entanglement}, Nature {\bf 468}, 769 (2010).

	J. Anders, and A. Winter, {\it Entanglement and separability of quantum harmonic oscillator systems at finite temperature}, Quantum Inf. Comput. {\bf 8}, 0245 (2008).

	J. Anders, {\it Thermal state entanglement in harmonic lattices}, Phys. Rev. A {\bf 77}, 062102 (2008).

	J.-T. Hsiang, and  B. L. Hu, {\it Hot entanglement? --  A nonequilibrium quantum field theory scrutiny}, Phys. Lett. B {\bf 750}, 396 (2015).   
  
  	J.-T. Hsiang, and B. L. Hu,  {\it Quantum entanglement at high temperatures? -- Bosonic systems in nonequilibrium steady state}, JHEP {\bf 11} (2015) 090. 

\bibitem{OL12}
	S. Olivares, \href{https://doi.org/10.1140/epjst/e2012-01532-4}{\textit{Quantum optics in the phase space: A tutorial on Gaussian states}}, Eur. Phys. J. Special Topics, \textbf{203}, 3 (2012).

\bibitem{CalHu87}  
	E. Calzetta, and B. L. Hu, {\it Closed time-path functional formalism in curved spacetime: Application to cosmological back-reaction problems}, Phys. Rev. D {\bf 35}, 495 (1987).

\bibitem{CamVer94} 
	A. Campos, and E. Verdaguer, {\it Semiclassical equations for weakly inhomogeneous cosmologies}, Phys. Rev. D, {\bf49}, 1861 (1994).

\bibitem{CanSci}
	P. Candelas, and D. W. Sciama, {\it Irreversible thermodynamics of black holes},  Phys. Rev. Lett. {\bf 38}, 1372 (1977); Erratum: Phys. Rev. Lett. {\bf 39}, 1640 (1977).
	
	D. W. Sciama, {\it Thermal and quantum Fluctuations in special and general relativity: an Einstein synthesis} in {\sl Centenario di Einstein} (Editrici Giunti Barbera Universitaria, 1979).

\bibitem{Mottola} 
	E. Mottola, {\it Quantum fluctuation-dissipation theorem for general relativity}, Phys. Rev. D {\bf 33}, 2136 (1986).

\bibitem{QFric}
	J. B. Pendry, {\it Shearing the vacuum - quantum friction}, J. Phys. Condens. Matter {\bf9}, 10301 (1997); {\it Quantum friction–fact or fiction?}, New J. of Phys. {\bf12},  033028 (2010).

	A. I. Volokitin, and B. N. J. Persson, {\it Near-field radiative heat transfer and noncontact friction}, Rev. Mod. Phys. {\bf79}, 1291 (2007).

	F. Intravaia, R. O. Behunin, and D. A. Dalvit, {\it Quantum friction and fluctuation theorems}, Phys. Rev. A {\bf89}, 050101 (2014).

	M. B. Far\'ias, C. D. Fosco, F. C. Lombardo, F. D. Mazzitelli, and A. E. Rubio L\'opez, {\it Functional approach to quantum friction: effective action and dissipative force}, Phys. Rev. D {\bf91}, 105020 (2015).

\bibitem{QFricNESS}
	 D. Reiche, F. Intravaia, J.-T. Hsiang, K. Busch, B. L. Hu, {\it Nonequilibrium thermodynamics of quantum friction}, Phys. Rev. A {\bf102}, 050203 (2020).


\end{thebibliography}
\end{document}